\documentclass[11pt,a4paper]{article}
\usepackage{jcappub}
\usepackage{multirow}
\usepackage{graphicx}
\usepackage{amsmath,amsfonts,amsthm}
\usepackage{array}
\usepackage{tcolorbox}
\usepackage{siunitx}
\usepackage{mdframed}
\definecolor{light-gray}{gray}{0.98}
\usepackage[normalem]{ulem}
\usepackage{newcommands}
\DeclareGraphicsExtensions{.png,.pdf}
\graphicspath{{./figs/}{./figs/}{./}}
\subheader{LAPTH-033/25, LUPM:25-001\\
\today}

\title{In-depth analysis of the clustering of dark matter particles around primordial black holes. Part II. Analytical prescriptions for spikes}
\author[a]{Julien Lavalle}
\author[b]{and Pierre Salati}
\affiliation[a]{Laboratoire Univers et Particules de Montpellier (LUPM), Universit\'e de Montpellier \& CNRS, Place Eug\`ene Bataillon, 34095 Montpellier Cedex 05, France}
\affiliation[b]{Laboratoire d'Annecy de Physique Th\'eorique (LAPTh), CNRS, Universit\'e Savoie Mont-Blanc, 9 Chemin de Bellevue, 74940 Annecy, France}
\emailAdd{lavalle@in2p3.fr}
\emailAdd{pierre.salati@lapth.cnrs.fr}

\abstract{Primordial black holes (PBHs) are very appealing dark matter (DM) candidates. It is highly plausible though, should they exist, that they would not make up all of the DM. Several studies showed that if the rest of DM is made of thermal particles, then these should accumulate around such PBHs, leading to the formation of very dense spikes in the radiation era. We contributed a detailed analytical study about this phenomenon, providing clear explanations as for the origin of scaling relations in the form of power-law density profiles with up to 3 different spectral indices, i.e. $3/4$, $3/2$, and $9/4$, and 4 asymptotic regimes.
Here, we further derive an approximate analytical solution that enables fast numerical predictions for the density profiles of these spikes. We also address the specific case of self-annihilating DM species and derive new approximate analytical formulae. Our approximate density yields the correct annihilation rate within $\pm 15\%$ precision.
We then focus on indirect detection in the cosmic microwave background and in extragalactic gamma-rays. We shed new and subtle light on how mutually exclusive PBHs and self-annihilating DM species can really be. In particular, the discovery of a population of sub-solar PBHs would set stringent constraints on the $s$-wave annihilation cross-section of these particles, a point so far missed in the literature.}

\arxivnumber{2511.16800}

\begin{document}
\maketitle
\vskip 1.0cm
\noindent
{\small This is the Accepted Manuscript version of an article accepted for publication in Journal of Cosmology and Astroparticle Physics as JCAP {\bf 05} (2026) 035.
\\
Neither SISSA Medialab Srl nor IOP Publishing Ltd is responsible for any errors or omissions in this version of the manuscript or any version derived from it. The Version of Record is available online at
{\tt https://doi.org/10.1088/1475-7516/2026/05/035}.
}

\newpage
\section{Introduction}
\label{sec:intro}

Primordial black holes (PBHs) have exerted a strong (and not only gravitational) attraction over the years since their formulation by Hawking~\cite{Hawking1971,CarrEtAl1974}, either as peculiar class of a stimulating theoretical object concentrating many open questions in theoretical physics (e.g.~\cite{Hawking1975,DvaliEtAl2020}), or as potential dark matter (DM) candidates \cite{Chapline1975}. After a significant decrease of interest in these objects as serious contenders for DM in the late 90's due to the accumulation of observational constraints (see e.g.~\cite{CarrEtAl2021,GreenEtAl2021} for reviews), in particular from microlensing searches (e.g.~\cite{TisserandEtAl2007,Green2016,PetacEtAl2022}), they gained popularity again after the discovery of gravitational waves (GW) originating in black-hole merger events rather aside from the typical mass range expected at the time \cite{BirdEtAl2016,SasakiEtAl2016}. Even though GW observations have turned into constraints since then \cite{RaidalEtAl2024,HuetsiEtAl2021}, part of this renewed craze is also somewhat related to the so far unsuccessful searches for new particles beyond the standard model at colliders or with other experimental or observational strategies, although the latter do not preclude at all their existence at the moment \cite{ArcadiEtAl2025,BernalEtAl2017,DiLuzioEtAl2020}.

\vskip 0.1cm
Even though PBH production mechanisms are usually rather fine-tuned \cite{ColeEtAl2023}, and accumulated constraints leave only a small mass window for PBHs to be all of the DM \cite{CarrEtAl2021,GreenEtAl2021}, contemplating the possibility that PBHs could still make a fraction of the DM is both exciting and legitimate, and actually turns out to be a very interesting theoretical question. Expressing it simply, to what extent can PBHs and other DM species co-exist? Indeed, even if subdominant in the matter content of the Universe, their co-existence as DM partners could lead to interesting phenomena, and might actually not be always possible. The role of PBHs as DM accreting objects and potential seeds of highly concentrated dark halos was envisaged in several early studies \cite{Ricotti2007,MackEtAl2007,RicottiEtAl2008a}, leading in the meantime to constraints if the other DM species (assumed unique for simplicity) would be self-annihilating particles \cite{LackiEtAl2010}---and later to studies of other potential signals through modifications of the waveforms expected in GW originating in black hole (BH) merger events \cite{BarausseEtAl2014,KavanaghEtAl2020}. In all of these studies, the details of the DM density profiles accumulated around BHs is key, and so is the reliability in their determination.

\vskip 0.1cm
A big step forward was made by Eroshenko \cite{Eroshenko2016} who realized that if the partner DM species is made of thermally produced particles that reach thermal equilibrium before kinetically decoupling from the plasma, then DM accretion could actually be determined on very precise footings by numerically integrating all particle orbits during the radiation domination era from kinetic decoupling on, allowing for exact (numerical) solutions in the description of this process. This very elegant method relies on two important facts: (i) thermal equilibrium fixes the initial conditions, and (ii) the local dynamics close to the BH is driven by the BH mass only\footnote{Assumption (ii) is discussed at the end of section~\ref{subsec:phase_diagram} and justified in appendix~\ref{append:M_halo_vs_M_BH}.}.
These solutions were found to be scale-invariant power-law density profiles $\propto r^{-\gamma}$, dubbed DM spikes owing to the impressively large DM densities accumulated around PBHs at such early times, further exhibiting a sharp decrease as a function of radius $r$. In a previous paper \cite{BoudaudEtAl2021}, we actually contributed a complete analytical derivation and physical understanding of the three power-law indices allowed for this accretion process, namely $\gamma=3/4$, $3/2$, and $9/4$, which are fully determined by the main input parameters of the model configuration. We also corrected for a misconception in the angular part of the integral that propagated in the literature, leading to potentially large numerical errors.

\vskip 0.1cm
The prediction of spikes in the mixed PBH-particle DM scenario is therefore fully established from theoretical grounds, and the corresponding density profiles can be determined very accurately insofar as the dynamics are fully controlled by the mass of the central PBH---a genuine linear problem as this holds. Simulations were carried out \cite{AdamekEtAl2019}, though in a very limited part of the available parameter space, which confirmed the trend of early numerical results. We note that non-linearities are still expected to manifest (i) in the outskirts of the spike region after secondary DM infall occurs during the matter domination era, i.e. when the accreted DM mass gets significant enough to backreact on itself, and (ii) in the case of self-annihilation in the central spike region. In the latter case, analytical solutions to such non-linearity can still be found if annihilation proceeds through $s$-wave processes \cite{Vasiliev:2007vh,Shapiro:2016ypb}, as will be discussed later in this paper. In any case, an important question raised by this mixed DM scenario in which the partner DM species can self-annihilate is whether or not it is allowed at all by current observations. After some early work based on approximate scaling relations \cite{RicottiEtAl2008a,LackiEtAl2010}, several studies using more or less controlled versions of Eroshenko's method have derived observational constraints inferred from extragalactic gamma-rays \cite{BoucennaEtAl2018,CarrEtAl2021a,ChandaEtAl2024} and from the cosmic microwave background (CMB) measurements \cite{GinesEtAl2022}. In this paper, we wish to revisit the way these constraints can be extracted on a more pedestrian and contemplative pace. Our goal is actually twofold: first, we want to revisit the relevant calculations in detail, providing analytical insights to back up both our understanding and numerical results; second, we want to come up with useful analytical approximations that would allow the interested reader to easily get numerical results correct up to a reasonably good precision, specializing in signal predictions for indirect DM searches in the extragalactic gamma-rays and in the CMB.

\vskip 0.1cm
The paper develops as follows. In section~\ref{sec:DM_profiles}, we first shortly revisit the scaling regimes featuring the postcollapse spike profiles, summarizing results we already derived in \cite{BoudaudEtAl2021} but with a renewed perspective focused on the present work and aimed at justifying our analytical approximations. Next, in section~\ref{sec:Gamma_BH}, we enter the discussion of the spike annihilation rate. We carefully design an analytical approximation scheme, and derive our main results which are presented in the form of approximate analytical formulae. Then, in section~\ref{sec:constraints}, we turn to a review of the general methodology employed to derive observational constraints using both extragalactic gamma-rays and CMB measurements as our main observables. We pay a particular attention to the notion of ``mutual exclusiveness'' when discussing the range of parameter space where PBHs and self-annihilating DM would be either allowed or forbidden, and point to subtleties that were not clearly identified in previous works. In particular, we show that the discovery by GW observatories of a population of sub-solar PBHs would set stringent constraints on the $s$-wave annihilation cross-section of thermal DM species, a point so far missed in the literature.
We emphasize that section~\ref{sec:constraints} is more about the general methodology, and not meant to derive accurate constraints through a rigorous statistical treatment of the data. We leave this to future work. Finally, we conclude in section~\ref{sec:conclusions}.

\section{Dark matter post-collapse profiles}
\label{sec:DM_profiles}

%
In our calculation of the DM post-collapse density profile, we consider the three following conditions for WIMP capture by a PBH:\\
{\bf (i)} feeling the gravitational pull of a neighbor PBH,\\
{\bf (ii)} being free streaming through the ambient plasma, non-relativistic, and collisionless,\\
{\bf (iii)} having a velocity smaller than the local escape velocity.\\
These assumptions are reasonable, although alternative mechanisms for particle capture within a spike may exist, although subdominant.

\vskip 0.1cm
To commence, DM particles must lie inside the so-called sphere of influence of the object, where they can feel its gravitational pull. In the early universe, space is actually filled with a dense plasma inside which the attraction of a PBH is overcome by the overall expansion beyond a critical distance $r_{\rm inf}$ identified with the radius of that sphere of influence. The relation between $r_{\rm inf}$ and the black hole mass $M_{\rm BH}$ takes the form\footnote{See also relation~(2.11) in ~\cite{BoudaudEtAl2021}.}~\cite{AdamekEtAl2019}
 \begin{equation}
M_{\rm BH} = \frac{16 \pi}{3 {\eta_{\rm ta}}} \, r_{\rm inf}^{3} \, \rho_{\rm tot} \,,
\label{eq:definition_r_inf_a}
\end{equation}
where the coefficient $\eta_{\rm ta}$ has been found equal to $1.086$ in~\cite{BoudaudEtAl2021}. As time goes on, the universe cools down, the energy density $\rho_{\rm tot}$ of the primordial plasma drops and the radius of influence $r_{\rm inf}$ increases.
Should they feel the PBH attraction, DM species must also move freely inside the ambient medium.
At early times, they collide onto the constituents of the primordial plasma. Outside the sphere of influence, they are pulled away from the black hole by the overall expansion. Within that region, they mostly feel the black hole attraction. As long as kinetic decoupling has not occurred, they frequently collide onto the plasma constituents. This generates a drag force that prevents the particles from falling onto the PBH. A stable spike cannot form in these conditions.
Inside the sphere of influence, the plasma also feels mostly the PBH attraction but expands insofar as its constituents are ultra-relativistic. This may not be correct very close to the Schwarzschild radius though.
Kinetic decoupling occurs when the exchange of energy between DM and ambient plasma becomes slower than the expansion itself. The plasma becomes transparent to the DM species. These stream freely afterwards and move unimpeded. Thermalization between the radiation bath and DM breaks at kinetic decoupling, leaving WIMP momenta decreasing in time as they only undergo redshift.
The last condition bears on DM velocities which should not exceed, at injection, the local escape velocity from the PBH.

\vskip 0.1cm
During the radiation dominated era, DM skirts start building up around PBHs as soon as DM undergoes kinetic decoupling. This occurs at plasma temperature $T_{\rm kd}$. We define the kinetic decoupling parameter $x_{\rm kd}$ as the ratio ${m_{\chi}}/{T_{\rm kd}}$, with $m_{\chi}$ the WIMP mass. At kinetic decoupling, the particles lying inside the sphere of influence of a PBH start falling onto it. As time passes on, layers lying farther away from the PBH start falling in turn, hence an onion like structure of the pre-collapse DM distribution whose profile is well approximated by\footnote{See also relation~(2.21) in ~\cite{BoudaudEtAl2021}.}
\begin{equation}
\rho_{i}({r}_{i}) \simeq \left\{
\begin{tabular}{ll}
$\rho_{i}^{\rm kd}$ & if ${r}_{i} \le {r}_{\rm kd} \,,$\\
$\rho_{i}^{\rm kd} \left( {{r}_{i}}/{{r}_{\rm kd}} \right)^{-9/4}$ & if ${r}_{\rm kd} \le {r}_{i} \le {r}_{\rm eq} \,.$
\end{tabular}
\label{eq:pre_collapse}
\right.
\end{equation}
At kinetic decoupling, the cosmological DM density is $\rho_{i}^{\rm kd}$ while the radius of the sphere of influence is ${r}_{\rm kd}$. At matter-radiation equality, that radius is ${r}_{\rm eq}$ and the accretion of DM species stops to be considered as the radiation era comes to an end.

\vskip 0.1cm
The pre-collapse DM velocities $\beta_{i}$ follow a Maxwellian distribution ${\cal F}_{\rm MB} \! \left( \beta_{i} | {r}_{i} \right)$ with dispersion velocity $\sigma_{i}$.
Notice that this may not be true very close to the black hole horizon, within the inner few Schwarzschild radii. In this narrow region, the plasma is heated by the ongoing accretion of matter onto the PBH, and the dispersion velocity of DM species may be larger than $\sigma_{i}$. Studying the behavior of the plasma and the DM species which it contains requires a hydrodynamical analysis which is beyond the scope of this work. Anyway, the region involved is so small that it does not contribute to the annihilation rate and does not affect the observational bounds derived in section~\ref{sec:constraints}.
Particles injected at radii ${r}_{i} \leq {r}_{\rm kd}$ have the 1D dispersion velocity of DM at kinetic decoupling, i.e. $\sigma_{\rm kd} = {x_{\rm kd}}^{-1/2}$. Those trapped farther away at radius ${r}_{i} > {r}_{\rm kd}$ start orbiting the PBH at a later cosmic time $t_{i}$. After kinetic decoupling occurred, their dispersion velocity $\sigma_{i}$ has been redshifted, hence a unique relation between $\sigma_{i}$, $t_{i}$ and eventually ${r}_{i}$ as discussed in~\cite{BoudaudEtAl2021}. According to Liouville's theorem, $\sigma_{i}$ scales with ${r}_{i}$ like ${\rho_{i}}^{1/3}$. At matter-radiation equality, the DM dispersion velocity has decreased down to $\sigma_{\rm \! eq}$.

\vskip 0.1cm
The initial state of a WIMP that feels for the first time the gravitational pull of a PBH is specified by the radius $r_{i}$, the velocity $\beta_{i}$ and the angle $\theta_{i} = ( \vec{\beta}_{i} , \vec{r}_{i} )$ between the injection and radial directions. From now on, it is convenient to define the reduced radius $\tilde{r}_{i}$ as the physical radius $r_{i}$ expressed in units of the PBH Schwarzschild radius $r_{\rm S} = {2 G M_{\rm BH}}/{c^{2}}$. Note also that velocities are expressed in units of the speed of light $c$.
The particle is trapped provided that the parameter $u \equiv \beta_{i}^{2} \tilde{r}_{i}$, describing the ratio of kinetic to potential energy, is less than $1$.
It reaches the radius $\tilde{r}$ if $u \geq 1 - X$, where $X$ stands for the ratio ${\tilde{r}_{i}}/{\tilde{r}}$, i.e. the destination potential to initial potential ratio. In the extreme case where $u = 1 - X$, the destination radius $\tilde{r}$ is the apocenter of a purely radial trajectory.
If both conditions are met, energy and angular momentum conservation implies that the radial velocity $\beta_{r}$ at radius $\tilde{r}$ fulfills the equation
\begin{equation}
\sin^{2} \theta_{i} \, + \, \left\{ \frac{\tilde{r}^{2}}{\tilde{r}_{i}^{2} \beta_{i}^{2}} \right\} \beta_{r}^{2} =
\frac{\tilde{r}^{2}}{\tilde{r}_{i}^{2}}
\left\{
1 + \frac{1}{\beta_{i}^{2}} \left( \frac{1}{\tilde{r}} - \frac{1}{\tilde{r}_{i}} \right)
\right\} \equiv 1 - {\cal Y}_{\rm m} \,.
\end{equation}
This relation is a definition for the quantity ${\cal Y}_{\rm m}$ which plays an important role in the post-collapse DM profile. For obscure reasons, ${\cal Y}_{\rm m}$ has been constrained to be only positive in the past literature. Allowing it to be also negative, as it should, is not only the correct way to define it, but is actually paramount to avoid large numerical errors~\cite{BoudaudEtAl2021}.

\vskip 0.1cm
The contribution of each WIMP to the minispike density inside the layer of radius $\tilde{r}$ and thickness $\mathrm{d} \tilde{r}$ is proportional to the fraction ${2 \mathrm{d}t}/{T_{\rm orb}}$ of the orbital time spent inside it. The duration $\mathrm{d}t$ depends on the radial velocity $\beta_{r}$, which depends itself on the injection angle $\theta_{i}$ and on the parameter ${\cal Y}_{\rm m}$. The orbital period $T_{\rm orb}$ follows Kepler's third law of planetary motion. At fixed $\tilde{r}_{i}$ and $\beta_{i}$, it does not depend on $\theta_{i}$ since it depends only on total energy, not on angular momentum. Equipped with all these notations, the post-collapse DM density at radius $\tilde{r}$ can be readily derived as the integral over the initial phase space\footnote{See also relation~(3.28) in ~\cite{BoudaudEtAl2021}.}
\begin{equation}
\rho(\tilde{r}) = \frac{4}{\tilde{r}} \iint
\tilde{r}_{i\,} \mathrm{d}\tilde{r}_{i\,} \rho_{i}(\tilde{r}_{i}) \times \mathrm{d}\beta_{i}^{2}\,{\cal F}_{\rm MB} \! \left( \beta_{i} | {r}_{i} \right) \times
\left\{ \frac{1}{\tilde{r}_{i}} - \beta_{i}^{2} \right\}^{3/2} \!\!\! \times
{\cal J}({\cal Y}_{\rm m}) \,,
\label{eq:post_collapse_density}
\end{equation}
where the function ${\cal J}({\cal Y}_{\rm m})$ is defined as the angular integral
\beq
{\cal J}({\cal Y}_{\rm m}) = 
{\int_{0}^{\theta_{i}^{0}}} \! \frac{\mathrm{d}(-\cos\theta_{i})}{\sqrt{\cos^{2}\!\theta_{i} - {\cal Y}_{\rm m}}} \,.
\label{eq:integral_cal_J}
\eeq
In the previous expression, the upper bound $\theta_{i}^{0}$ is equal to ${\rm arccos}(\sqrt{{\cal Y}_{\rm m}})$ if ${\cal Y}_{\rm m} \geq 0$, and to ${\pi}/{2}$ if ${\cal Y}_{\rm m} \leq 0$. A straighforward calculation yields
\beq
{\cal J}({\cal Y}_{\rm m}) \! = \! \ln \left( 1 + \sqrt{1 - {\cal Y}_{\rm m}} \right) \, - \, \frac{1}{2} \ln \left| {\cal Y}_{\rm m} \right| \,.
\label{eq:definition_cal_J}
\eeq
The lower bound $\theta_{i} = 0$ of integral~(\ref{eq:integral_cal_J}) corresponds to radial trajectories, i.e. very ellipitical orbits whose pericenters lie close to the BH. Since our approach is developped in the framework of Newtonian mechanics, such orbits are allowed. But in the framework of general relativity, some of these trajectories cross the BH horizon. The DM particles following these get trapped by the central object and disappear. This effect, which should be negligible when the species are injected far from the BH, may be significant close to the Schwarzschild radius. We defer to a future work its analysis. Insofar as subsequent DM annihilation is eventually expected to erase the post-collapse density profile, we do not expect in fine a significant effect.

\vskip 0.1cm
In spite of its simplicity, the numerical integration of relation~(\ref{eq:post_collapse_density}) turns out to be tricky. The angular integral ${\cal J}$ diverges logarithmically when ${\cal Y}_{\rm m} = 0$. The Gauss-Legendre method is by far the most stable way to compute the post-collapse DM density $\rho(\tilde{r})$ but is time consuming. That is why we have devised a fast and accurate way to get $\rho(\tilde{r})$, enabling further studies of the cosmological and astrophysical implications of DM minispikes around PBHs to be rapidly performed.

\subsection{Asymptotic regimes}
\label{subsec:slopes}

As featured in Fig.~4 of \cite{BoudaudEtAl2021}, the post-collapse DM profiles exhibit a rich diversity of behaviors depending on radius $\tilde{r}$ and PBH mass $M_{\rm BH}$. The minispikes extend from the Schwarzschild radius $\tilde{r}_{\rm S} \equiv 1$ up to the radius of influence at matter-radiation equality $\tilde{r}_{\rm eq} \propto M_{\rm BH}^{-2/3}$.
The DM profiles are power laws $\rho(\tilde{r}) \propto \tilde{r}^{- \gamma}$ where the slope $\gamma$ can take three distinct values, i.e. $\gamma = 3/4$, $3/2$ and $9/4$. As showed in the analysis presented in~\cite{BoudaudEtAl2021}, four distinct asymptotic behaviors can be identified, on which we will build our approximation. For each of these regimes, the expression of the DM profile is very simple, and involves a specific integral ${\cal I}^{\rm asy}$ running over a portion of the $(X , u)$ phase space. As mentioned above, WIMP injection is described by the parameters $u \equiv \beta_{i}^{2} \tilde{r}_{i}$ and $X \equiv {\tilde{r}_{i}}/{\tilde{r}}$, while the injection angle $\theta_{i}$ is integrated out in function ${\cal J}$ whose argument ${\cal Y}_{\rm m}$ can be expressed in terms of $X$ and $u$ as
\beq
{\cal Y}_{\rm m} = 1 - \frac{1}{uX} + \frac{1}{X^{2}} \! \left( \frac{1}{u} - 1 \right) .
\eeq

\vskip 0.1cm
\noindent $\bullet$ {\bf The caustic regime $\gamma = 3/4$}\\
This regime dominates close to the PBH, for radii $\tilde{r}$ much smaller than the kinetic decoupling parameter $x_{\rm kd}$. The dominant contribution to the minispike density comes from the accumulation of orbits around those whose periastrons and apoastrons are respectively located at $\tilde{r}$ and $\tilde{r}_{i}$, and for which the angular integral ${\cal J}$ diverges. These trajectories are piled up on $\tilde{r}$ where they create a caustic. The slope $3/4$ translates therefore some kind of resonant behavior where the DM density at $\tilde{r}$ is built essentially from trajectories for which this radius behaves like an attractor.
The post-collapse DM density can be expressed as
\beq
\rho_{3/4}(\tilde{r}) =
\frac{{\cal A}_{3/4}}{\tilde{r}^{3/4}}
\;\;\;\text{with}\;\;\;
{\cal A}_{3/4} = \sqrt{\frac{2}{\pi^{3}}} \, \dfrac{\rho_{i}^{\rm kd}}{\sigma_{\rm kd}^{3/2}} \, \zeta^{3/4} \, {\cal I}_{3/4}^{\rm asy} \,,
\label{eq:rho_asymptotic_slope_3_4}
\eeq
where $\rho_{i}^{\rm kd}$ and $\sigma_{\rm kd}$ stand respectively for the cosmological density and dispersion velocity of DM at kinetic decoupling. The parameter $\zeta$ is equal to $2 \, \Gamma(7/4)^{4/3} \simeq 1.78713$
while the phase space integral ${\cal I}_{3/4}^{\rm asy}$ is defined as
\beq
{\cal I}_{3/4}^{\rm asy} =  \frac{2}{3}
\int_{- \infty}^{1} \mathrm{d}{\cal Y}_{\rm m} \, \frac{{\cal J}({\cal Y}_{\rm m})}{(1 - {\cal Y}_{\rm m})^{5/4}} \simeq 4.18879 \,.
\eeq
The coefficient ${\cal A}_{3/4}$ does not depend on $M_{\rm BH}$, hence a universal behavior of the DM density once expressed as a function of the reduced radius $\tilde{r}$.

\vskip 0.1cm
\noindent $\bullet$ {\bf The Keplerian regime $\gamma = 3/2$}\\
For light PBHs and large radii $\tilde{r}$, WIMPs are injected with velocities large with respect to the local escape speed at $\tilde{r}_{i}$. As a consequence, the DM species that are captured have a flat distribution in velocity space. The integral~(\ref{eq:post_collapse_density}) over phase space simplifies into
\beq
\rho_{3/2}(\tilde{r}) = \frac{{\cal A}_{3/2}}{\tilde{r}^{3/2}}
\;\;\;\text{where}\;\;\;
{\cal A}_{3/2} = \sqrt{\frac{2}{\pi^{3}}} \, \dfrac{\rho_{i}^{\rm kd}}{\sigma_{\rm kd}^{3}} \, {\cal I}_{3/2}^{\rm asy} \,.
\label{eq:rho_asymptotic_slope_3_2}
\eeq
The phase space integral ${\cal I}_{3/2}^{\rm asy}$ can be expressed as
\beq
{\cal I}_{3/2}^{\rm asy} = \int_{0}^{+\infty} \!\! {\dfrac{\mathrm{d}X}{X^{3/2}}}
\int_{u_{\rm inf}}^{1} \! \mathrm{d}u \; (1 - u)^{3/2} \, {\cal J} \simeq
1.04720 \,,
\label{eq:I_3_2_total_a}
\eeq
where $u_{\rm inf} = {\rm max}\{ 0 , (1-X) \}$.
In this regime, the only scale dependence is set by the orbital period $T_{\rm orb}$. The post-collapse density $\rho(\tilde{r})$ scales like ${1}/{T_{\rm orb}}$, which is in turn proportional to $\tilde{r}^{-3/2}$ according to Kepler's third law of celestial mechanics, hence the slope $\gamma = 3/2$.
Notice that the coefficient ${\cal A}_{3/2}$ does not depend on $M_{\rm BH}$, as in the $3/4$ caustic regime. The post-collapse DM profile $\rho(\tilde{r})$ is also universal.

\vskip 0.1cm
\noindent $\bullet$ {\bf Radial infall starting at kinetic decoupling $\gamma = 3/2$}\\
For the heaviest PBHs, i.e. above the critical value $M_{2}$ defined hereafter, and at intermediate radii $\tilde{r}$ smaller than the radius of influence $\tilde{r}_{\rm kd}$ at kinetic decoupling, the density profile also follows a slope $3/2$, although for completely different reasons. In this regime of heavy black hole masses, the WIMP dispersion velocity at injection is everywhere smaller than the local escape speed. The post-collapse density results from the radial infall of the DM population starting already at kinetic decoupling. In this limit, we infer the power law
\beq
\rho'_{3/2}(\tilde{r}) = \frac{{\cal A}_{3/2}^{\prime}}{\tilde{r}^{3/2}}
\;\;\;\text{where}\;\;\;
{\cal A}_{3/2}^{\prime} = \sqrt{\frac{2}{\pi^{3}}} \, \rho_{i}^{\rm kd} \, \tilde{r}_{\rm kd}^{3/2} \, {\cal I}_{3/2}^{\rm asy \, \prime} \,.
\label{eq:rho_asymptotic_slope_3_2_prime}
\eeq
The phase space integral ${\cal I}_{3/2}^{\rm asy \, \prime}$ is showed in~\cite{BoudaudEtAl2021} to be well approximated by ${2} \sqrt{2 \pi} \simeq 5.01326$.
In this regime, the profile is no longer universal. The coefficient ${\cal A}_{3/2}^{\prime}$ depends actually on PBH mass through $\tilde{r}_{\rm kd}$, and scales like ${1}/{M_{\rm BH}}$.

\vskip 0.1cm
\noindent $\bullet$ {\bf Radial infall starting after kinetic decoupling $\gamma = 9/4$}\\
For PBHs with masses $M_{\rm BH}$ larger than the critical value $M_{1}$ defined hereafter, the dispersion velocity of DM at injection can be smaller than the local escape speed in the outer regions of the minispike, \ie~for radii $\tilde{r}$ larger than the radius of influence $\tilde{r}_{\rm kd}$ at kinetic decoupling. The post-collapse density at $\tilde{r}$ results in that case from the radial infall of the successive layers of the onion-like initial distribution located above $\tilde{r}$. The DM profile is given by
\beq
\rho_{9/4}(\tilde{r}) = \sqrt{\frac{2}{\pi^{3}}} \; \rho_{i}(\tilde{r}) \, {\cal I}_{9/4}^{\rm asy} \,.
\label{eq:rho_slope_9_4}
\eeq
In this expression, $\rho_{i}(\tilde{r})$ is the cosmological pre-collapse DM density at the precise moment when $\tilde{r}$ becomes the radius of influence of the PBH. Neglecting the evolution in time of the thermodynamical coefficients $g_{\rm eff}$ and $h_{\rm eff}$ that come into play in the calculation of the effective number of degrees of freedom of the energy and entropy densities of the primordial plasma, we can approximate $\rho_{i}(\tilde{r})$ by the second relation of Eq.~(\ref{eq:pre_collapse}) which yields the scaling ${\tilde{r}}^{-9/4}$ and the slope $9/4$.
In appendix~\ref{append:rho_i_vs_tilde_r_0}, we go beyond this approximation and derive the exact relation between $\rho_{i}$ and $\tilde{r}$.
The phase space integral is equal to
\beq
{\cal I}_{9/4}^{\rm asy} = \sqrt{2 \pi}
\left\{ \int_{1}^{+ \infty} \frac{\mathrm{d}X}{X^{5/4}} \frac{1}{\sqrt{X - 1}} \equiv B(3/4 , 1/2) \! \right\} =
\frac{8_{\,}\pi^{2}}{\Gamma(1/4)^{2}} \simeq 6.00658 \,.
\eeq
%
In the previous expression, the notation $B(p , q)$ refers to the beta function, also called the Euler integral of the first kind.
At fixed $\tilde{r}$, the post-collapse DM density $\rho(\tilde{r})$ scales approximately like $\tilde{r}_{\rm kd}^{9/4} \propto {1}/{M_{\rm BH}^{3/2}}$. Like in the previous regime, profiles are no longer universal.

\vskip 0.1cm
Two critical values for the PBH mass $M_{\rm BH}$ can be defined, as briefly mentioned in the previous review of the various asymptotic regimes.
The mass $M_{1}$ is the maximal value below which the DM average velocity $\bar{\beta}_{i}$ at injection is everywhere {\it larger} than the local escape speed from the radius of influence $\tilde{r}_{i}$.
The mass $M_{2}$ is the exact opposite. It corresponds to the minimal value of $M_{\rm BH}$ above which the DM average velocity $\bar{\beta}_{i}$ at injection is everywhere {\it smaller} than the local escape speed from the radius $\tilde{r}_{i}$.
Quantitative definitions of $M_{1}$ and $M_{2}$ have already been given in~\cite{BoudaudEtAl2021}. In this article, we adopt a more phenomenological point of view and derive these critical values from the phase diagram in section~\ref{subsec:phase_diagram}.

\vskip 0.1cm
The post-collapse DM profile $\rho(\tilde{r})$ depends on how the PBH mass $M_{\rm BH}$ compares with $M_{1}$ and $M_{2}$, hence three mass domains.

\vskip 0.1cm
\noindent {\bf (i)} Very light PBHs -- $M_{\rm BH} < M_{1}$.
The inner regions are dominated by a universal profile with slope $3/4$ (caustic regime) while the outer regions follow a universal profile with slope $3/2$ (Keplerian regime).

\vskip 0.1cm
\noindent {\bf (ii)} Heavy PBHs -- $M_{1} < M_{\rm BH} < M_{2}$.
The behavior of the DM density in the inner regions of the minispike is the same as before, while the outer regions undergo radial infall and follow a non-universal profile with slope $9/4$. The heavier the PBH, the thicker the region dominated by radial infall.

\vskip 0.1cm
\noindent {\bf (iii)} Heaviest PBHs -- $M_{2} < M_{\rm BH}$.
Black holes are so heavy that radial infall occurs as early as kinetic decoupling and implies regions located well below the radius $\tilde{r}_{\rm kd}$. Very small radii are still dominated by the caustic regime with slope $3/4$. As $\tilde{r}$ increases, the minispike profile departs from this universal behavior and follows the radial infall regimes with slopes $3/2$ below $\tilde{r}_{\rm kd}$ and $9/4$ above.

\subsection{Construction of the phase diagram}
\label{subsec:phase_diagram}

In this article, we give analytical prescriptions for a fast derivation of the post-collapse DM density profile $\rho(\tilde{r})$. We build an approximation from the asymptotic behaviors reviewed above. An efficient calculation requires to determine rapidly which of these regimes applies to the particular values of PBH mass and radius under scrutiny.
This is where the phase diagram comes into play. In this section, we construct a map in the $(\tilde{r} , M_{\rm BH})$ plane where we delineate the various domains inside which each asymptotic regime dominates. Such a map will also allow us to get a better understanding of the minispike behavior once DM self-annihilations are taken into account in section~\ref{sec:Gamma_BH}.

\begin{figure}[t!]
\centering
\includegraphics[width=0.70\textwidth]{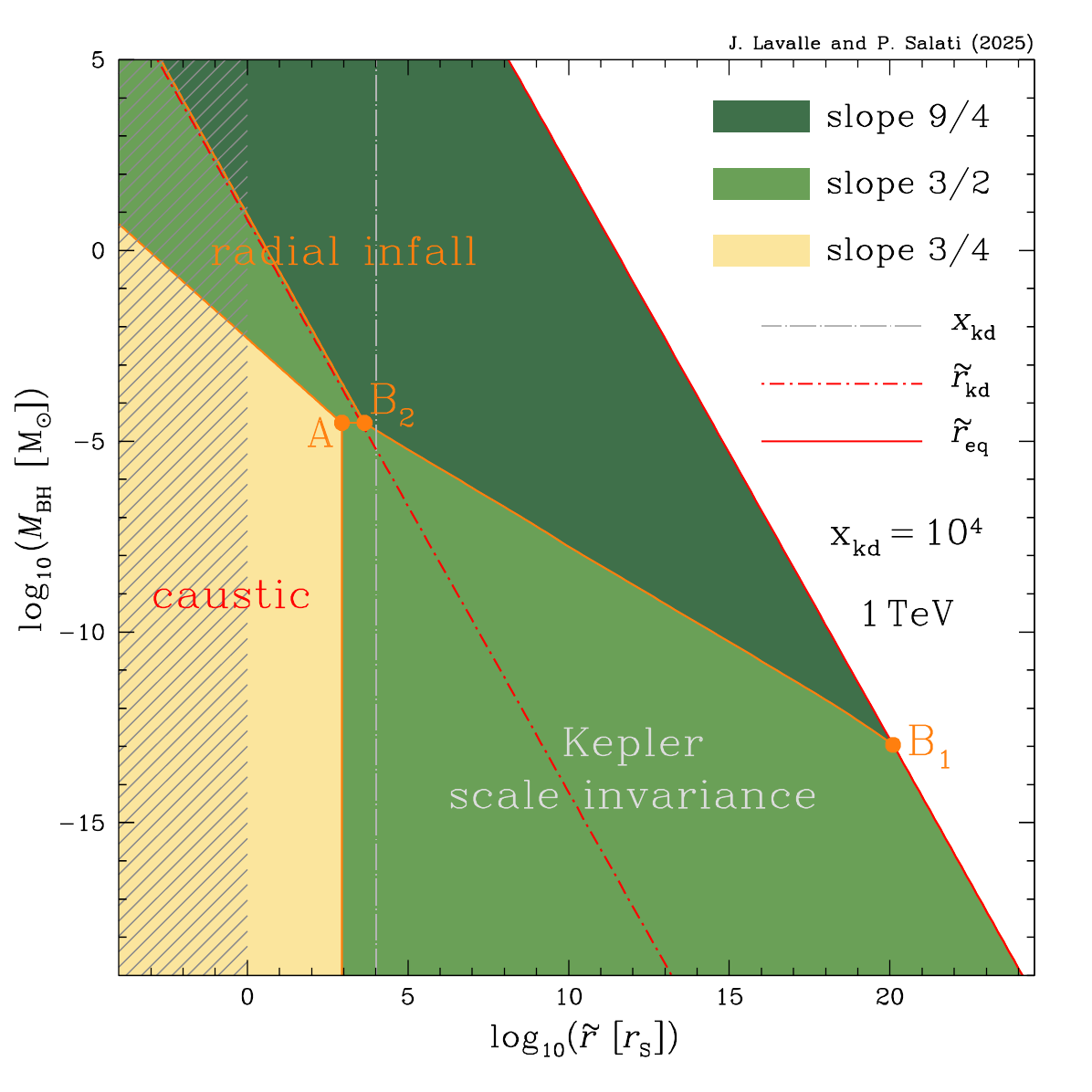}
\caption{The post-collapse DM density follows a power law whose radial index depends on radius~$\tilde{r}$ and PBH mass $M_{\rm BH}$. The regions of prevalence for each particular index are plotted in this phase diagram with different colors. From the center of the DM halo to its outskirts, the slope is respectively equal to $3/4$, $3/2$ and, for heavy objects, $9/4$. This phase diagram corresponds to a DM mass $m_{\chi}$ of 1~TeV and a thermal decoupling parameter $x_{\rm kd}$ of $10^{4}$. The shaded gray vertical band lies below the Schwarzschild radius.
\label{fig:phase_diagram}}
\end{figure}

\vskip 0.1cm
The formation of minispikes around PBHs starts at WIMP kinetic decoupling and goes on until matter-radiation equality, after which the process is no longer considered.
After equality, the accreted DM masses could actually grow beyond the PBH seed masses, such that non-linearities would be expected in the building up of these outer halos which could no longer be described by our analytical calculation. However, the very central parts of the spikes that we accurately describe are so dense that they can hardly be affected by this external dynamics.
Accordingly, in the $(\tilde{r} , M_{\rm BH})$ plane of Fig.~\ref{fig:phase_diagram}, the phase diagram extends rightward up to the radius of influence $\tilde{r}_{\rm eq}$ at matter-radiation equality. This radius represents the outer boundary of minispikes and corresponds to the solid red line of the plot.
Borrowing from the Planck collaboration~\cite{AghanimEtAl2020} the cosmological parameters $\Omega_{\rm b}h^{2} = 0.02237$, $\Omega_{\rm dm}h^{2} = 0.1200$ and $z_{\rm eq}$ = 3,402 while using a cosmic microwave background temperature of $2.72548 \; {\rm K}$~\cite{Fixsen2009}, we can use relation~(\ref{eq:definition_r_inf_a}) to get
\beq
\tilde{r}_{\rm eq}(m) = \tilde{r}_{\rm eq}^{0} \, m^{-2/3}
\;\;\;\text{where}\;\;\;
m \equiv {M_{\rm BH}}/{{\rm M}_{\odot}} \,.
\label{eq:definition_r_tilde_eq}
\eeq
At matter-radiation equality, the reduced radius of influence of a $1 \; {\rm M}_{\odot}$ black hole is found to be $ \tilde{r}_{\rm eq}^{0} = 2.87 \times 10^{11}$.
Since the Schwarzschild radius of such an object is $r_{\rm S}^{0} = 2.95 \; {\rm km}$, the corresponding physical radius $r_{\rm eq}^{0}$ is equal to $8.49 \times 10^{11} \; {\rm km} \simeq 5,67 \; {\rm AU}$.
At that time, the plasma temperature is $T_{\rm eq} = \text{9,275} \; {\rm K} \equiv 7.99 \times 10^{-10} \; {\rm GeV}$,
and the cosmological DM density is $\rho_{i}^{\rm eq} \simeq 8.88 \times 10^{-20} \; {\rm g \, cm^{-3}}$.

\vskip 0.1cm
More generally, notice that the physical radii of influence increase like $m^{1/3}$ while the reduced radii decrease like $m^{-2/3}$. In the $(\tilde{r} , M_{\rm BH})$ plane, the reduced radius of influence $\tilde{r}_{\rm inf}$ at cosmological time $t_{i}$ obeys the equation
\beq
m^{2} \, \tilde{r}_{\rm inf}^{3} \left\{ \! \frac{\rho_{\rm tot}(t_{i})}{\rho_{\rm S}^{0}} \! \right\} = 1 \,,
\label{eq:definition_r_inf_b}
\eeq
where $\rho_{\rm tot}$ is the energy density of the plasma. This density can be expressed in units of the benchmark value
\beq
\rho_{\rm S}^{0} = \frac{3 \eta_{\rm ta}}{16 \pi} \, \frac{1 \; {\rm M}_{\odot}}{(r_{\rm S}^{0})^{3}} \simeq
5.00 \times 10^{15} \; {\rm g \, cm^{-3}} \,.
\label{eq:rho_S_0}
\eeq
We find that $\tilde{r}_{\rm inf}$ is actually proportional to $m^{-2/3}$ with a scaling factor depending on cosmological time $t_{i}$, or alternatively on plasma temperature $T_{i}$. We will come back to this point very shortly.

\vskip 0.1cm
Last but not least, we will hereafter define the boundary between domains a and b as the location in the phase diagram where the corresponding asymptotic expressions for the DM density are equal
\beq
\rho_{\rm a}(\tilde{r} , m) = \rho_{\rm b}(\tilde{r} , m) \,.
\eeq
This equation yields a unique relation between radius and PBH mass which is represented by an orange line in the $(\tilde{r} , M_{\rm BH})$ plane of Fig.~\ref{fig:phase_diagram}.
Before we introduce the notion of triple point, we emphasize that even though asymptotic regimes allow us to identify three different slopes for the spike profile, the slope of index 3/2 has two different origins, and therefore characterizes two different areas in the phase diagram of \citefig{fig:phase_diagram}. In the following, we will distinguish these two 3/2 slopes when defining contact points or contact regions, hence the definition of two different triple points below.

\vskip 0.1cm
\noindent $\bullet$ {\bf Triple point A and critical mass $M_{2}$}\\
To start our exploration of the phase diagram, we must first define the properties of the DM particle, \ie~its mass $m_{\chi}$ and kinetic decoupling parameter $x_{\rm kd} = {m_{\chi}}/{T_{\rm kd}}$. In the illustration of Fig.~\ref{fig:phase_diagram}, we have selected a 1~TeV WIMP with a kinetic decoupling temperature of 100~MeV. Calculating the plasma energy density at that time and using Eq.~(\ref{eq:definition_r_inf_b}) yields, for a $1 \; {\rm M}_{\odot}$ black hole, the radius of influence $\tilde{r}_{\rm kd}^{0} = 3.42$, which physically amounts to $10.1 \; {\rm km}$.

\vskip 0.1cm
At the triple point A of the phase diagram, both regimes with slope $3/2$ (Keplerian and radial infall), and the caustic regime with slope $3/4$, have same asymptotic DM densities
\beq
\rho_{3/4}(\tilde{r}_{\rm A}) = \rho_{3/2}(\tilde{r}_{\rm A}) = \rho'_{3/2}(\tilde{r}_{\rm A} , m_{\rm A}) \,.
\label{eq:def_point_A}
\eeq
Recasting this relation as
\beq
\frac{{\cal A}_{3/4}}{{\tilde{r}_{\rm A}}^{3/4}} =
\frac{{\cal A}_{3/2}}{{\tilde{r}_{\rm A}}^{3/2}} =
\frac{{\cal A}_{3/2}^{\prime}(m_{\rm A})}{{\tilde{r}_{\rm A}}^{3/2}} \,,
\eeq
we readily infer from Eqs.~(\ref{eq:rho_asymptotic_slope_3_4}), (\ref{eq:rho_asymptotic_slope_3_2}) and (\ref{eq:rho_asymptotic_slope_3_2_prime}) the coordinates of A in the $(\tilde{r} , M_{\rm BH})$ plane
\beq
\tilde{r}_{\rm A} = \frac{x_{\rm kd}}{\zeta}
\left\{ \frac{{\cal I}_{3/2}^{\rm asy}}{{\cal I}_{3/4}^{\rm asy}} \right\}^{\! 4/3}
\;\text{and}\;\;\;
m_{\rm A} \equiv \frac{M_{\rm A}}{{\rm M}_{\odot}} =
\frac{{\cal I}_{3/2}^{\rm asy \, \prime}}{{\cal I}_{3/2}^{\rm asy}}
\left\{ \frac{\tilde{r}_{\rm kd}^{0}}{x_{\rm kd}} \right\}^{\! 3/2} \,.
\label{eq:point_A}
\eeq

\vskip 0.1cm
In the previous expression, the first equality is obtained by identifying $\rho_{3/4}$ with $\rho_{3/2}$. The boundary between the asymptotic regimes with slopes $3/4$ (caustic) and $3/2$ (Keplerian) corresponds in the phase diagram of Fig.~\ref{fig:phase_diagram} to a vertical line that goes all the way up to point A, where it stops. Its position does only depend on the kinetic decoupling parameter $x_{\rm kd}$ insofar as the ratio ${\tilde{r}_{\rm A}}/{x_{\rm kd}}$ is always equal to $8.81 \times 10^{-2}$.

\vskip 0.1cm
The second identity is derived by equating $\rho_{3/2}$ (Keplerian) with $\rho'_{3/2}$ (radial infall) and yields the PBH mass $M_{\rm A}$. It also defines the frontier between these asymptotic regimes, which appears in Fig.~\ref{fig:phase_diagram} as a horizontal line connecting points A and B$_{2}$, where B$_2$ is yet another triple point to be discussed in more detail later. That border separates what we call the heaviest PBHs from the others. That is why we are tempted to identify the corresponding critical mass $M_{2}$ with $M_{\rm A}$. In the case of Fig.~\ref{fig:phase_diagram}, we find a numerical value of $3.03 \times 10^{-5} \; {\rm M_{\odot}}$.

\vskip 0.1cm
Although this definition may seem ad hoc, we insist that it corresponds to the physical prescription given in section~\ref{subsec:slopes}. Actually, around a PBH with mass $M_{2} \equiv M_{\rm A}$, the radius of the sphere of influence at kinetic decoupling is
\beq
\tilde{r}_{\rm kd}(m_{2}) = \tilde{r}_{\rm kd}^{0} \, m_{2}^{-2/3} \,.
\eeq
The second identity of Eq.~(\ref{eq:point_A}) can be recast into
\beq
\sigma_{\rm kd}^{2} \tilde{r}_{\rm kd}(m_{2}) \equiv \frac{\tilde{r}_{\rm kd}(m_{2})}{x_{\rm kd}} =
\left\{ \frac{{\cal I}_{3/2}^{\rm asy}}{{\cal I}_{3/2}^{\rm asy \, \prime}} \right\}^{\! 2/3} \simeq 0.352 \,.
\label{eq:r_tld_kd_of_m2}
\eeq
This expression is universal and does not depend on DM properties. Notice that a WIMP injected at radius $r_{i}$ with initial velocity $v_{i}$ is characterized by the kinetic-to-potential energy ratio
\beq
\frac{K_{i}}{|U_{i}|} = \frac{{v_{i}^{2}}/{2}}{{G M_{\rm BH}}/{r_{i}}} = \beta_{i}^{2} \tilde{r}_{i} \equiv u \,.
\eeq
Replacing the velocity $\beta_{i}$ by its root mean square $\bar{\beta}_{i} \equiv \sqrt{3}_{} \sigma_{i}$, with $\sigma_{i}$ the local 1D dispersion velocity, we conclude that most WIMPs in the surroundings of a PBH are captured provided that
\beq
\frac{\overline{{K}_{i}}}{|U_{i}|} = 3_{} \sigma_{i}^{2} \tilde{r}_{i} \leq 1 \,.
\eeq
The kinetic-to-potential energy ratio is maximal on the surface of the sphere of influence at kinetic decoupling. For a PBH with critical mass $M_{2}$, this yields
\beq
3_{} \sigma_{\rm kd}^{2} \tilde{r}_{\rm kd}(m_{2}) \simeq 1.056 \,.
\eeq
The ratio of the WIMP average velocity to the local escape speed, both taken at the radius $\tilde{r}_{\rm kd}$, is equal to $1.028$. This result, obtained essentially from phenomenological arguments based on asymptotic expressions of the minispike density profile, is very close to the value of $1$ implied by the physical definition of $M_{2}$ given in section~\ref{subsec:slopes}. This remarkable agreement makes us confident that our approach based on the phase diagram is robust.

\vskip 0.1cm
The frontier $\tilde{r}_{\rm t}$ between the asymptotic regimes with slopes $3/4$ (caustic) and $3/2$ (radial infall) is given by the condition
\beq
\rho_{3/4}(\tilde{r}_{\rm t}) = \rho'_{3/2}(\tilde{r}_{\rm t} , m)
\;\;\;\text{or alternatively by}\;\;\;
\frac{{\cal A}_{3/4}}{\tilde{r}_{\rm t}^{3/4}} =
\frac{{\cal A}_{3/2}^{\prime}(m)}{\tilde{r}_{\rm t}^{3/2}} \,.
\eeq
The PBH mass must be larger than $M_{2}$ for the radial infall regime with slope $3/2$ to exist. The last expression yields the radius of transition
\beq
\tilde{r}_{\rm t} = \tilde{r}_{\rm A} \left( {m}/{m_{2}} \right)^{- 4/3} \equiv
\tilde{r}_{\rm A} \left\{ \! \frac{M_{\rm BH}}{M_{2}} \! \right\}^{\! - 4/3} .
\label{eq:tr_3_4_to _3_2_radial_infall}
\eeq
As featured in Fig.~\ref{fig:phase_diagram}, this radius becomes unphysical at very large PBH masses as it lies in the shaded gray vertical band below the Schwarzschild radius.
%
%

\vskip 0.1cm
\noindent $\bullet$ {\bf Definition of $\upsilon^{\prime}$ and triple point B$_{2}$}\\
The transition between the two radial infall asymptotic regimes, with respective slopes $3/2$ and $9/4$, corresponds to the line in the phase diagram where
\beq
\rho'_{3/2}(\tilde{r}_{\rm t} , m) = \rho_{9/4}(\tilde{r}_{\rm t} , m) \,.
\label{eq:transition_3_2_p_to_9_4_a}
\eeq
Neglecting the evolution of the thermodynamical coefficients $g_{\rm eff}$ and $h_{\rm eff}$ around the plasma temperature $T_{\rm kd}$ at kinetic decoupling, we can use the second relation of Eq.~(\ref{eq:pre_collapse}) to calculate the cosmological DM density $\rho_{i}(\tilde{r})$ at injection. This yields the asymptotic density
\beq
\rho_{9/4}(\tilde{r} , m) = \sqrt{\frac{2}{\pi^{3}}} \; {\cal I}_{9/4}^{\rm asy} \; \rho_{i}^{\rm kd}
\left\{ \frac{\tilde{r}_{\rm kd}}{\tilde{r}} \right\}^{\! 9/4} \,,
\label{eq:rho_slope_9_4_approx}
\eeq
where the dependence on PBH mass $m$ goes through $\tilde{r}_{\rm kd}$. Relation~(\ref{eq:transition_3_2_p_to_9_4_a}) yields the transition radius
\beq
\tilde{r}_{\rm t} = \upsilon \, \tilde{r}_{\rm kd}
\;\;\;\text{where}\;\;\;
\upsilon = \left\{ \frac{{\cal I}_{9/4}^{\rm asy}}{{\cal I}_{3/2}^{\rm asy \, \prime}} \right\}^{\! 4/3} \!\! \simeq 1.273 \,.
\eeq
In appendix~\ref{append:upsilon_prime}, we go a step further and take into account the dependence of $g_{\rm eff}$ and $h_{\rm eff}$ with plasma temperature. We show that the coefficient $\upsilon$ has to be replaced by the new value $\upsilon^{\prime}$ which depends now on kinetic decoupling temperature $T_{\rm kd}$. We show that $\upsilon^{\prime}$ lies between $1$ and $\upsilon$. The radius of transition $\tilde{r}_{\rm t}$ between the radial infall asymptotic regimes can be more correctly expressed as
\beq
\tilde{r}_{\rm t}(m) = \upsilon^{\prime} \, \tilde{r}_{\rm kd} = \upsilon^{\prime} \, \tilde{r}_{\rm kd}^{0} \, m^{-2/3} \,.
\label{eq:transition_3_2_p_to_9_4_b}
\eeq
It is very convenient to tabulate once and for all the coefficient $\upsilon^{\prime}$ as a function of $T_{\rm kd}$. In the case of Fig.~\ref{fig:phase_diagram}, where $T_{\rm kd}$ is equal to $100 \; {\rm MeV}$, we find $\upsilon^{\prime} \simeq 1.261$, very close to $\upsilon$.

\vskip 0.1cm
At the triple point B$_{2}$ of the phase diagram, the radial infall regimes with slope $3/2$ and $9/4$, as well as the Keplerian regime, have same asymptotic post-collapse DM densities
\beq
\rho_{3/2}(\tilde{r}_{\rm B_{2}}) = \rho'_{3/2}(\tilde{r}_{\rm B_{2}} , m_{\rm B_{2}}) =
\rho_{9/4}(\tilde{r}_{\rm B_{2}} , m_{\rm B_{2}}) \,.
\label{eq:def_point_B2}
\eeq
The first identity yields the mass $M_{\rm B_{2}} = M_{\rm A} \equiv M_{2}$. In the configuration of Fig.~\ref{fig:phase_diagram}, we already found a value of $3.03 \times 10^{-5} \; {\rm M_{\odot}}$ for $M_{2}$.
The last identity in Eq.~(\ref{eq:def_point_B2}) yields the radius
\beq
\tilde{r}_{\rm B_{2}} = \upsilon^{\prime} \, \tilde{r}_{\rm kd}^{0} \, m_{2}^{-2/3} \,.
\eeq
Numerically, we find a ratio ${\tilde{r}_{\rm B_{2}}}/{x_{\rm kd}}$ of $0.444$.
In the phase diagram, the frontier between the asymptotic regimes with slopes $3/2$ (Keplerian and radial infall) is represented by the orange horizontal segment connecting points A and B$_{2}$.

\vskip 0.1cm
\noindent $\bullet$ {\bf Critical mass $M_{1}$ and point B$_{1}$}\\
The transition between the Keplerian regime with slope $3/2$ and the radial infall regime with slope $9/4$ corresponds to the condition
\beq
\rho_{3/2}(\tilde{r}_{\rm t}) = \rho_{9/4}(\tilde{r}_{\rm t} , m) \,.
\label{eq:transition_3_3_to_9_4_a}
\eeq
As discussed above, the Keplerian regime exists only for PBH masses smaller than $M_{2}$. Using the exact expression for the cosmological DM density is mandatory, but time consuming. We present in appendix~\ref{append:Lambda} a method allowing for a fast and accurate derivation of the radius of transition $\tilde{r}_{\rm t}$ for any values of the DM parameters $m_{\chi}$ and $T_{\rm kd}$, and for any given PBH mass $m = {M_{\rm BH}}/{{\rm M}_{\odot}}$.

\vskip 0.1cm
Using expressions~(\ref{eq:rho_asymptotic_slope_3_2}) and (\ref{eq:rho_slope_9_4}) for the asymptotic densities, as well as relation~(\ref{eq:rho_i_vs_r_tilde_EQ}) for the cosmological DM density $\rho_{i}^{\rm t}$ at the cosmological time $t_{\rm t}$ and plasma temperature $T_{\rm t}$ for which the radius of influence of the PBH is precisely equal to $\tilde{r}_{\rm t}$, the condition fulfilled by the radius of transition can be straightforwardly written as
\beq
\frac{\tilde{r}_{\rm t}}{\tilde{r}_{\rm eq}} =
\Lambda \,
\left\{ \frac{g_{\rm eff}(T_{\rm eq})}{g_{\rm eff}(T_{\rm t})} \right\} \!
\left\{ \frac{h_{\rm eff}(T_{\rm t})}{h_{\rm eff}(T_{\rm eq})} \right\}^{\! 4/3}
\;\text{where}\;\;
\Lambda = \left( {m_{1}}/{m} \right)^{4/3} \,.
\label{eq:transition_3_3_to_9_4_b}
\eeq
In the previous expression, $m$ denotes the PBH mass while the benchmark value $m_{1}$ may be defined as
\beq
m_{1} =
\left\{ \frac{{\cal I}_{9/4}^{\rm asy}}{{\cal I}_{3/2}^{\rm asy}} \right\}
\left\{ \frac{\tilde{r}_{\rm eq}^{0}}{x_{\rm kd}} \right\}^{\! 3/2} \!
\left\{ \frac{a_{\rm kd}}{a_{\rm eq}} \right\}^{\! 3}.
\label{eq:definition_m_1_a}
\eeq
The mass $m_{1}$ depends only on the DM parameters through $x_{\rm kd}$ and $a_{\rm kd}$.
The ratio of the cosmological DM densities ${\rho_{i}^{\rm eq}}$ (at matter-radiation equality) and ${\rho_{i}^{\rm kd}}$ (at kinetic decoupling) appears naturally in the derivation of $m_{1}$. We have replaced it by the ratio of the corresponding scale factors $a_{\rm eq}$ and $a_{\rm kd}$ insofar as
${\rho_{i}^{\rm eq}}/{\rho_{i}^{\rm kd}} = {a_{\rm kd}^{3}}/{a_{\rm eq}^{3}} < 1$. This allows for a more intuitive interpretation of $m_{1}$. At this stage, several remarks are in order.

\vskip 0.1cm
A minispike extends up to the radius of influence $\tilde{r}_{\rm eq}$ of its PBH taken at matter-radiation equality. In expression~(\ref{eq:transition_3_3_to_9_4_b}), the radius $\tilde{r}_{\rm t}$, at which the slope of the post-collapse radial profile transitions from $3/2$ (Keplerian) to $9/4$ (radial infall), cannot exceed that outer boundary.
Since the reference temperature $T_{\rm t}$ is by construction larger than $T_{\rm eq}$, the scaling with degrees of freedom goes approximately like
\beq
\left\{ \frac{g_{\rm eff}(T_{\rm eq})}{g_{\rm eff}(T_{\rm t})} \right\} \!
\left\{ \frac{h_{\rm eff}(T_{\rm t})}{h_{\rm eff}(T_{\rm eq})} \right\}^{\! 4/3} \! \simeq
\left\{ \frac{h_{\rm eff}(T_{\rm t})}{h_{\rm eff}(T_{\rm eq})} \right\}^{\! 1/3} \geq 1 \,.
\eeq
Therefore, only values of $\Lambda \leq 1$ yield acceptable physical solutions respecting $\tilde{r}_{\rm t} \leq \tilde{r}_{\rm eq}$, i.e. contained within the spike.
As a consequence, the transition toward the radial infall regime with slope $9/4$ occurs only if the PBH mass $m$ is larger than $m_{1}$. For smaller masses, this asymptotic regime never appears. We are therefore led to identify $m_{1}$, as defined in Eq.~(\ref{eq:definition_m_1_a}), with the critical mass $M_{1}$ discussed in section~\ref{subsec:slopes}, setting $m_{1} \equiv {M_{1}}/{{\rm M}_{\odot}}$.
In the case of Fig.~\ref{fig:phase_diagram}, we find a mass $M_{1}$ of $1.097 \times 10^{-13} \; {\rm M}_{\odot}$.

\vskip 0.1cm
At the critical mass $M_{1}$, the coefficient $\Lambda$ is equal to $1$. Relation~(\ref{eq:transition_3_3_to_9_4_b}) yields a transition at the minispike boundary with
\beq
\tilde{r}_{\rm t}(m_{1}) = \tilde{r}_{\rm eq}(m_{1}) = \tilde{r}_{\rm eq}^{0} \, m_{1}^{-2/3} \,.
\eeq
The radius $\tilde{r}_{\rm t}(m_{1}) \equiv \tilde{r}_{\rm B_{1}}$ and mass $m_{1} \equiv m_{\rm B_{1}}$ define the coordinates of B$_{1}$ in the phase diagram. This point lies on the line of the outer boundary $\tilde{r}_{\rm eq}$.
Since the curve connecting B$_{1}$ to B$_{2}$ in Fig.~\ref{fig:phase_diagram} indicates the frontier between the Keplerian and radial infall regimes, we could have also derived the coordinates of B$_{1}$ by requiring that condition~(\ref{eq:transition_3_3_to_9_4_a}) is fulfilled at the minispike boundary $\tilde{r}_{\rm eq}$. This would have led to the same critical mass
\beq
m_{1} =
\left\{ \frac{{\cal I}_{9/4}^{\rm asy}}{{\cal I}_{3/2}^{\rm asy}} \right\}
\left\{ \frac{\rho_{i}^{\rm eq}}{\rho_{i}^{\rm kd}} \right\}
\left\{ \frac{\tilde{r}_{\rm eq}^{0}}{x_{\rm kd}} \right\}^{\! 3/2} \!
\propto T_{\rm kd}^{-5/2} m_{\chi}^{-3/2} \,.
\label{eq:definition_m_1_b}
\eeq

\vskip 0.1cm
Finally, although the mass $m_{1}$ has been derived essentially from graphical considerations pertaining to the phase diagram, it closely corresponds to the physical prescription given in section~\ref{subsec:slopes}. As showed in~\cite{BoudaudEtAl2021}, the kinetic-to-potential energy ratio of DM particles injected after kinetic decoupling, and hence beyond $\tilde{r}_{\rm kd}$, is a decreasing function of injection radius $\tilde{r}_{i}$ where\footnote{The quantity $\bar{u}_{i} \equiv \sigma_{i}^{2} \tilde{r}_{i}$ has already been introduced in relation~(3.30) of~\cite{BoudaudEtAl2021}. It represents the typical extension of the Gaussian distribution of initial WIMP velocities in the velocity triangle.}
\beq
\frac{\overline{{K}_{i}}}{|U_{i}|} = 3_{} \sigma_{i}^{2} \tilde{r}_{i} \propto \frac{1}{\sqrt{\tilde{r}_{i}}} \,.
\eeq
This ratio reaches a minimal value of $3_{} \sigma_{\rm eq}^{2} \tilde{r}_{\rm eq}$ at matter-radiation equality. After kinetic decoupling, the DM dispersion velocity scales like the inverse of the scale factor. Relation~(\ref{eq:definition_m_1_a}) can be converted into
\beq
\sigma_{\rm eq \,}^{2} \tilde{r}_{\rm eq}(m_{1}) = \left\{ \frac{{\cal I}_{3/2}^{\rm asy}}{{\cal I}_{9/4}^{\rm asy}} \right\}^{\! 2/3}
\;\text{where}\;\;
\tilde{r}_{\rm eq}(m_{1}) \equiv \tilde{r}_{\rm eq}^{0} \, m_{1}^{-2/3}
\;\;\text{while}\;\;
\frac{\sigma_{\rm eq}}{\sigma_{\rm kd}} \equiv \frac{a_{\rm kd}}{a_{\rm eq}}\,.
\eeq
At $M_{1}$, the minimal value of the kinetic-to-potential energy ratio is precisely equal to
\beq
\frac{\overline{{K}_{\rm eq}}}{|U_{\rm eq}|} = 3_{} \sigma_{\rm eq}^{2} \tilde{r}_{\rm eq} \equiv
3 \left\{ \frac{{\cal I}_{3/2}^{\rm asy}}{{\cal I}_{9/4}^{\rm asy}} \right\}^{\! 2/3} \simeq 0.936 \,.
\eeq
This value is universal and falls very close to $1$, in agreement with our physical definition of $M_{1}$. For that particular mass, the kinetic-to-potential energy ratio of WIMPs at injection is expected to be equal to $1$ at the boundary of the minispike. It exceeds unity everywhere else. For PBHs lighter than $M_{1}$, the DM average velocity $\bar{\beta}_{i}$  at injection is everywhere larger than the local escape speed\footnote{This is strictly true for radii larger than $\tilde{r}_{\rm kd}$ as well as for a large portion of the sphere of influence at kinetic decoupling. There is nevertheless an inessential region very close to the PBH where the potential energy always dominates over kinetic energy.}.

\vskip 0.1cm
Equipped with all these notations, we have drawn in Fig.~\ref{fig:phase_diagram} the phase diagram for a 1~TeV DM particle whose kinetic decoupling parameter $x_{\rm kd}$ is equal to $10^{4}$. Kinetic decoupling occurs at a temperature $T_{\rm kd}$ of 100~MeV.
The construction starts by drawing in the $(\tilde{r} , M_{\rm BH})$ plane the radius of influence $\tilde{r}_{\rm eq}$ at matter-radiation equality as a function of PBH mass. We get the solid red line which represents the surface of DM minispikes as well as the boundary of the phase diagram.
The radius of influence $\tilde{r}_{\rm kd}$ at kinetic decoupling is represented by the dotted short-dashed red line. For both lines, $\tilde{r}$ varies as $m^{-2/3}$ with $m = {M_{\rm BH}}/{\rm M_{\odot}}$.

\vskip 0.1cm
The next step is to draw the dotted long-dashed gray vertical line along which $\tilde{r}$ is set equal to the kinetic decoupling parameter $x_{\rm kd}$.
The intersection of that vertical line with the $\tilde{r}_{\rm kd}$ line falls always near the critical points A and B$_{2}$.
According to relation~(\ref{eq:r_tld_kd_of_m2}), the radius of influence $\tilde{r}_{\rm kd}(m_{2})$ at the critical mass $m_{2} \equiv {M_{2}}/{\rm M_{\odot}}$ is roughly equal to ${x_{\rm kd}}/{3}$. A and B$_{2}$ are located along the same horizontal line at PBH mass $M_{2}$. Both points are very close to the $x_{\rm kd}$ and $\tilde{r}_{\rm kd}$ lines. We actually showed above that
\beq
\frac{\tilde{r}_{\rm A}}{x_{\rm kd}} \simeq 0.0881
\;\;\;\text{while}\;\;\;
\frac{\tilde{r}_{\rm A}}{\tilde{r}_{\rm kd}(m_{2})} \simeq 0.250 \,.
\eeq
These ratios are universal and do not depend on WIMP parameters. The radius at B$_{2}$ fulfills similar relations with
\beq
\frac{\tilde{r}_{\rm B_{2}}}{\tilde{r}_{\rm kd}(m_{2})} = \upsilon^{\prime} \simeq 1.261
\;\;\;\text{implying that}\;\;\;
\frac{\tilde{r}_{\rm B_{2}}}{x_{\rm kd}} \simeq 0.444 \,.
\eeq
The parameter $\upsilon^{\prime}$ depends now on kinetic decoupling temperature $T_{\rm kd}$ but is always bounded by $1$ and $\upsilon = 1.273$.

\vskip 0.1cm
The critical point B$_{1}$ lies on the $\tilde{r}_{\rm eq}$ solid red line. The frontier between the Keplerian region with slope $3/2$ (light-green) and the radial infall domain with slope $9/4$ (dark-green) extends below $M_{2}$. The radius $\tilde{r}_{\rm t}$ at which the transition occurs is the solution of Eq.~(\ref{eq:transition_3_3_to_9_4_b}). Neglecting the variations of the thermodynamical coefficients $g_{\rm eff}$ and $h_{\rm eff}$, we find that $\tilde{r}_{\rm t}$ is proportional to $\tilde{r}_{\rm eq} \Lambda \propto m^{-2}$, hence the straight line connecting B$_{2}$ to B$_{1}$ in Fig.~\ref{fig:phase_diagram}. A closer inspection would reveal tiny wiggles insofar as $g_{\rm eff}$ and $h_{\rm eff}$ do vary with temperature during the radiation era.

\vskip 0.1cm
The various asymptotic regions with their partition lines can be readily inferred from the prescriptions worked out in this section. The light-brown, light-green and dark-green areas correspond respectively to slopes $3/4$, $3/2$ and $9/4$. The domain with slope $3/2$ can be split into a radial infall part above $M_{2}$ and the Keplerian zone below.
%
\begin{figure}[h!]
\centering
\includegraphics[width=0.495\textwidth]{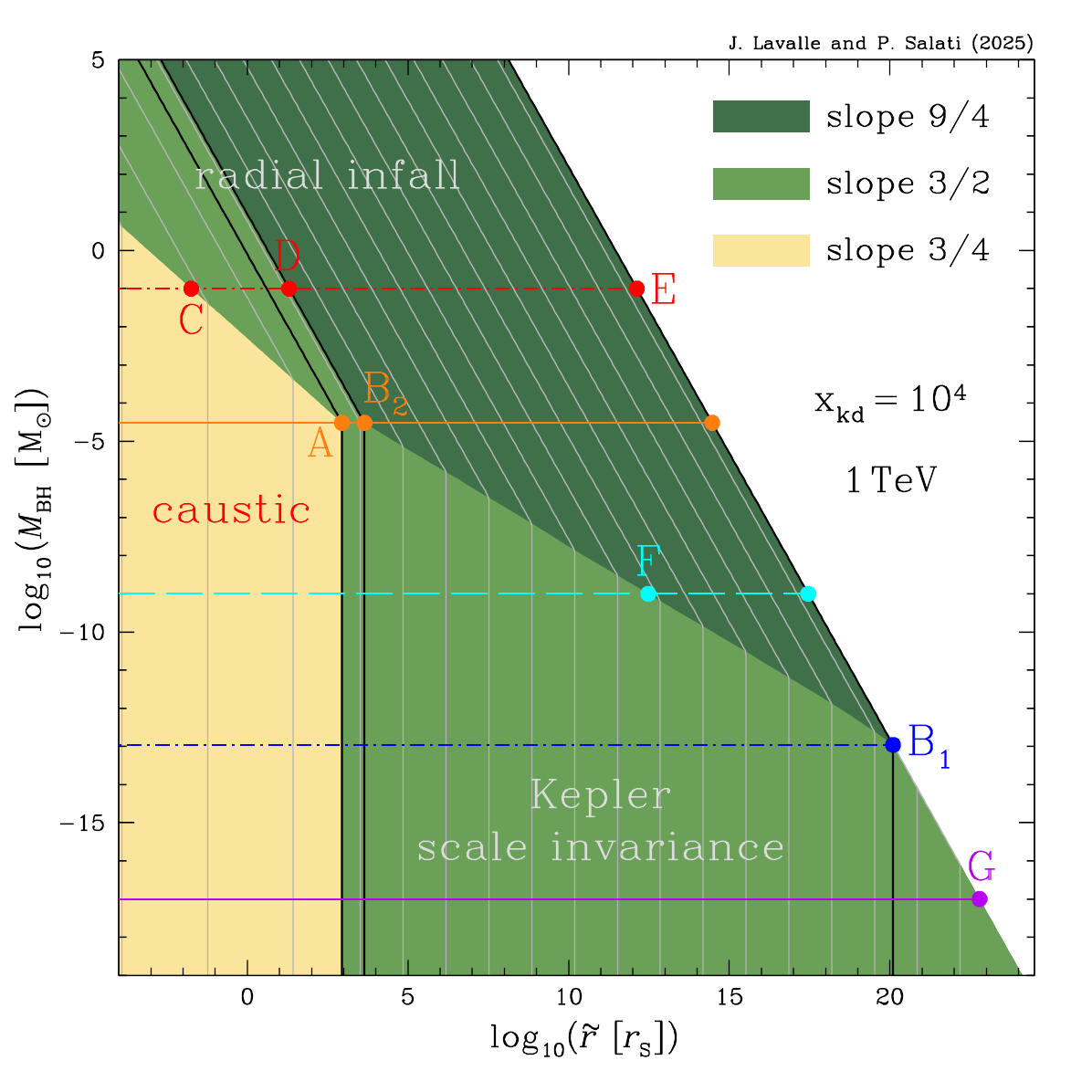}
\includegraphics[width=0.495\textwidth]{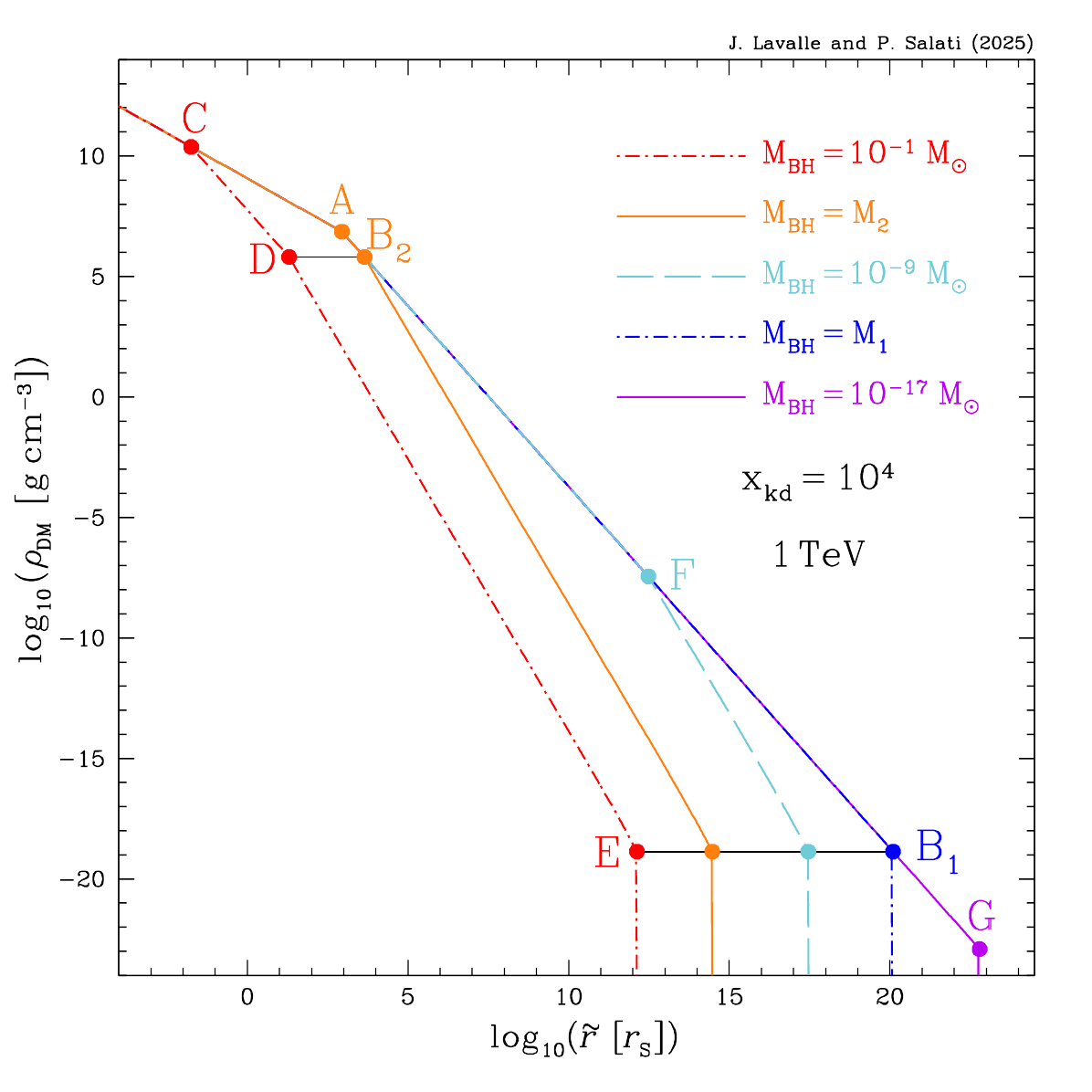}
\caption{In the {\bf left panel}, the phase diagram of Fig.~\ref{fig:phase_diagram} has been reproduced. Each thin gray line corresponds to a specific value of the post-collapse DM density which, from left to right, has been varied from $10^{12}$ down to $10^{-22} \; {\rm g \, cm^{-3}}$ by decrements of $10^{2}$.
The critical points A, B$_{2}$ and B$_{1}$  lie along the solid black iso-density lines where the DM density is respectively equal to $7.28 \times 10^{6}$, $6.44 \times 10^{5}$ and $1.36 \times 10^{-19} \; {\rm g \, cm^{-3}}$.
%
%
In the {\bf right panel}, the DM density profiles have been plotted versus reduced radius $\tilde{r}$ for different values of the PBH mass. These are coded with color and line type as indicated. Each profile of the right panel corresponds to a particular horizontal line in the left panel.
A few remarkable points are also featured in both panels and discussed in the text.
\label{fig:iso_rho_lines_and_DM_profiles}}
\end{figure}
%

\vskip 0.1cm
To get a deeper insight in the structure of DM minispikes, we have explored in Fig.~\ref{fig:iso_rho_lines_and_DM_profiles} how the density profiles change with PBH mass. With same DM parameters as before, we have first selected the critical values $M_{1}$ (dotted short-dashed blue) and $M_{2}$ (solid orange). These delineate three mass ranges. We have then chosen a value inside each of these intervals and considered the masses $10^{-17} \; {\rm M_{\odot}}$ (solid purple), $10^{-9}\; {\rm M_{\odot}}$ (long-dashed cyan) and $0.1 \; {\rm M_{\odot}}$ (dotted short-dashed red). Each of the density profiles plotted in the right panel correspond to a particular horizontal line in the phase diagram of the left panel. A few remarkable points, including the critical points A, B$_{2}$ and B$_{1}$, are also plotted in both panels for pedagogical purposes and are discussed hereafter.

\vskip 0.1cm
Like on a hiking map, the thin gray lines of the phase diagram are iso-density contours. In the caustic (light-brown) and Keplerian (light-green) regions, the lines are vertical insofar as the DM profile is universal once the post-collapse density is expressed as a function of the reduced radius $\tilde{r}$. All the profiles that we have considered lie at some point along the same curve which connects points C, A and G. This behavior occurs as long as the corresponding horizontal slices in the phase diagram cross the caustic and Keplerian regions. When they enter the radial infall domains, profiles depart from the universal curve. In these regions, the iso-density lines are leaning leftward with $\tilde{r} \propto m^{-2/3}$.

\vskip 0.1cm
The prototypical examples featured in Fig.~\ref{fig:iso_rho_lines_and_DM_profiles} nicely summarize the rich dynamics of DM minispikes presented at the end of section~\ref{subsec:slopes}.
Below $M_{1}$, the density profile is always universal. As $\tilde{r}$ increases, the slope is $3/4$ and switches to $3/2$ at point A. It stays constant until the surface is reached at $\tilde{r}_{\rm eq}$, i.e. at point G in the case of the $10^{-17} \; {\rm M_{\odot}}$ PBH of the solid purple curve. Increasing that mass would make G closer to B$_{1}$.

\vskip 0.1cm
Between $M_{1}$ and $M_{2}$, the density profile departs from the universal behavior when the radial infall regime with slope $9/4$ sets in. This occurs at point F in the case of the $10^{-9} \; {\rm M_{\odot}}$ PBH of the long-dashed cyan curve. In the left panel of Fig.~\ref{fig:iso_rho_lines_and_DM_profiles}, F sits at the border between the Keplerian (light-green) and radial infall (dark-green) domains. In the right panel, the profile leaves the universal curve at F and falls steeply with slope $9/4$.
Varying the PBH mass from $M_{1}$ to $M_{2}$ would make point F move from B$_{1}$ up to B$_{2}$. This two extreme situations are respectively featured by the dotted short-dashed blue and the solid orange curves.

\vskip 0.1cm
The dotted short-dashed red curve highlights the case of the heaviest PBHs. As is clear in the left panel of Fig.~\ref{fig:iso_rho_lines_and_DM_profiles}, the frontier between the caustic (light-brown) and radial infall (light-green) regions is no longer vertical above $M_{2}$. In our example, the transition between these asymptotic regimes occurs at point C\footnote{In the plot, point C lies in the unphysical region below the Schwarzschild radius. Should we have taken a lighter PBH, with mass $10^{-3} \; {\rm M_{\odot}}$ for instance, the transition from $3/4$ to $3/2$ would have turned physical.} where the slope increases from $3/4$ to $3/2$. A new transition occurs at point D between the two radial infall regimes and the slope jumps from $3/2$ to $9/4$. Point E corresponds to the surface of the minispike.
Varying the PBH mass from $M_{2}$ upward would make point C slip away from A along the universal curve and would shift the entire profile to the left.

\vskip 0.1cm
Before embarking for a comparison between numerical and semi-analytical results, it is appropriate at this stage to check the validity of a key assumption for our derivation of the post-collapse DM density profile. In section~\ref{sec:intro}, we have assumed that {\sl the local dynamics close to the BH is driven by the BH mass only.}
In appendix~\ref{append:M_halo_vs_M_BH} we have thoroughly examined this question and calculated the mass $M_{\rm halo}$ of the DM spike.
We conclude that this condition holds. The mass $M_{\rm halo}$ is negligible with respect to $M_{\rm BH}$ for PBHs lighter than $M_{1}$, which increases as WIMP mass $m_{\chi}$ decreases as shown in relation~(\ref{eq:definition_m_1_b}). For heavier black holes, the mass of the spike amounts to a significant fraction (67\%) of $M_{\rm BH}$ only close to the surface, an inessential region for DM annihilations. Besides, the post-collapse density in this case undergoes everywhere a negligible variation when $M_{\rm halo}$ is taken into account on top of $M_{\rm BH}$.

\subsection{Numerical vs approximate semi-analytical results}
\label{subsec:num_vs_approx}

The aim of this work is to construct an acceptable semi-analytical approximation for the DM density inside minispikes. The calculation should be fast and yield a value close to the result of the lengthy numerical integration~(\ref{eq:post_collapse_density}).
To accelerate the code, a few tables can be pre-calculated as explained in appendix~\ref{append:fast_code}, together with the thermodynamical coefficients $g_{\rm eff}$ and $h_{\rm eff}$.

\vskip 0.1cm
The post-collapse DM density $\rho$ depends on reduced radius $\tilde{r}$, PBH mass $M_{\rm BH}$ and DM parameters.
The WIMP mass $m_{\chi}$ and decoupling parameter $x_{\rm kd}$ need first to be specified. This allows to derive the properties of the phase diagram, i.e. the critical masses $M_{1}$ and $M_{2}$ as well as the coordinates in the $(\tilde{r} , M_{\rm BH})$ plane of the particular points A, B$_{1}$ and B$_{2}$. Expressions for these quantities have been thoroughly discussed in section~\ref{subsec:phase_diagram}.
%
\begin{figure}[h!]
\centering
\includegraphics[width=0.495\textwidth]{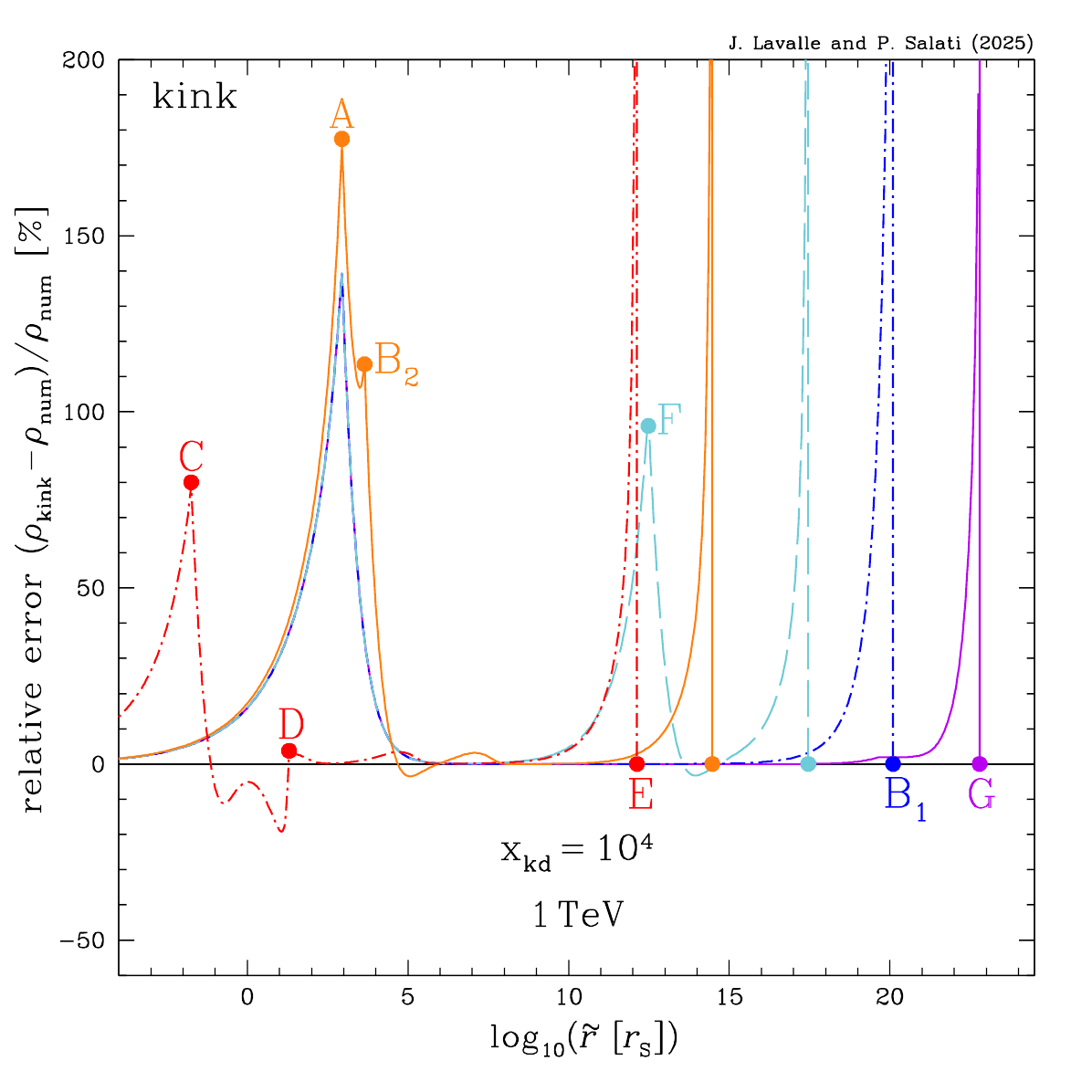}
\includegraphics[width=0.495\textwidth]{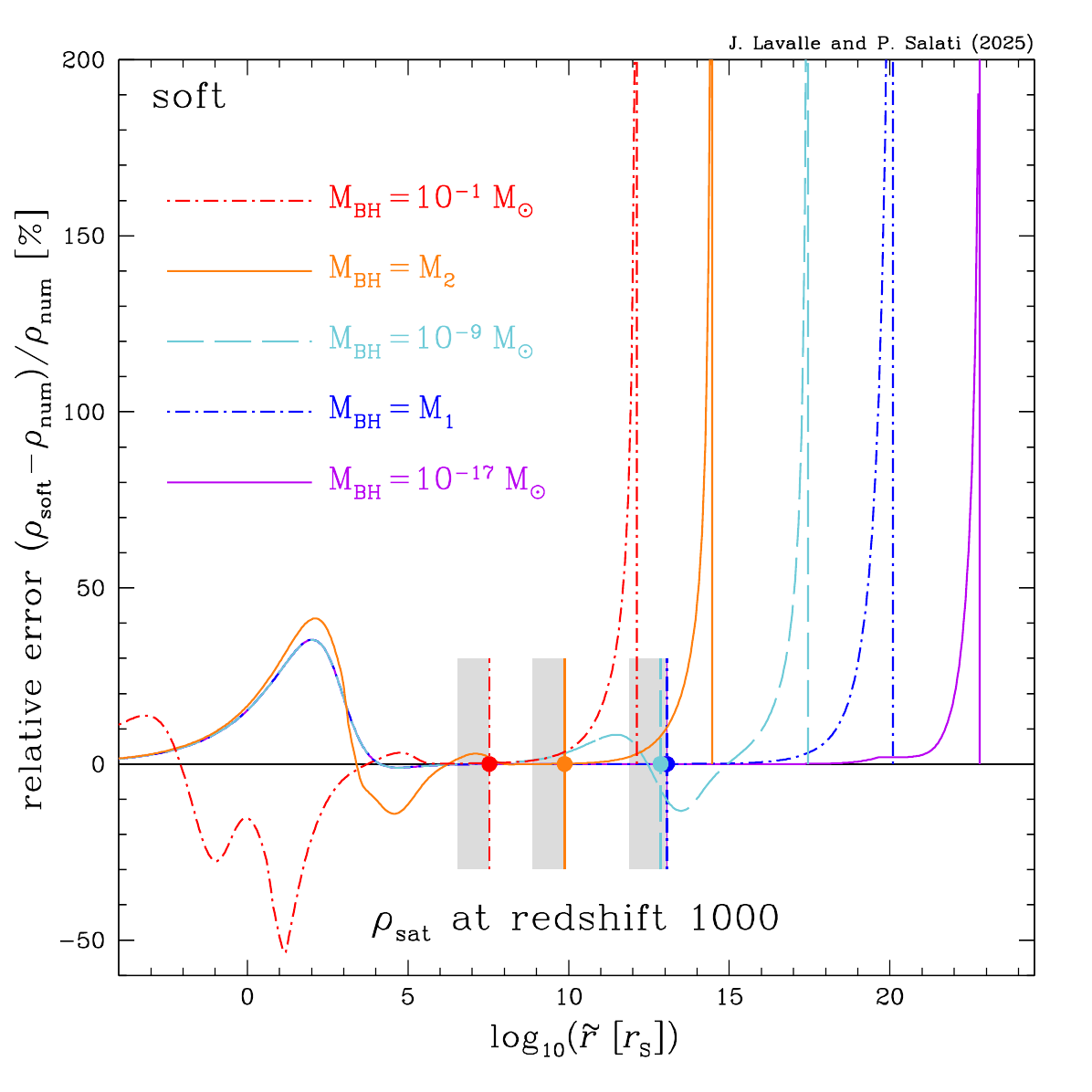}
\caption{The relative difference between the approximate post-collapse DM density and the exact numerical result is featured as a function of reduced radius $\tilde{r}$. The DM parameters are the same as for the previous figures. We have also considered the same PBH masses as in Fig.~\ref{fig:iso_rho_lines_and_DM_profiles} with the same code for colors and line type.
The {\bf left panel} is devoted to the kink approximation. The sharp peaks which the curves exhibit correspond to changes in the asymptotic regime occuring at the very same points A, B$_{2}$, C, D and F as in Fig.~\ref{fig:iso_rho_lines_and_DM_profiles}.
The soft approximation is featured in the {\bf right panel}. For each PBH mass, the colored vertical edges of the light-gray bands and their associated points indicate the radii below which the post-collapse DM density is larger than the saturation density $\rho_{\rm sat}$ taken at a redshift $z$ of $10^{3}$ for a thermal DM annihilation cross-section. To the left of these edges, the DM distribution has been erased by annihilation and the profiles have been flattened.
\label{fig:errors_kink_soft}}
\end{figure}
%

As a first approximation, dubbed kink, we simply use the asymptotic expressions in their respective domains of prevalence. The code determines the PBH mass regime by comparing $M_{\rm BH}$ to $M_{1}$ and $M_{2}$. The prescription is as follows.
\renewcommand{\labelitemi}{$-$}
\begin{itemize}
\item{If $M_{\rm BH} \le M_{1}$, the density is given by relation~(\ref{eq:rho_asymptotic_slope_3_4}) below $\tilde{r}_{\rm A}$, and by~(\ref{eq:rho_asymptotic_slope_3_2}) above.}
\item{If the PBH mass lies between $M_{1}$ and $M_{2}$, the same algorithm can be used. Furthermore, somewhere between $\tilde{r}_{\rm B_{2}}$ and $\tilde{r}_{\rm eq}$, the profile enters the radial infall regime with slope $9/4$. A fast derivation of the radius at which this transition occurs is presented in appendix~\ref{append:Lambda}. The post-collapse density $\rho$ is then given by the asymptotic expression~(\ref{eq:rho_slope_9_4}), which makes use of the cosmological DM density $\rho_{i}$. The latter can be fastly derived as explained in appendix~\ref{append:rho_i_vs_tilde_r_0}.}
\item{For the heaviest PBHs, i.e. above $M_{2}$, three asymptotic regimes come successively into play as $\tilde{r}$ increases. The transition radius between the caustic (slope $3/4$) and radial infall (slope $3/2$) regions is now given by Eq.~(\ref{eq:tr_3_4_to _3_2_radial_infall}).
In this intermediate region, the density fulfills Eq.~(\ref{eq:rho_asymptotic_slope_3_2_prime}).
Close to $\tilde{r}_{\rm kd}$, the radial infall regime changes as the slope switches from $3/2$ to $9/4$. Relation~(\ref{eq:transition_3_2_p_to_9_4_b}) yields the exact location of the transition. Beyond that point, we recover the asymptotic expression~(\ref{eq:rho_slope_9_4}). As explained in appendix~\ref{append:upsilon_prime}, the coefficient $\upsilon^{\prime}$ can be rapidly calculated once $T_{\rm kd}$ is defined.}
\end{itemize}
Using the asymptotic expressions makes the approximated density $\rho_{\rm kink}$ continuous, but not its derivative at the borders between domains of prevalence where $\rho_{\rm kink}$ presents sharp breaks by construction.
In the left panel of Fig.~\ref{fig:errors_kink_soft}, we have plotted the relative difference between the kink approximation and the result of the numerical integration~(\ref{eq:post_collapse_density}), for the five profiles of Fig.~\ref{fig:iso_rho_lines_and_DM_profiles}. The code for color and line type is the same, as well as DM parameters. Notice that the vertical scale is linear and extends from $- 60\%$ to $+ 200\%$, while the radius $\tilde{r}$ varies logarithmically over more than 28 decades.

\vskip 0.1cm
As anticipated, the kink approximation overestimates the numerical result at the transition points. The curves exhibit sharp peaks precisely at the same points A, B$_{2}$, C, D and F as in Fig.~\ref{fig:iso_rho_lines_and_DM_profiles} where the asymptotic regime changes and the density profile softens. The relative error is significant. It reaches $80\%$ at C for a $0.1 \; {\rm M_{\odot}}$ PBH. It exceeds $100\%$ at A whatever the PBH mass. The two peaks of the solid orange curve at A and B$_{2}$ are noticeable. The peak at F corresponds to the onset of the radial infall regime in the intermediate case of a $10^{-9}\; {\rm M_{\odot}}$ PBH.

\vskip 0.1cm
To suppress the peaks and obtain a DM density closer to the numerical result, we have considered a new approximation, dubbed soft. Once the PBH mass is defined, the various asymptotic regimes {a} are identified with their respective densities $\rho_{\rm a}$. The density $\rho_{\rm soft}$ fulfills the definition
\beq
\frac{1}{\rho_{\rm soft}} = {\displaystyle \sum_{\rm regimes \, a}} \; \frac{1}{\rho_{\rm a}} \,.
\label{eq:soft_approximation_DM_rho}
\eeq
The relative error associated to $\rho_{\rm soft}$ is presented in the right panel of Fig.~\ref{fig:errors_kink_soft}. The peaks have disappeared and the agreement between approximation and numerical integration has improved. As showed by the long-dashed cyan curve, the peak at F has vanished. The errors at A and B$_{2}$ have considerably decreased and do not exceed $40\%$ now. The dotted short-dashed red curve exhibits wiggles though, and indicates that $\rho_{\rm soft}$ undershoots the numerical result by a factor of 2 at a radius $\tilde{r}$ of order $10$.

\vskip 0.1cm
The soft approximation yields very decent results at radii $\tilde{r}$ below $10^{-2}$. In this region, the caustic regime dominates. To check if the post-collapse DM density is well approximated by relation~(\ref{eq:rho_asymptotic_slope_3_4}), the numerical integration~(\ref{eq:post_collapse_density}) has been performed with injection radius $\tilde{r}_{i}$ varying from $0$ up to $\tilde{r}_{\rm eq}$. According to the right panel of Fig.~\ref{fig:errors_kink_soft}, the relative error in that range is less than $15\%$.
It might be worth emphasizing that for the particular values which we have selected for $m_{\chi}$ and $x_{\rm kd}$, the transition between the caustic (slope $3/4$) and radial infall (slope $3/2$) regimes occurs at unphysical radii for PBHs heavier than $0.01 \; {\rm M_{\odot}}$, but can move to larger radii for larger values of $x_{\rm kd}$.

\vskip 0.1cm
An actual calculation should nevertheless take into account the existence of a horizon around each PBH. The numerical integration of relation~(\ref{eq:post_collapse_density}) must be performed from an injection radius $\tilde{r}_{i}$ of at least $1$. We expect our approximations to break down at low radii.
This might be a problem insofar as the annihilation rates $\Gamma_{\rm BH}$ of minispikes, which we are aiming at in this work, depend sensitively on the DM density at small radii where it is the largest.

\vskip 0.1cm
We are rescued by the ongoing DM annihilations taking place inside minispikes as the universe expands. The DM density is erased and cannot exceed the so-called saturation value $\rho_{\rm sat}$ defined in section~\ref{subsec:phase_diagram_truncation}. The limits on PBHs and DM species which are discussed in section~\ref{sec:constraints} are based on observations of the post-recombination universe. At the epochs under scrutiny, annihilation has considerably erased the post-collapse DM profiles. Our approximations are to be used in the outer parts of minispikes, where the DM density is less than $\rho_{\rm sat}$.
At a redshift $z$ of $10^{3}$ for instance, a thermal DM annihilation cross-section of $3 \times 10^{-26} \; {\rm cm^{3} \, s^{-1}}$ translates into the saturation density
\beq
\rho_{\rm sat} =
4.97 \times 10^{-9} \; {\rm g \, cm^{-3}} \,.
\eeq
%
%
%
%
%
%
In the right panel of Fig.~\ref{fig:errors_kink_soft}, the colored vertical edges of the light-gray bands indicate, for each PBH mass, the saturation radius $\tilde{r}_{\rm sat}$ below which the minispike density has been erased down to $\rho_{\rm sat}$. A precise determination of the post-collapse DM density is therefore of no use below $\tilde{r}_{\rm sat}$ whereas it is mandatory above it. In this external region, the error is always small, except close to the surface at $\tilde{r}_{\rm eq}$.

\vskip 0.1cm
In both panels, the relative error diverges actually at the outer boundaries of minispikes. Both approximations yield there a non-vanishing result while the numerical integration returns a much smaller value, sometimes set to $0$ by brute force. We expect the relative error to explode at $\tilde{r}_{\rm eq}$, hence the spikes which both panels display. Although spectacular, these divergences are harmless in most cases. Close to the surface, the DM density is so small that it does not contribute significantly to $\Gamma_{\rm BH}$ in general.
A detailed inspection of the right panel shows nevertheless that the dotted short-dashed blue spike at $M_{1}$ is particularly thick compared to the others. This point will be clarified at the end of section~\ref{subsec:Gamma_BH_comparison} where the annihilation rates calculated with the numerical integral~(\ref{eq:post_collapse_density}) and the soft approximation~(\ref{eq:soft_approximation_DM_rho}) are compared.

\section{Annihilation rate of the dark matter minispikes}
\label{sec:Gamma_BH}

DM minispikes around PBHs have kept the memory of how dense the universe was during the radiation era, especially at kinetic decoupling. DM species inside these structures are highly packed, and if made of self-annihilating particles, are subject to strong annihilation. The rate of annihilation of a PBH minispike is the integral over its volume of the local DM annihilation rate
\beq
\Gamma_{\rm BH} = {\displaystyle \int} \mathrm{d}^{3}{\mathbf{r}}
\left\{ \frac{1}{2} \asv n^{2} \right\} \! ({\mathbf{r}}) \,.
\label{eq:Gamma_BH_a}
\eeq
The integral runs over the sphere of influence at matter-radiation equality while the domain located below the horizon does not contribute. Here we assume that all of non-PBH DM is made of self-annihilating Majorana fermions, hence the factor ${1}/{2}$. In this article, we only deal with $s$-wave annihilation, whose thermally averaged cross-section $\asv$ does not depend on DM velocity. As this quantity is constant throughout the minispike, it can be factored out of integral~(\ref{eq:Gamma_BH_a}). The DM number density $n = {\rho}/{m_{\chi}}$ does depend on position $\mathbf{r}$ through the density profile $\rho(\tilde{r})$ and WIMP mass $m_{\chi}$.
Under these conditions, the annihilation rate simplifies into the spherical integral
\beq
\Gamma_{\rm BH} = \frac{1}{2} \asv \, r_{\rm S}^{3} {\displaystyle \int_{\tilde{r}_{\rm min}}^{\tilde{r}_{\rm max}}} 4 \pi \tilde{r}^{2} \mathrm{d}{\tilde{r}} \,
\left\{ \frac{\rho(\tilde{r})}{m_{\chi}} \right\}^{\! 2},
\label{eq:Gamma_BH_b}
\eeq 
where $r_{\rm S}$ is the PBH Schwarzschild radius. The integral runs from $\tilde{r}_{\rm min} \sim 1$ up to $\tilde{r}_{\rm max} \equiv \tilde{r}_{\rm eq}$.

\vskip 0.1cm
Because of annihilations, the density profile $\rho(\tilde{r})$ departs from the distribution resulting from the sole collapse of DM onto PBHs. The latter has been calculated by many authors~\cite{Eroshenko2016,AdamekEtAl2019,CarrEtAl2021a,BoudaudEtAl2021}, and section~\ref{sec:DM_profiles} is devoted to the presentation of a tractable approximation.
But as soon as DM self-annihilates inside minispikes, the density profiles of these structures are eroded and their densest regions can be significantly depleted.

\vskip 0.1cm
The actual process is complicated to model and should, in principle, be included in the overall dynamical evolution of the spike and regulate the high density accumulated during the radiation era. This formally leads to a non-linear equation. Particles that were initially injected on a given trajectory annihilate as they move across the innermost regions. The larger the density, the stronger the annihilation. We expect their population to be depleted as time goes on. The contribution of each elliptical orbit to the overall DM distribution decreases, and so does the DM density. This leads in turn to a non-trivial variation of the rate of annihilation. To summarize, the DM density and its rate of destuction are entangled through a complex interplay between all orbits.
Note also that the annihilation of DM starts in principle as soon as minispikes form at kinetic decoupling.
A complete treatment of this destruction mechanism is beyond the scope of this paper. It requires a dedicated analysis taking into account the specificities of DM collapse onto PBHs during the radiation era. We note that the existence of an annihilation plateau has been discussed in~\cite{Vasiliev:2007vh,Shapiro:2016ypb}, although in a quite different context.

\subsection{Saturation density and truncation in the phase diagram}
\label{subsec:phase_diagram_truncation}

\begin{figure}[h!]
\centering
\includegraphics[width=0.495\textwidth]{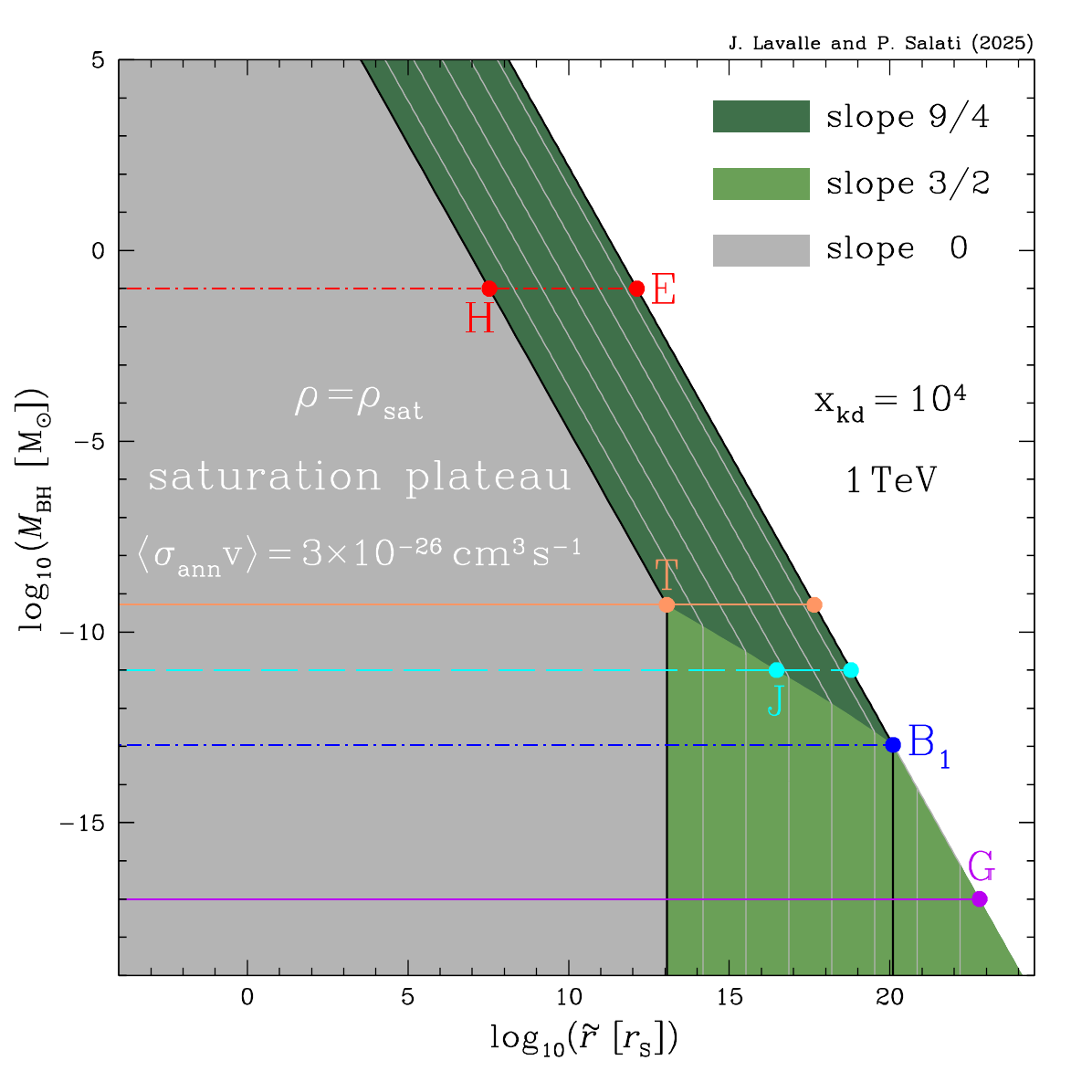}
\includegraphics[width=0.495\textwidth]{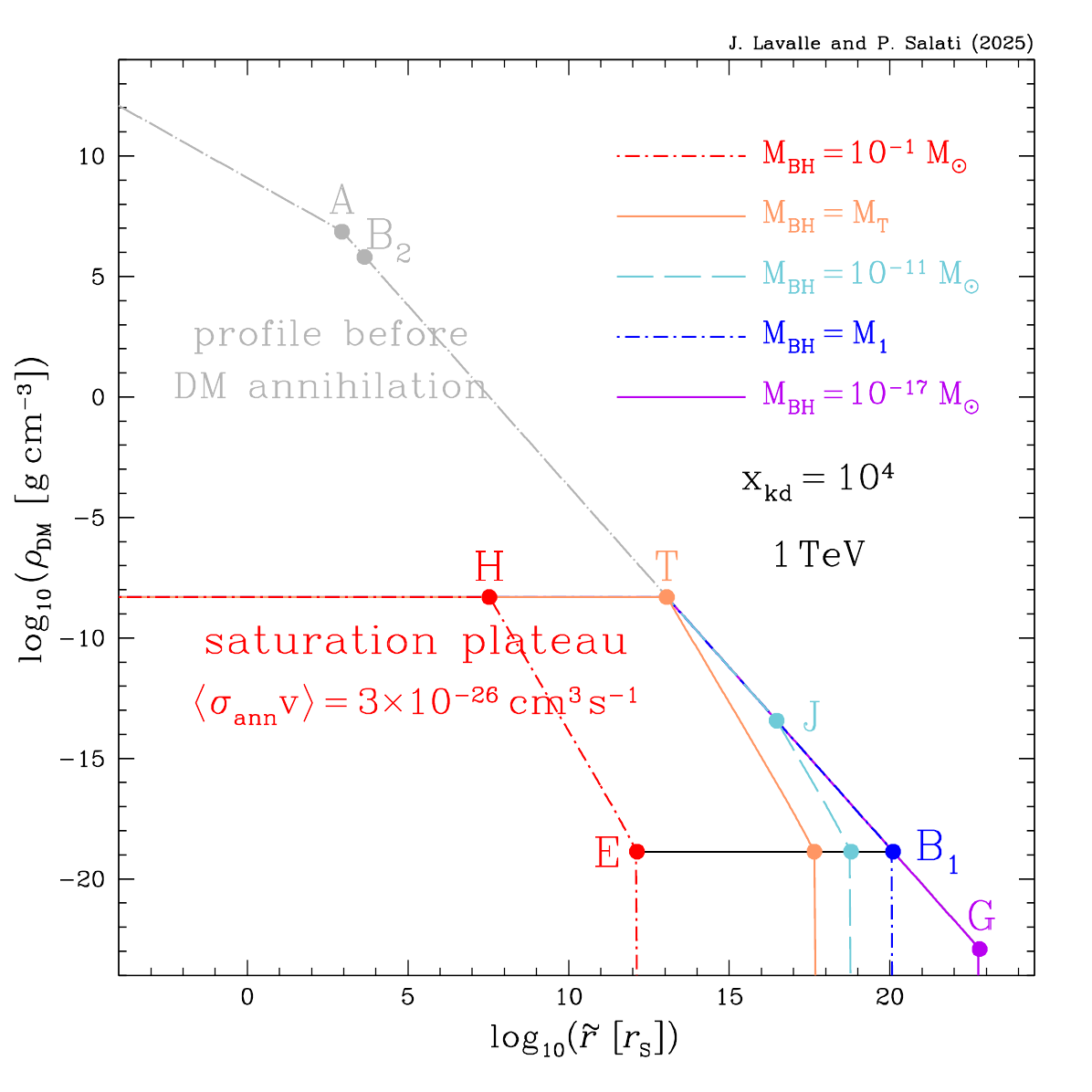}
\caption{Same panels as in Fig.~\ref{fig:iso_rho_lines_and_DM_profiles} where DM annihilation is now included.
In the {\bf left panel}, the inner region of the phase diagram is replaced by a saturation plateau where the DM density is equal to the constant $\rho_{\rm sat}$. For a redshift $z$ of $10^{3}$ and a thermal DM annihilation cross-section, we find a value of $4.97 \times 10^{-9} \; {\rm g \, cm^{-3}}$.
The solid black iso-density line that runs through points T and H corresponds to $\rho_{\rm sat}$, and features the border $\tilde{r}_{\rm sat}$ of the saturation plateau.
The critical point B$_{1}$  lies along the solid black iso-density line where the DM density is equal to $1.36 \times 10^{-19} \; {\rm g \, cm^{-3}}$.
%
%
In the {\bf right panel}, the DM density profiles have been plotted versus reduced radius $\tilde{r}$ for different values of PBH mass. These are coded with color and line type as indicated. Each profile of the right panel corresponds to a particular horizontal line in the left panel. DM annihilation has erased the post-collapse profiles and flattened them into the saturation plateau
A few remarkable points are also featured in both panels and discussed in the text.
\label{fig:saturated_PD_and_sat_profiles}}
\end{figure}

An annihilation plateau has been so far considered in most publications as a reasonable proxy for the inner density profile of DM clumps undergoing annihilation after their formation.
The underlying assumption is that DM particles move along a direction that is perpendicular to the gradient of their density. If so, we can identify the partial and total time derivatives of the DM density and write
\beq
\frac{{\rm d} \rho}{{\rm d}t} = \frac{\partial \rho}{\partial t} + \mathbf{v} \! \cdot \! \mathbf{\nabla}\rho \equiv \frac{\partial \rho}{\partial t}
\;\;\;\text{insofar as}\;\;\;
\mathbf{v} \perp  \mathbf{\nabla}\rho \,.
\label{eq:total_partial_d_on_dt}
\eeq
The DM density $\rho$ decreases as a result of WIMP annihilations and evolves according to
\beq
\frac{{\rm d} \rho}{{\rm d}t} = m_{\chi} \frac{{\rm d} n}{{\rm d}t} =
- 2 \times \left\{ \frac{1}{2} \asv n^{2} \right\} \times m_{\chi}
\;\;\;\text{with}\;\;\;
\rho = n m_{\chi} \,.
\label{eq:evolution_rho_a}
\eeq
Two particles are destroyed in a single annihilation, hence the extra factor of $2$ in the previous equality. Assuming spherical symmetry while combining expressions~(\ref{eq:total_partial_d_on_dt}) and (\ref{eq:evolution_rho_a}) yields
\beq
\frac{\partial}{\partial t} \left\{ \rho(t , \tilde{r}) \right\} = - \frac{\asv}{m_{\chi}} \left\{ \rho(t , \tilde{r}) \right\}^{2} \,,
\label{eq:evolution_rho_b}
\eeq
where $\tilde{r} \equiv {r}/{r_{\rm S}}$ is the reduced radius. This equation is strictly correct if WIMPs move along circular orbits (see above).
Its straightforward solution can be written as
\beq
\frac{1}{\rho(t_{2} , \tilde{r})} = \frac{1}{\rho(t_{1} , \tilde{r})} + {\displaystyle \int_{t_{1}}^{t_{2}}} \frac{\asv}{m_{\chi}} \, {\rm d}t \,.
\eeq
In the case of $s$-wave annihilation, the average cross-section $\asv$ is a constant independent of WIMP velocity and we get
\beq
\frac{1}{\rho(t_{2} , \tilde{r})} = \frac{1}{\rho(t_{1} , \tilde{r})} + \frac{1}{\rho_{\rm sat}} \,.
\eeq
The so-called saturation density appears, that is defined as
\beq
\rho_{\rm sat} = \frac{m_{\chi}}{\asv \Delta t}
\;\;\;\text{where}\;\;\;
\Delta t = t_{2} - t_{1} \,.
\label{eq:definition_rho_sat_a}
\eeq
We will herefater identify time $t_{1}$ with the age $t_{\rm eq}$ of the universe at the end of the radiation era, when minispikes have completely formed. Time $t_{2}$ is the epoch of the physical process under scrutiny in  section~\ref{sec:constraints}, i.e. the injection of energy in the post-recombination plasma, or the recent extragalactic emission of gamma-rays.
We will therefore identify $\rho(t_{1} , \tilde{r})$ with the post-collapse density profile without DM annihilation $\rho(\tilde{r})$ as derived by~\cite{Eroshenko2016,AdamekEtAl2019,CarrEtAl2021a,BoudaudEtAl2021}. This initial profile is eroded by annihilations to become, at time $t > t_{\rm eq}$, equal to\footnote{We stress that neglecting annihilation processes during spike formation provides an excellent approximation as long as we are interested in annihilation signals at much later times, $t\gg t_{\rm eq}$, which is the case for both CMB distortion and extragalactic gamma-ray signals. In other cases, one should actually include self-annihilation effects starting from the formation stage, which would hinder the derivation of analytical solutions.}
\beq
\rho(t , \tilde{r}) = \frac{\rho_{\rm sat} \, \rho(\tilde{r})}{\rho_{\rm sat} + \rho(\tilde{r})} \simeq \left\{
\begin{tabular}{ll}
$\rho_{\rm sat}$ & if $\rho(\tilde{r}) \gg \rho_{\rm sat} \,,$\\
$\rho(\tilde{r})$ & if $\rho(\tilde{r}) \ll \rho_{\rm sat} \,.$
\end{tabular}
\right.
\label{eq:rho_after_DM_annihilation_a}
\eeq
The transition between these regimes occurs at the saturation radius $\tilde{r}_{\rm sat}$ where both densities $\rho(\tilde{r})$ and $\rho_{\rm sat}$ are equal. The inner regions of minispikes are erased and replaced by a plateau with density $\rho_{\rm sat}$ extending to $\tilde{r}_{\rm sat}$ as showed in the left panel of Fig.~\ref{fig:saturated_PD_and_sat_profiles}. The impact on the phase diagram of the DM annihilations taking place between matter-radiation equality and a redshift $z$ of $10^{3}$  is featured for the same WIMP parameters as in Fig.~\ref{fig:iso_rho_lines_and_DM_profiles}.
In the dark-gray region extending to the left of the solid black iso-density line that crosses points T and H, the DM density has be flattened to its saturation value $\rho_{\rm sat}$ of $4.97 \times 10^{-9} \; {\rm g \, cm^{-3}}$.
%
%
The effect is even more dramatic in the right panel. The post-collapse universal profile, which corresponds to the dotted long-dashed dark-gray curve, is now replaced by a horizontal segment featuring the saturation plateau. For radii larger than $\tilde{r}_{\rm sat}$, the post-collapse DM profiles are not modified, as a quick inspection of the differences between Figs.~\ref{fig:iso_rho_lines_and_DM_profiles} and \ref{fig:saturated_PD_and_sat_profiles} would show.

\vskip 0.1cm
In table~\ref{tab:rho_sat_vs_rho_B1_B2}, the saturation density has been calculated for two values of the redshift and for several DM parameters covering a large spectrum of possibilities. The annihilation cross-section has been set equal to its benchmark value of $3 \times 10^{-26} \; {\rm cm^{3} \, s^{-1}}$. We remark that $\rho_{\rm sat}$ is always bounded by the post-collapse DM densities $\rho_{\rm B_{1}}$ and $\rho_{\rm B_{2}}$ at critical points B$_{1}$ and B$_{2}$ of the phase diagram. This allows to derive a simple approximation for the annihilation rate $\Gamma_{\rm BH}$ in the next section. The last line of the table corresponds to the plots of Fig.~\ref{fig:saturated_PD_and_sat_profiles}.
%
\begin{table}[h!]
\begin{center}
{\begin{tabular}{|c|c|c|c|c|c|}
\hline
$x_{\rm kd}$ & $m_{\chi}$ & $\rho_{\rm B_{1}}$ & $\rho_{\rm sat}$ at $z=0$ & $\rho_{\rm sat}$ at $z=1000$ & $\rho_{\rm B_{2}}$ \\
\hline
\hline
$10^{2}$ & 1 MeV & $1.355 \times 10^{-19}$ & $1.363 \times 10^{-19}$ & $4.974 \times 10^{-15}$ & $1.543 \times 10^{-7}$ \\
%
%
\hline
$10^{2}$ & 1 GeV & $1.355 \times 10^{-19}$ & $1.363 \times 10^{-16}$ & $4.974 \times 10^{-12}$ & $4.223 \times 10^{2}$ \\
%
%
\hline
$10^{2}$ & 1 TeV & $1.355 \times 10^{-19}$ & $1.363 \times 10^{-13}$ & $4.974 \times 10^{-9}$ & $3.385 \times 10^{12}$ \\
%
%
\hline
\hline
$10^{4}$ & 1 MeV & $1.355 \times 10^{-19}$ & $1.363 \times 10^{-19}$ & $4.974 \times 10^{-15}$ & $1.547 \times 10^{-13}$ \\
%
%
\hline
$10^{4}$ & 1 GeV & $1.355 \times 10^{-19}$ & $1.363 \times 10^{-16}$ & $4.974 \times 10^{-12}$ & $1.917 \times 10^{-4}$ \\
%
%
\hline
$10^{4}$ & 1 TeV & $1.355 \times 10^{-19}$ & $1.363 \times 10^{-13}$ & $4.974 \times 10^{-9}$ & $6.439 \times 10^{5}$ \\
%
%
\hline
\end{tabular}}
\end{center}
\caption{The saturation density $\rho_{\rm sat}$ taken at two different redshifts is compared to the post-collapse DM densities at points B$_{1}$ and B$_{2}$ of the phase diagram $(\tilde{r} , M_{\rm BH})$ (see Fig.~\ref{fig:phase_diagram}) for different kinetic decoupling parameters $x_{\rm kd}$ and DM masses $m_{\chi}$. Densities are expressed in ${\rm g \, cm^{-3}}$. The annihilation cross-section $\asv$ has been set equal to its thermal value of $3 \times 10^{-26} \; {\rm cm^{3} \, s^{-1}}$. The saturation densities in which we are interested in section~\ref{sec:constraints} are always bounded by $\rho_{\rm B_{1}}$ and $\rho_{\rm B_{2}}$.
\label{tab:rho_sat_vs_rho_B1_B2}}
%
\end{table}
%

\subsection{Behavior of the annihilation rate}
\label{subsec:Gamma_BH_behavior}

The saturation density $\rho_{\rm sat}$ enters in the definition of the annihilation rate~(\ref{eq:Gamma_BH_b}) of minispikes which we modify into
\begin{equation}
\Gamma_{\rm BH} = \frac{1}{2} \asv \left\{ \frac{\rho_{\rm sat}}{m_{\chi}} \right\}^{2} \! r_{\rm S}^{3} \; {\cal J}_{\rm BH}
\;\;\;\text{where}\;\;\;
{\cal J}_{\rm BH} \equiv
{\displaystyle \int_{\tilde{r}_{\rm min}}^{\tilde{r}_{\rm eq}}} 4 \pi \tilde{r}^{2} {\rm d}\tilde{r}
\left\{ \frac{\rho(t , \tilde{r})}{\rho_{\rm sat}} \right\}^{2}.
\label{eq:Gamma_BH_c}
\end{equation}
The radial integral ${\cal J}_{\rm BH}$ runs from the lower bound $\tilde{r}_{\rm min}$ up to the surface of the minispike at $\tilde{r}_{\rm eq}$. The latter radius depends on PBH mass as given in Eq.~(\ref{eq:definition_r_tilde_eq}). The saturation density is defined by relation~(\ref{eq:definition_rho_sat_a}) in which $\Delta t = t_{\rm U}(z) - t_{\rm eq}$, with $t_{\rm eq}$ and $t_{\rm U}(z)$ respectively the age of the universe at matter-radiation equality and at redshift $z$.
We defer to the next section the numerical estimate of ${\cal J}_{\rm BH}$ using the soft approximation for the post-collapse DM density, and concentrate here on the derivation of a simplistic approximation which will prove to be very helpful to understand the observational bounds discussed in section~\ref{sec:constraints}.

\vskip 0.1cm
To commence, we disregard the effect of the PBH horizon and set $\tilde{r}_{\rm min}$ equal to $0$. In appendix~\ref{append:more_correct_Gamma_BH}, we give more correct expressions where $\tilde{r}_{\rm min}$ is set to $1$ since minispikes extend only outside the Schwarzschild radii of their central objects\footnote{Rigorously, relativistic corrections are expected close to the Schwarzschild radius of a PBH, a region where our calculation cannot be considered as reliable as elsewhere.}. The correction turns out to be very small, except for super-heavy black holes.
Then, as showed in table~\ref{tab:rho_sat_vs_rho_B1_B2}, the saturation density is in general bounded by $\rho_{\rm B_{1}}$ and $\rho_{\rm B_{2}}$, the densities at critical points B$_{1}$ and B$_{2}$ of the phase diagram. This is actually the case in the panels of Fig.~\ref{fig:saturated_PD_and_sat_profiles} which we will use hereafter as a pedagogical example.

\vskip 0.1cm
In the left panel, point T lies in the phase diagram at the intersection of the saturation plateau boundary (solid black line) with the frontier separating the Keplerian region with slope $3/2$ (light-green) from the radial infall domain with slope $9/4$ (dark-green). The coordinates of T in the $(\tilde{r} , M_{\rm BH})$ plane fulfill the conditions
\beq
\rho_{\rm sat} = \rho_{3/2}(\tilde{r}_{\rm T}) = \rho_{9/4}(\tilde{r}_{\rm T} , m_{\rm T})
\;\;\;\text{where}\;\;\;
m_{\rm T} \equiv {M_{\rm T}}/{\rm M_{\odot}} \,.
\eeq
Once the saturation density $\rho_{\rm sat}$ has been derived, the first equality yields the reduced radius
\beq
\tilde{r}_{\rm T} = x_{\rm kd} \left\{ \! \frac{2^{1/3}}{\pi} \! \right\}
\left\{ {\cal I}_{3/2}^{\rm asy} \; \frac{\rho_{i}^{\rm kd}}{\rho_{\rm sat}} \right\}^{\! 2/3}.
\label{eq:r_tilde_T_a}
\eeq
Deriving the mass $M_{\rm T}$ requires first to use Eq.~(\ref{eq:rho_slope_9_4}) to get the cosmological DM density $\rho_{i}$ corresponding to $\rho_{\rm sat}$. From the prescriptions of appendix~\ref{append:tilde_r_0_vs_rho_i}, it is straightforward to get the radius $\tilde{r}_{\rm T}^{0}$ of the sphere of influence of a $1 \; {\rm M_{\odot}}$ PBH at the exact moment $t_{i}$ at which the cosmological DM density is equal to $\rho_{i}$. The mass $M_{\rm T}$ is given by
\beq
m_{\rm T} = \left( {\tilde{r}_{\rm T}^{0}}/{\tilde{r}_{\rm T}} \right)^{3/2} \,.
\label{eq:definition_m_T_a}
\eeq
This mass $M_{\rm T}$ together with the critical mass $M_{1}$ delineate three regions which, in Fig.~\ref{fig:saturated_PD_and_sat_profiles}, are exemplified by the values $10^{-17}$ (solid purple), $10^{-11}$ (long-dashed cyan) and $0.1 \; {\rm M_{\odot}}$ (dotted short-dashed red) for $M_{\rm BH}$.

\vskip 0.1cm
\noindent $\bullet$ {\bf $\rho_{\rm B_{1}} \leq \rho_{\rm sat} \leq \rho_{\rm B_{2}}$ and $M_{\rm BH} \leq M_{1}$}\\
A close inspection of the solid purple curve shows that the density is flat up to $\tilde{r}_{\rm T}$ (saturation plateau) and decreases as a power law with slope $3/2$ above. The radial integral ${\cal J}_{\rm BH}$ can be split into two parts
\beq
{\cal J}_{\rm BH} =
{\displaystyle \int_{0}^{\tilde{r}_{\rm T}}} 4 \pi \tilde{r}^{2} {\rm d}\tilde{r}
+
{\displaystyle \int_{\tilde{r}_{\rm T}}^{\tilde{r}_{\rm eq}}} 4 \pi \tilde{r}^{2} {\rm d}\tilde{r}
\left\{ \frac{\tilde{r}_{\rm T}}{\tilde{r}} \right\}^{\! 3}.
\eeq
This yields
\beq
{\cal J}_{\rm BH} = \frac{4}{3} \pi_{\,} \tilde{r}_{\rm T}^{3} \, {\cal K}_{\rm BH}
\;\;\;\text{with}\;\;\;
{\cal K}_{\rm BH} = 1 + 3 \ln \! \left\{ \frac{\tilde{r}_{\rm eq}}{\tilde{r}_{\rm T}} \right\}.
\label{eq:K_BH_small_mass_a}
\eeq

\vskip 0.1cm
\noindent $\bullet$ {\bf $\rho_{\rm B_{1}} \leq \rho_{\rm sat} \leq \rho_{\rm B_{2}}$ and $M_{1} \leq M_{\rm BH} \leq M_{\rm T}$}\\
This intermediate case corresponds to the long-dashed cyan curve. In the left panel of Fig.~\ref{fig:saturated_PD_and_sat_profiles}, point J lies on the demarcation line separating the Keplerian (slope $3/2$) and radial infall (slope $9/4$) regions. The reduced radius $\tilde{r}_{\rm J}$ can be readily derived using the method described in section~\ref{subsec:phase_diagram} and appendix~\ref{append:Lambda} based on the calculation of $\Lambda = ({m_{1}}/{m})^{4/3}$. Three parts contribute now to ${\cal J}_{\rm BH}$. The profile is flat up to $\tilde{r}_{\rm T}$. It decreases with slope $3/2$ between T and J. Above $\tilde{r}_{\rm J}$ and up to the surface at $\tilde{r}_{\rm eq}$, the radial index is $9/4$. The radial integral can be expressed as
\beq
{\cal J}_{\rm BH} =
{\displaystyle \int_{0}^{\tilde{r}_{\rm T}}} 4 \pi \tilde{r}^{2} {\rm d}\tilde{r}
+
{\displaystyle \int_{\tilde{r}_{\rm T}}^{\tilde{r}_{\rm J}}} 4 \pi \tilde{r}^{2} {\rm d}\tilde{r}
\left\{ \frac{\tilde{r}_{\rm T}}{\tilde{r}} \right\}^{\! 3}
+
{\displaystyle \int_{\tilde{r}_{\rm J}}^{\tilde{r}_{\rm eq}}} 4 \pi \tilde{r}^{2} {\rm d}\tilde{r}
\left\{ \frac{\tilde{r}_{\rm T}}{\tilde{r}_{\rm J}} \right\}^{\! 3}
\left\{ \frac{\tilde{r}_{\rm J}}{\tilde{r}} \right\}^{\! 9/2},
\eeq
which leads to
\beq
{\cal J}_{\rm BH} = \frac{4}{3} \pi_{\,} \tilde{r}_{\rm T}^{3} \, {\cal K}_{\rm BH}
\;\;\;\text{where}\;\;\;
{\cal K}_{\rm BH} = 3 + 3 \ln \! \left\{ \frac{\tilde{r}_{\rm J}}{\tilde{r}_{\rm T}} \right\} -
2 \left\{ \frac{\tilde{r}_{\rm J}}{\tilde{r}_{\rm eq}} \right\}^{\! 3/2}.
\label{eq:K_BH_medium_mass}
\eeq

\vskip 0.1cm
\noindent $\bullet$ {\bf $\rho_{\rm B_{1}} \leq \rho_{\rm sat} \leq \rho_{\rm B_{2}}$ and $M_{\rm T} \leq M_{\rm BH}$}\\
The prototypical example of this configuration is featured by the dotted short-dashed red curve. The profile is flat up to the border of the saturation plateau at H, while it decrases above with slope $9/4$ up to the surface at E. In the left panel, point H lies on the iso-density solid black line with density $\rho_{\rm sat}$. The reduced radii at points H and T are related through
\beq
{\tilde{r}_{\rm H}}/{\tilde{r}_{\rm T}} \equiv ({M_{\rm BH}}/{M_{\rm T}})^{-2/3} \,.
\eeq
Point E lies at the surface of the minispike at $\tilde{r}_{\rm eq}$. Once the radii $\tilde{r}_{\rm H}$ and $\tilde{r}_{\rm E} \equiv \tilde{r}_{\rm eq}$ have been determined, the radial integral ensues
\beq
{\cal J}_{\rm BH} =
{\displaystyle \int_{0}^{\tilde{r}_{\rm H}}} 4 \pi \tilde{r}^{2} {\rm d}\tilde{r}
+
{\displaystyle \int_{\tilde{r}_{\rm H}}^{\tilde{r}_{\rm E}}} 4 \pi \tilde{r}^{2} {\rm d}\tilde{r}
\left\{ \frac{\tilde{r}_{\rm H}}{\tilde{r}} \right\}^{\! 9/2}.
\eeq
We can recast it into
\beq
{\cal J}_{\rm BH} = \frac{4}{3} \pi_{\,} \tilde{r}_{\rm T}^{3} \, {\cal K}_{\rm BH}
\;\;\;\text{with}\;\;\;
{\cal K}_{\rm BH} = \left\{ \frac{M_{\rm T}}{M_{\rm BH}} \right\}^{\! 2}
\left\{ 3 - 2 \left( {\tilde{r}_{\rm H}}/{\tilde{r}_{\rm E}} \right)^{3/2} \right\}.
\label{eq:K_BH_large_mass_a}
\eeq

\vskip 0.1cm
\noindent
Increasing the PBH mass from $M_{1}$ to $M_{\rm T}$ makes point J shift from B$_{1}$ to T in the left panel of Fig.~\ref{fig:saturated_PD_and_sat_profiles}. In this regime, the integral ${\cal K}_{\rm BH}$ is defined by Eq.~(\ref{eq:K_BH_medium_mass}). Notice the continuity of this expression with the definitions~(\ref{eq:K_BH_small_mass_a}) and (\ref{eq:K_BH_large_mass_a}) at the transition points B$_{1}$ and T.

\vskip 0.1cm
Although the saturation density never exceeds the post-collapse density $\rho_{\rm B_{2}}$ in what follows, it may occasionally be smaller than $\rho_{\rm B_{1}}$. As showed in table~\ref{tab:rho_sat_vs_rho_B1_B2}, $\rho_{\rm sat}$ is close to $\rho_{\rm B_{1}}$ for a 1~MeV DM species. Increasing the annihilation cross-section $\asv$ by a factor of $2$ over its thermal value would make $\rho_{\rm sat}$ drop below $\rho_{\rm B_{1}}$. If so, two possibilities arise, depending on PBH mass. In this configuration, the saturation radius $\tilde{r}_{\rm T}$ is still defined by Eq.~(\ref{eq:r_tilde_T_a}).
We remark that now, at the transition mass $M_{\rm T}$, the radius of the minispike is precisely equal to $\tilde{r}_{\rm T}$. This translates into the definition
\beq
m_{\rm T} = \left( { \tilde{r}_{\rm eq}^{0}}/{\tilde{r}_{\rm T}} \right)^{3/2}
\;\;\;\text{where}\;\;\;
m_{\rm T} \equiv {M_{\rm T}}/{\rm M_{\odot}} \,.
\label{eq:definition_m_T_b}
\eeq
In the panels of Fig.~\ref{fig:saturated_PD_and_sat_profiles}, point T would lie between B$_{1}$ and G along the $\tilde{r}_{\rm eq}$ line, with $M_{\rm T}$ less than $M_{1}$.

\vskip 0.1cm
\noindent $\bullet$ {\bf $\rho_{\rm sat} \leq \rho_{\rm B_{1}}$ and $M_{\rm BH} \leq M_{\rm T}$}\\
The saturation plateau extends up to $\tilde{r}_{\rm T}$ above which the density decreases with slope $3/2$ until the surface is reached at $\tilde{r}_{\rm eq}$. This case has already been examined. The radial integral ${\cal J}_{\rm BH}$ is given by Eq.~(\ref{eq:K_BH_small_mass_a}). We can go a step further and take advantage of relations~(\ref{eq:definition_r_tilde_eq}) and (\ref{eq:definition_m_T_b}) to show that
\beq
{\cal J}_{\rm BH} = \frac{4}{3} \pi_{\,} \tilde{r}_{\rm T}^{3} \, {\cal K}_{\rm BH}
\;\;\;\text{with}\;\;\;
{\cal K}_{\rm BH} = 1 + 2 \ln \! \left\{ \frac{M_{\rm T}}{M_{\rm BH}} \right\}.
\label{eq:K_BH_small_mass_b}
\eeq

\vskip 0.1cm
\noindent $\bullet$ {\bf $\rho_{\rm sat} \leq \rho_{\rm B_{1}}$ and $M_{\rm T} \leq M_{\rm BH}$}\\
The minispike is homogeneous with density $\rho_{\rm sat}$. The radial integral ${\cal J}_{\rm BH}$ is readily given by
\beq
{\cal J}_{\rm BH} =
{\displaystyle \int_{0}^{\tilde{r}_{\rm eq}}} 4 \pi \tilde{r}^{2} {\rm d}\tilde{r} = \frac{4}{3} \pi_{\,} \tilde{r}_{\rm eq}^{3} \,,
\eeq
which translates into
\beq
{\cal J}_{\rm BH} = \frac{4}{3} \pi_{\,} \tilde{r}_{\rm T}^{3} \, {\cal K}_{\rm BH}
\;\;\;\text{with}\;\;\;
{\cal K}_{\rm BH} = \left\{ \frac{M_{\rm T}}{M_{\rm BH}} \right\}^{\! 2}.
\label{eq:K_BH_large_mass_b}
\eeq
At the transition mass $M_{\rm T}$, both definitions~(\ref{eq:K_BH_small_mass_b}) and (\ref{eq:K_BH_large_mass_b}) yield the same value of $1$ for ${\cal K}_{\rm BH}$. We leave it to the interested reader to show that definitions~(\ref{eq:K_BH_small_mass_a}), (\ref{eq:K_BH_medium_mass}) and (\ref{eq:K_BH_large_mass_a}) boil down to Eq.~(\ref{eq:K_BH_small_mass_b}) and (\ref{eq:K_BH_large_mass_b}) when the saturation density $\rho_{\rm sat}$ is equal to $\rho_{\rm B_{1}}$.

\vskip 0.1cm
We also remark that for very light black holes, way lighter than the solar mass scale, the integral ${\cal K}_{\rm BH}$ is a slowly varying function of PBH mass which becomes eventually dominated by $- 2 \ln (m)$ where $m \equiv {M_{\rm BH}}/{\rm M_{\odot}}$ (with $m\ll 1$ here). Conversely, in the heavy PBH regime, ${\cal K}_{\rm BH}$ decreases like $m^{-2}$. This will have important consequences on the bounds derived in section~\ref{sec:constraints}. In particular, at fixed DM parameters, we anticipate that the annihilation rate $\Gamma_{\rm BH}$ scales like $-m^{3} \ln(m)$ in the low (highly subsolar) PBH mass regime and like $m$ for heavy objects owing to the $r_{\rm S}^{3} \propto m^{3}$ term in the definition~(\ref{eq:Gamma_BH_c}).

\vskip 0.1cm
In appendix~\ref{append:more_correct_Gamma_BH_slope_0}, we take into account the Schwarzschild radius in the minispike structure, and calculate integral~(\ref{eq:Gamma_BH_c}) by setting the lower bound $\tilde{r}_{\rm min}$ equal to $1$. New expressions for ${\cal K}_{\rm BH}$ are given, which are very close to those derived in this section. In Fig.~\ref{fig:saturated_PD_and_sat_profiles} for instance, point T lies at radius $\tilde{r}_{\rm T} = 1.14 \times 10^{13}$ and mass $M_{\rm T} = 5.15 \times 10^{-10} \; {\rm M_{\odot}}$.
For PBH mass smaller than $M_{\rm T}$, the correction to ${\cal K}_{\rm BH}$ is equal to ${1}/{\tilde{r}_{\rm T}^{3}} = 6.83 \times 10^{-40}$ and is vanishingly small.
For heavy objects, the correction to ${\cal K}_{\rm BH}$ starts to be sensible when $\tilde{r}_{\rm H}$ approaches~$1$, i.e. for masses larger than $M_{\rm T\,} \tilde{r}_{\rm T}^{3/2} \simeq 1.97 \times 10^{10} \; {\rm M_{\odot}}$.
As showed in Eq.~(\ref{eq:M_BH_maximum}), super-heavy black holes have a masse cut-off at $1.54 \times 10^{17} \; {\rm M_{\odot}}$, a value above which ${\cal K}_{\rm BH}$ vanishes. This bound is close to the amount of matter enclosed in the causal horizon at matter-radiation equality.

\vskip 0.1cm
As DM particles orbit around the central PBH, they explore different regions of the minispike while continuously annihilating with each other. The actual density profile results from the intricate interactions of all bound trajectories. In this context, some analyses~\cite{Vasiliev:2007vh,Shapiro:2016ypb} indicate that the saturation plateau should be replaced by a weak cusp with slope lying between $0$ and $1/2$.
These publications explore the formation of black holes at the centers of already existing DM clumps during the matter dominated era. These clumps are dragged in turn by the collapsing central objects and compressed through adiabatic contraction, yielding very dense spikes which are subsequently erased by self-annihilations. In~\cite{Shapiro:2016ypb}, the DM profile in the spike is evolved with time by integrating numerically the Botlzmann equation while assuming a power law in energy for the initial phase space distribution function. A weak cusp is showed to develop, which gradually extends.

\vskip 0.1cm
In this article, the context is different since DM collapses during the radiation era around already existing PBHs. The initial phase space density is also completely determined, with a velocity distribution far from spherical. A dedicated analysis is well beyond the scope of this work though. Results will be derived hereafter by modeling DM annihilations through the formation of a saturation plateau which grows as time goes on.
In appendix~\ref{append:more_correct_Gamma_BH_slope_1_on_2}, we nevertheless explore the effect on ${\cal K}_{\rm BH}$ of replacing the saturation plateau by a weak cusp with slope $1/2$. The inner profile becomes
\beq
\rho(\tilde{r} \leq \tilde{r}_{\rm sat}) = \rho_{\rm sat} \left( {\tilde{r}_{\rm sat}}/{\tilde{r}} \right)^{1/2} \,.
\eeq
We remark that ${\cal K}_{\rm BH}$ increases at most by 50\% and does not change much. Replacing the saturation plateau by the weak cusp of~\cite{Vasiliev:2007vh,Shapiro:2016ypb} implies a moderate increase of the annihilation rate~$\Gamma_{\rm BH}$ and makes us confident about our procedure.

\subsection{Numerical results}
\label{subsec:Gamma_BH_comparison}

The calculation of ${\cal J}_{\rm BH}$ makes use of the time evolving spike density $\rho(t , \tilde{r})$. It is performed numerically, using for the post-collapse DM density $\rho(\tilde{r})$ either the result of integral~(\ref{eq:post_collapse_density}) or the soft approximation~(\ref{eq:soft_approximation_DM_rho}). The former procedure is time consuming, whereas the latter method allows to rapidly derive fairly accurate results as showed at the end of this section. A saturation plateau is always assumed.

\begin{figure}[h!]
\centering
\includegraphics[width=0.495\textwidth]{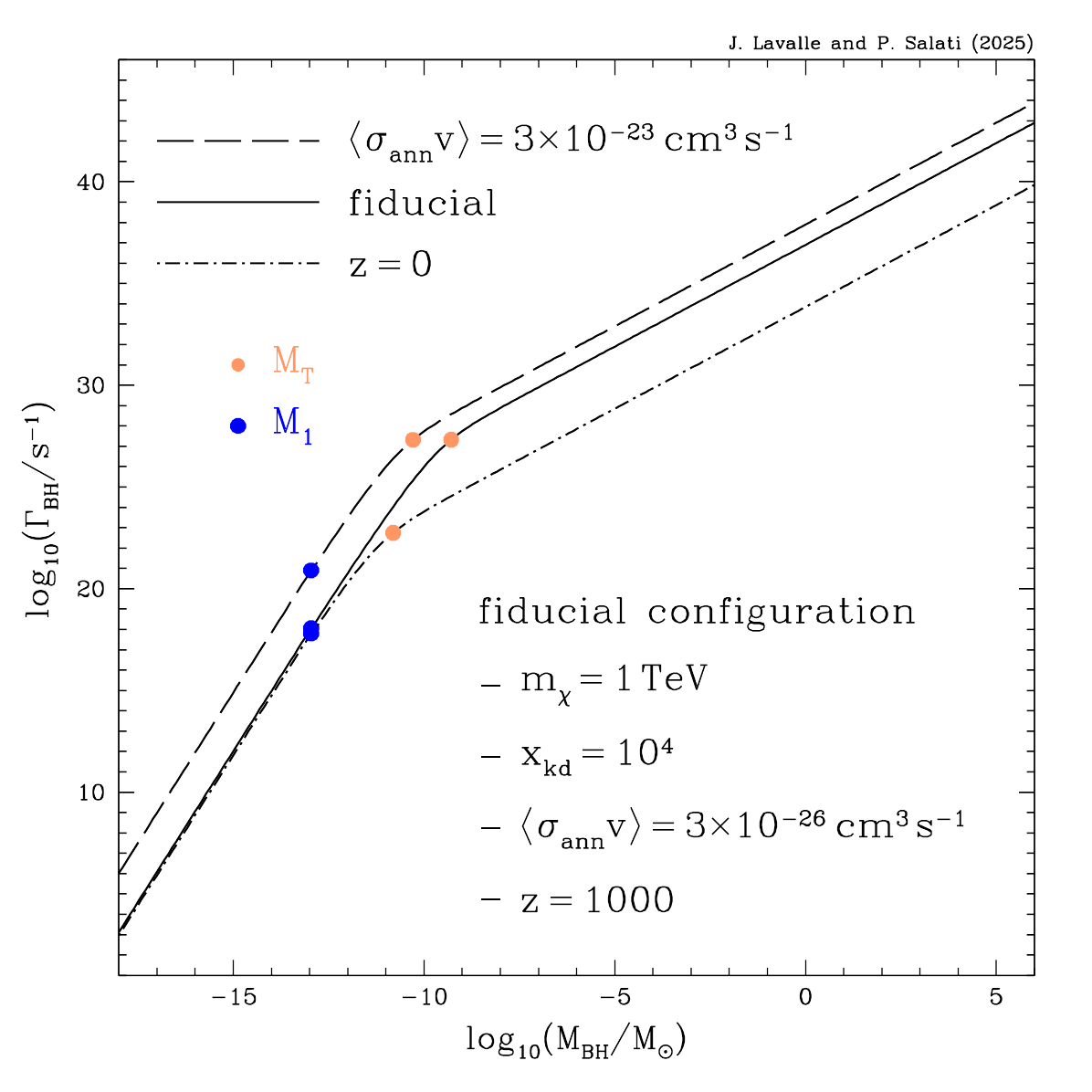}
\includegraphics[width=0.495\textwidth]{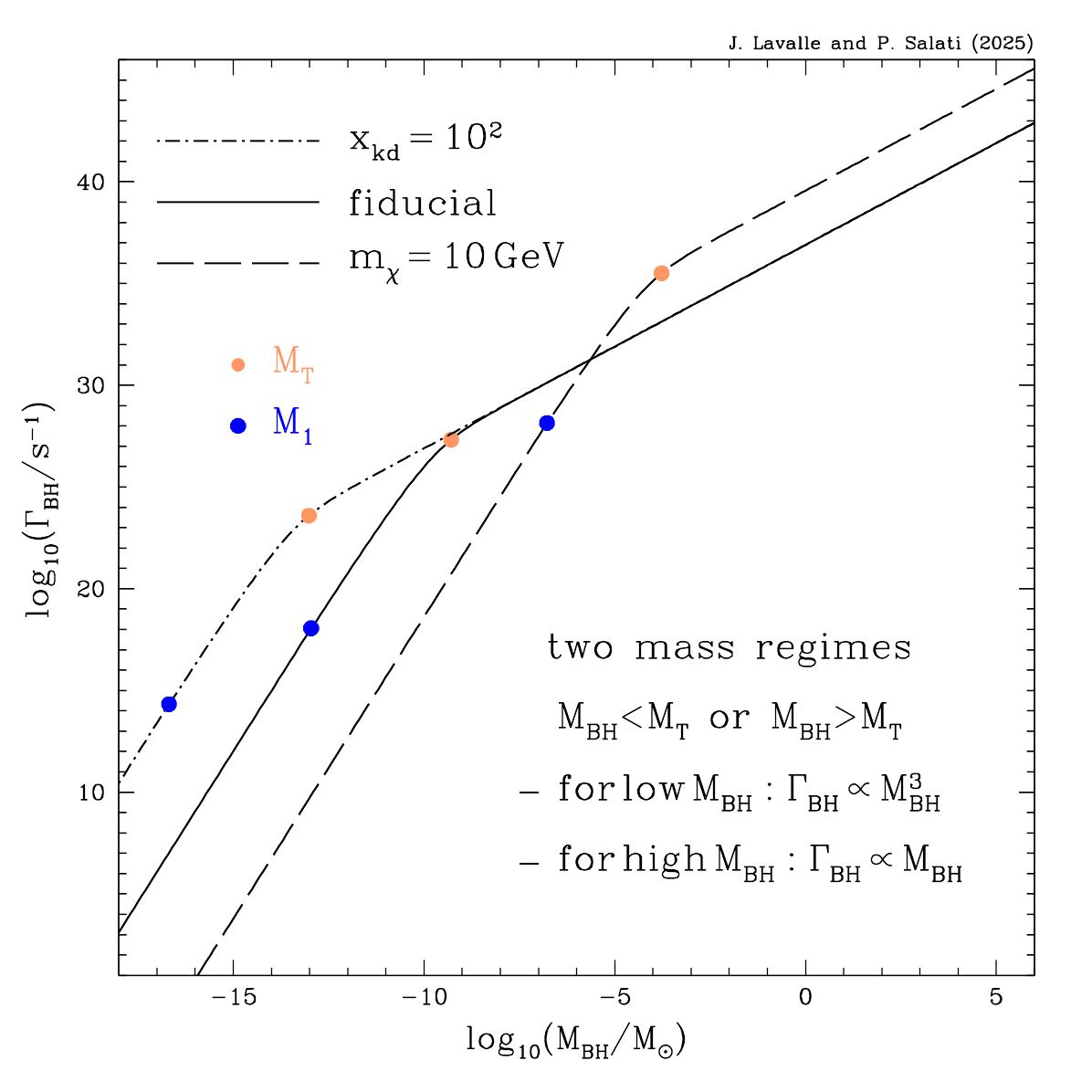}
\caption{The DM annihilation rate of minispikes is plotted as a function of PBH mass. The solid line corresponds to the fiducial case as specified in the left panel, while the other curves feature each the effect on $\Gamma_{\rm BH}$ of modifying one of the parameters of the model.
In the {\bf left panel}, the redshift is decreased from $10^{3}$ (post-recombination epoch) to $0$ (today). The dotted short-dashed line shifts downward only at high PBH mass. The long-dashed curve shows how increasing the annihilation cross-section $\asv$ by three orders of magnitude acts on $\Gamma_{\rm BH}$.
In the {\bf right panel}, decreasing the kinetic decoupling parameter $x_{\rm kd}$ from $10^{4}$ down to $10^{2}$ makes the annihilation rate grow only at low masses (dotted short-dashed). Varying the WIMP mass $m_{\chi}$ from $1 \; {\rm TeV}$ down to $10 \; {\rm GeV}$ induces an increase of $\Gamma_{\rm BH}$ at high mass but a decrease at low mass (long-dashed).
All the curves indicate that the annihilation rate increases with $M_{\rm BH}$ as a power law with index $3$ at low mass and $1$ at high mass.
These behaviors are further discussed in the text.
These curves have been derived using the soft approximation~(\ref{eq:soft_approximation_DM_rho}) with corrections~(\ref{eq:correction_soft}) to calculate the DM post-collapse density $\rho(\tilde{r})$, and eventually the time dependent spike density $\rho(t , \tilde{r})$. Eq.~(\ref{eq:post_collapse_density}) would have yielded the same results, as showed in Fig.~\ref{fig:error_Gamma_BH_various_models}.
\label{fig:Gamma_BH_vs_DM_parameters}}
\end{figure}

\vskip 0.1cm
The solid black curves in the panels of Fig.~\ref{fig:Gamma_BH_vs_DM_parameters} feature the variations of $\Gamma_{\rm BH}$ as a function of PBH mass in the case of our fiducial configuration with a $1 \; {\rm TeV}$ particle,  a kinetic decoupling parameter $x_{\rm kd}$ of $10^{4}$ and an annihilation cross-section $\asv$ equal to its thermal value of $3 \times 10^{-26} \; {\rm cm^{3} \, s^{-1}}$. The minispike annihilation rate is calculated at a redshift $z$ of $10^{3}$.
The effect of varying only one of these parameters is featured by the other curves. In the left panel, we explore how the solid line is modified by a change of redshift (dotted short-dashed) or by an increase of $\asv$ (long-dashed). In the right panel, we decrease the kinetic decoupling parameter (dotted short-dashed) or the WIMP mass (long-dashed).

\vskip 0.1cm
We observe that all curves exhibit the same behavior, with a characteristic power-law increase of the annihilation rate with PBH mass. We notice the existence of two distinct regimes. The spectral index is $3$ at low masses and switches to $1$ for heavy objects. As showed by the orange dots, the transition takes place approximately at $M_{\rm T}$. This mass is calculated for each set of parameters together with $\Gamma_{\rm BH}$, and positioned on the corresponding curve. All curves feature an inflection at the orange dots.
For completeness, we have also included the blue dots at which the PBH mass is equal to $M_{1}$.
%
%

\vskip 0.1cm
The annihilation rate exhibits a rich diversity of behaviors.
Changing the redshift for instance does not affect $\Gamma_{\rm BH}$ at low PBH mass, as showed by the dotted short-dashed line of the left panel. This is a remarkable result. Should DM minispikes be discovered around light black holes, we anticipate that their absolute luminosities would remain constant as the universe expands. This absence of variability would be a smoking gun signature for such objects insofar as other sources, like stars, are expected to evolve with redshift.

\vskip 0.1cm
Another noticeable property is featured by the dotted short-dashed curve of the right panel. Modifying the kinetic decoupling parameter $x_{\rm kd}$ has no effect on $\Gamma_{\rm BH}$ for heavy black holes. In this regime, the annihilation rate depends furthermore linearly on PBH mass. We anticipate that the limits on the contribution of PBHs to the astronomical DM, derived in section~\ref{subsec:f_PBH}, depend only on WIMP mass, and neither on $x_{\rm kd}$ nor on $M_{\rm BH}$.
In summary, to understand this complex phenomenology, we need to derive scaling relations for the annihilation rate as a function of model parameters in both the low and the high PBH mass regimes.

\vskip 0.1cm
\noindent $\bullet$ {\bf $M_{\rm BH} \leq M_{\rm T}$}\\
To commence, we remark that the extension $\tilde{r}_{\rm sat}$ of the saturation plateau, in reduced coordinates, does not vary with PBH mass as long as $M_{\rm BH} \leq M_{\rm T}$. This is true whether $\rho_{\rm sat}$ is larger or smaller than $\rho_{\rm B_{1}}$. In the left panel of Fig.~\ref{fig:saturated_PD_and_sat_profiles} for instance, we observe that $\tilde{r}_{\rm sat}$ is equal to $\tilde{r}_{\rm T}$ below $M_{\rm T}$, and is constant. Injecting relations~(\ref{eq:K_BH_small_mass_a}), (\ref{eq:K_BH_medium_mass}) or (\ref{eq:K_BH_small_mass_b}) in definition~(\ref{eq:Gamma_BH_c}) yields the same expression
\beq
\Gamma_{\rm BH} = \frac{1}{2} \asv \left\{ \frac{\rho_{\rm sat}}{m_{\chi}} \right\}^{2} \! r_{\rm S}^{3} \times
\frac{4}{3} \pi_{\,} \tilde{r}_{\rm T}^{3} \times {\cal K}_{3/2} \,,
\label{eq:Gamma_BH_d}
\eeq
where ${\cal K}_{3/2}$ is identified with ${\cal K}_{\rm BH}$.
Refined expressions for  ${\cal K}_{3/2}$ are collected in appendix~\ref{append:more_correct_Gamma_BH_slope_0} when the Schwarzschild radius of the black hole is taken into account. More specifically, ${\cal K}_{3/2}$ is defined by~(\ref{eq:K_BH_medium_mass_rS_on}) when $\rho_{\rm sat} \geq \rho_{\rm B_{1}}$ and $M_{1} \leq M_{\rm BH} \leq M_{\rm T}$. Alternatively, expression~(\ref{eq:K_BH_small_mass_a_rS_on}) can be used if $\rho_{\rm sat} \geq \rho_{\rm B_{1}}$ and $M_{\rm BH} \leq M_{1}$, or if $\rho_{\rm sat} \leq \rho_{\rm B_{1}}$ and $M_{\rm BH} \leq M_{\rm T}$. In the latter situation, the mass $M_{\rm T}$ is smaller than $M_{1}$, and ${\cal K}_{3/2}$ is also given by~(\ref{eq:K_BH_small_mass_b_rS_on}). Parameter  ${\cal K}_{3/2}$ is of order ${\cal O}(1)$.

\vskip 0.1cm
At low PBH mass, the saturation plateau is surrounded by a DM layer with slope $3/2$, and the saturation density scales like
\beq
\rho_{\rm sat} \propto {\tilde{r}_{\rm T}}^{-3/2}
\;\;\;\text{hence}\;\;\;
\rho_{\rm sat}^{2} \, {\tilde{r}_{\rm T}}^{3} \propto {\rm constant.}
\label{eq:scaling_saturation_sphere_3_2}
\eeq
Using Eq.~(\ref{eq:r_tilde_T_a}) to replace $\tilde{r}_{\rm T}$ in relation~(\ref{eq:Gamma_BH_d}) makes the saturation density $\rho_{\rm sat}$ cancel out. 
The cosmological DM density at kinetic decoupling $\rho_{i}^{\rm kd}$ appears, which we express as a function of $m_{\chi}$, $x_{\rm kd}$ and the effective number $h_{\rm eff}(T_{\rm kd})$ of entropic degrees of freedom at that time (a mere rescaling of today's DM density by the appropriate ratio of scale factors converted into a ratio of temperatures, which obviously involves the associated number of entropic degrees of freedom, see e.g.~ \cite{BinetruyEtAl1984a,KolbEtAl1990}). Tossing all these ingredients together, we derive the scaling relation
\beq
\Gamma_{\rm BH}^{(3/2)} = \left( 1.03 \times 10^{66} \; {\rm s^{-1}} \right) {\cal K}_{3/2}
\left\{ \frac{h_{\rm eff}(T_{\rm kd})}{h_{\rm eff}(T_{\rm eq})} \right\}^{\! 2}
{\cal S} \, \mu^{4} \, x_{\rm kd}^{-3} \, m^{3} \,,
\label{eq:scaling_G_BH_3_2_a}
\eeq
where ${\cal S}$ is the annihilation cross-section expressed in units of its thermal value, $\mu$ is the WIMP mass in TeV and $m$ is the PBH mass in solar units. The effective number of entropic degrees of freedom at matter-radiation equality is denoted by $h_{\rm eff}(T_{\rm eq})$.
We notice that the duration $\Delta t = t_{\rm U}(z) - t_{\rm eq}$ has disappeared. The minispike annihilation rate does not depend on redshift. As time goes on, the density $\rho_{\rm sat}$ of the saturation sphere decreases, but its radius $\tilde{r}_{\rm T}$ increases in such a way that it compensates for that effect as showed by the scaling~(\ref{eq:scaling_saturation_sphere_3_2}).
According to relation~(\ref{eq:scaling_G_BH_3_2_a}), increasing $\asv$ makes the solid curve of the left panel shift upward, while decreasing $m_{\chi}$ in the right panel makes it shift downward. Scaling~(\ref{eq:scaling_G_BH_3_2_a}) allows us to understand the results displayed in Fig.~\ref{fig:Gamma_BH_vs_DM_parameters} in the low PBH mass regime.

\vskip 0.1cm
For heavy black holes, the extension $\tilde{r}_{\rm sat}$ of the saturation plateau decreases with PBH mass like $M_{\rm BH}^{-2}$. Two cases must be distinguished depending on how the saturation density $\rho_{\rm sat}$ compares with $\rho_{\rm B_{1}}$.

\vskip 0.1cm
\noindent $\bullet$ {\bf $M_{\rm BH} \geq M_{\rm T}$ and $\rho_{\rm sat} \geq \rho_{\rm B_{1}}$}\\
Replacing ${\cal J}_{\rm BH}$ by expression~(\ref{eq:K_BH_large_mass_a}) in its definition~(\ref{eq:Gamma_BH_c}), we can write the annihilation rate as
\beq
\Gamma_{\rm BH} = \frac{1}{2} \asv \left\{ \frac{\rho_{\rm sat}}{m_{\chi}} \right\}^{2} \! r_{\rm S}^{3} \times
\frac{4}{3} \pi_{\,} \tilde{r}_{\rm T}^{3} \times \left\{ \frac{M_{\rm T}}{M_{\rm BH}} \right\}^{\! 2}
{\cal K}_{9/4} \,.
\label{eq:Gamma_BH_e}
\eeq
The coefficient ${\cal K}_{9/4}$ can be derived from relations~(\ref{eq:K_BH_large_mass_a_rS_on}) and (\ref{eq:K_BH_very_large_mass_a_rS_on}) depending on whether $\tilde{r}_{\rm H}$ is larger, or smaller, than $1$. This yields
\begin{equation}
{\cal K}_{9/4} = \left\{
\begin{tabular}{ll}
$3 - \left( {1}/{\tilde{r}_{\rm H}^{3}} \right) - 2 \left( {\tilde{r}_{\rm H}}/{\tilde{r}_{\rm E}} \right)^{3/2}$ & if $\tilde{r}_{\rm H} \geq 1 \,,$\\
&\\
$2 \, \tilde{r}_{\rm H}^{3/2} \left\{ 1 - \left( {1}/{\tilde{r}_{\rm eq}} \right)^{3/2} \right\}$ & if $\tilde{r}_{\rm H} \leq 1 \,.$
\end{tabular}
\label{eq:K_9_4_very_large_mass_a_rS_on}
\right.
\end{equation}
For super-heavy black holes, the minispike radius may become smaller than the Schwarzschild radius and ${\cal K}_{9/4}$ vanishes.

\vskip 0.1cm
In this regime, the transition mass $m_{\rm T} \equiv {M_{\rm T}}/{\rm M_{\odot}}$ is given by~(\ref{eq:definition_m_T_a}). The saturation radius $\tilde{r}_{\rm T}$ cancels out from Eq.~(\ref{eq:Gamma_BH_e}). It is replaced by the radius of influence $\tilde{r}_{\rm T}^{0}$ of a solar mass black hole at the exact time $t_{i}$ when the cosmological DM density is equal to
\beq
\rho_{i} = \sqrt{\frac{\pi^{3}}{2}} \, \frac{\rho_{\rm sat}}{{\cal I}_{9/4}^{\rm asy}}
\;\;\;\text{where}\;\;\;
\rho_{\rm sat} \equiv \rho_{9/4}(\tilde{r}_{\rm T} , m_{\rm T}) \equiv \rho_{9/4}(\tilde{r}_{\rm T}^{0} , 1) \,.
\eeq
The saturation sphere is surrounded by a DM layer with slope $9/4$. 
Once the cosmological density $\rho_{i}$ has been determined, the reduced radius $\tilde{r}_{\rm T}^{0}$ can be readily obtained as explained in appendix~\ref{append:tilde_r_0_vs_rho_i}. Both quantities fulfill the relation
\beq
{\cal C}_{2}(T_{i}) =
\left\{ \frac{\tilde{r}_{\rm T}^{0}}{\tilde{r}_{\rm eq}^{0}} \right\}
\left\{ \frac{\rho_{i}}{\rho_{i}^{\rm eq}} \right\}^{\! 4/9}
\;\;\text{with}\;\;\;
{\cal C}_{2}(T_{i}) \equiv
\left\{ \frac{g_{\rm eff}(T_{\rm eq})}{g_{\rm eff}(T_{i})} \right\}^{\! 1/3}
\left\{ \frac{h_{\rm eff}(T_{i})}{h_{\rm eff}(T_{\rm eq})} \right\}^{\! 4/9} \!,
\eeq
where $T_{i}$ denotes the temperature of the primordial plasma at time $t_{i}$ while $\rho_{i}^{\rm eq}$ is the DM density at matter-radiation equality. In appendix~\ref{append:tilde_r_0_vs_rho_i}, we explain how $\tilde{r}_{\rm T}^{0}$ can be derived from $\rho_{i}$, together with $T_{i}$ and ${\cal C}_{2}$.
Replacing $\tilde{r}_{\rm T}^{0}$ by its expression in terms of $\rho_{\rm sat}$ and ${\cal C}_{2}$ in Eq.~(\ref{eq:Gamma_BH_e}), it is straightforward to establish the new scaling relation for the minispike annihilation rate
\beq
\Gamma_{\rm BH}^{(9/4)} = \left( 6.26 \times 10^{34} \; {\rm s^{-1}} \right) {\cal K}_{9/4} \; {\cal C}_{2}^{3} \;
{\cal S}^{1/3} \, {\mu}^{-4/3} \, \tau^{-2/3} \, m \,,
\label{eq:scaling_G_BH_9_4_a}
\eeq
where $\tau$ denotes the duration $\Delta t$ expressed in units of $10^{15} \; {\rm s}$. Several remarks are in order.
\renewcommand{\labelitemi}{$-$}
\begin{itemize}
\item{The annihilation rate scales now linearly with PBH mass. In Fig.~\ref{fig:Gamma_BH_vs_DM_parameters}, this new scaling appears clearly on all curves above the orange dots.}
\item{As already discussed, the kinetic decoupling parameter $x_{\rm kd}$ does not come into play in the annihilation rate~(\ref{eq:scaling_G_BH_9_4_a}), in agreement with the behavior of the dotted short-dashed curve of the right panel.}
\item{The annihilation rate is a decreasing function of the duration $\tau$. As time goes on, minispikes around massive objects have a declining absolute luminosity. This explains why the high-mass part of the dotted short-dashed curve of the left panel shifts downward when the redshift decreases from $10^{3}$ down to $0$.}
\item{As featured by the long-dashed curve of the right panel above the orange dot, the lighter the DM species, the larger the annihilation rate. Conversely, this rate decreases at low PBH mass. All curves are shifted rightward and upward in the $(M_{\rm BH} , \Gamma_{\rm BH})$ plane as $m_{\chi}$ is decreased.}
\item{We finally observe that the rate $\Gamma_{\rm BH}$ is an increasing function of the annihilation cross-section $\asv$.}
\end{itemize}

\vskip 0.1cm
\noindent $\bullet$ {\bf $M_{\rm BH} \geq M_{\rm T}$ and $\rho_{\rm sat} \leq \rho_{\rm B_{1}}$}\\
In this configuration, the saturation plateau extends up to the surface of the minispike. The saturation radius $\tilde{r}_{\rm sat}$ can be identified with the influence radius $\tilde{r}_{\rm eq}$ of the object at matter-radiation equality. The radius $\tilde{r}_{\rm T}$ is still defined by~(\ref{eq:r_tilde_T_a}) while the transition mass $m_{\rm T} \equiv {M_{\rm T}}/{\rm M_{\odot}}$ is now given by~(\ref{eq:definition_m_T_b}) so that
\beq
\tilde{r}_{\rm sat} \equiv \tilde{r}_{\rm eq}(m) = \tilde{r}_{\rm T} \left( {m}/{m_{\rm T}} \right)^{-2/3} \,.
\eeq
Using expressions~(\ref{eq:K_BH_large_mass_b}) and (\ref{eq:K_BH_large_mass_b_rS_on}) in definition~(\ref{eq:Gamma_BH_c}), we can write the annihilation rate as
\beq
\Gamma_{\rm BH} = \frac{1}{2} \asv \left\{ \frac{\rho_{\rm sat}}{m_{\chi}} \right\}^{2} \! r_{\rm S}^{3} \times
\frac{4}{3} \pi_{\,} \tilde{r}_{\rm T}^{3} \times \left\{ \frac{M_{\rm T}}{M_{\rm BH}} \right\}^{\! 2}
{\cal K}_{0} \,,
\label{eq:Gamma_BH_f}
\eeq
where ${\cal K}_{0} = 1 - \left( {1}/{\tilde{r}_{\rm eq}} \right)^{3}$.
The scaling of the annihilation rate as a function of the parameters of the model can be described by
\beq
\Gamma_{\rm BH}^{(0)} = \left( 4.26 \times 10^{46} \; {\rm s^{-1}} \right) {\cal K}_{0} \;
{\cal S}^{-1} \, \tau^{-2} \, m \,.
\label{eq:scaling_G_BH_0_a}
\eeq
The annihilation rate is still linear in PBH mass. Its decline with duration $\Delta t$ is faster than in the previous configuration where $\rho_{\rm sat} \geq \rho_{\rm B_{1}}$.
But the most important difference is that $\Gamma_{\rm BH}$ is now a decreasing function of the annihilation cross-section, in contrast with the other situations. We anticipate that the annihilation rate should exhibit at some point a maximum depending on $M_{\rm BH}$ and $\asv$. Setting the other parameters fixed while keeping in mind that increasing $\asv$ always makes the saturation density drop by virtue of relation~(\ref{eq:definition_rho_sat_a}), we can distinguish two cases.
%
\begin{enumerate}
\item{If $M_{\rm BH}$ is larger than $M_{1}$, the annihilation rate increases with $\asv$ as long as $\rho_{\rm sat}$ exceeds $\rho_{\rm B_{1}}$. When $\rho_{\rm sat}$ becomes smaller than $\rho_{\rm B_{1}}$, the transition mass $M_{\rm T}$ gets smaller than $M_{1}$. We are in the situation where the scaling~(\ref{eq:scaling_G_BH_0_a}) applies, with $\Gamma_{\rm BH}$ decreasing now with $\asv$. The maximum of $\Gamma_{\rm BH}$ as a function of $\asv$ is therefore reached when $\rho_{\rm sat}$ and $\rho_{\rm B_{1}}$ are equal, i.e. at the critical cross-section
\beq
{\asv}_{\rm crit} = \frac{m_{\chi}}{\rho_{\rm B_{1}} \Delta t} = (1.32 \times 10^{-17} \; {\rm cm^{3} \, s^{-1}}) \, \mu \, \tau^{-1} \,.
\label{eq:sig_v_max_universal}
\eeq
The numerical value derived here does not depend on PBH mass. Notice that the post-collapse DM density $\rho_{\rm B_{1}}$ is a constant as showed in table~\ref{tab:rho_sat_vs_rho_B1_B2}. The critical cross-section ${\asv}_{\rm crit}$ increases with WIMP mass and decreases as cosmic time goes on.
}
\item{If the PBH mass is smaller than $M_{1}$, the maximum of $\Gamma_{\rm BH}$ as a function of $\asv$ is now reached when the transition mass $M_{\rm T}$ becomes equal to $M_{\rm BH}$. The corresponding saturation density can be obtained from~(\ref{eq:r_tilde_T_a}) and (\ref{eq:definition_m_T_b}), with the masses $m_{\rm T}$ and $m$ set equal. Defining the annihilation cross-section at the peak by the ratio ${m_{\chi}}/{\left( \rho_{\rm sat} \Delta t \right)}$ leads to the scaling relation
\beq
{\asv}_{\rm crit} = (5.94 \times 10^{-36} \; {\rm cm^{3} \, s^{-1}}) \,
\left\{ \frac{h_{\rm eff}(T_{\rm eq})}{h_{\rm eff}(T_{\rm kd})} \right\} \, \mu^{-2} \, x_{\rm kd}^{3/2} \, \tau^{-1} \, m^{-1} \,.
\label{eq:sig_v_max_normalization}
\eeq
We remark that this expression yields the same result as Eq.~(\ref{eq:sig_v_max_universal}) when $M_{\rm BH}$ is set equal to $M_{1}$.
}
\end{enumerate}

\subsection{Using the soft approximation for the annihilation rate}
\label{subsec:soft_rho_and_Gamma_BH}

\begin{figure}[h!]
\centering
\includegraphics[width=0.495\textwidth]{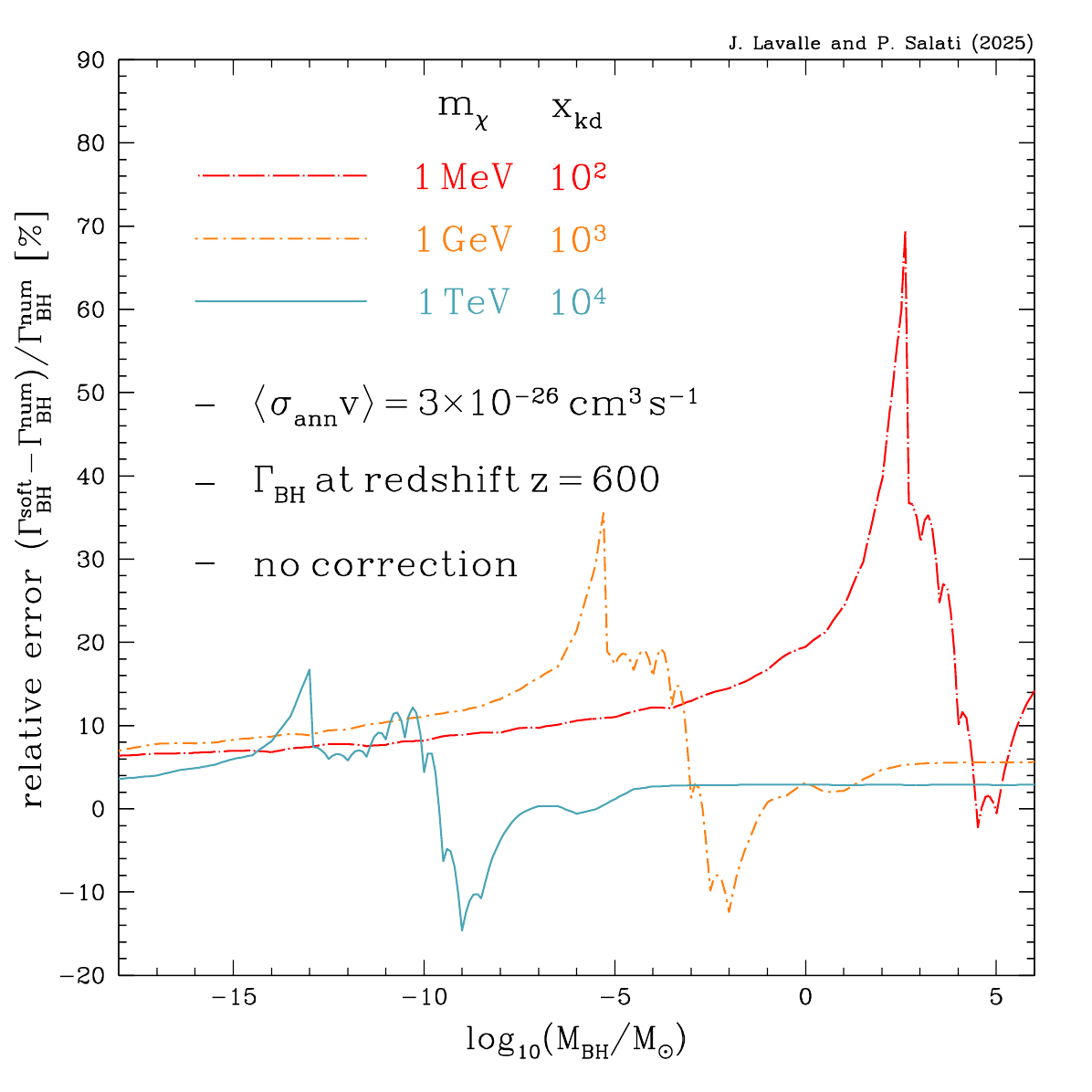}
\includegraphics[width=0.495\textwidth]{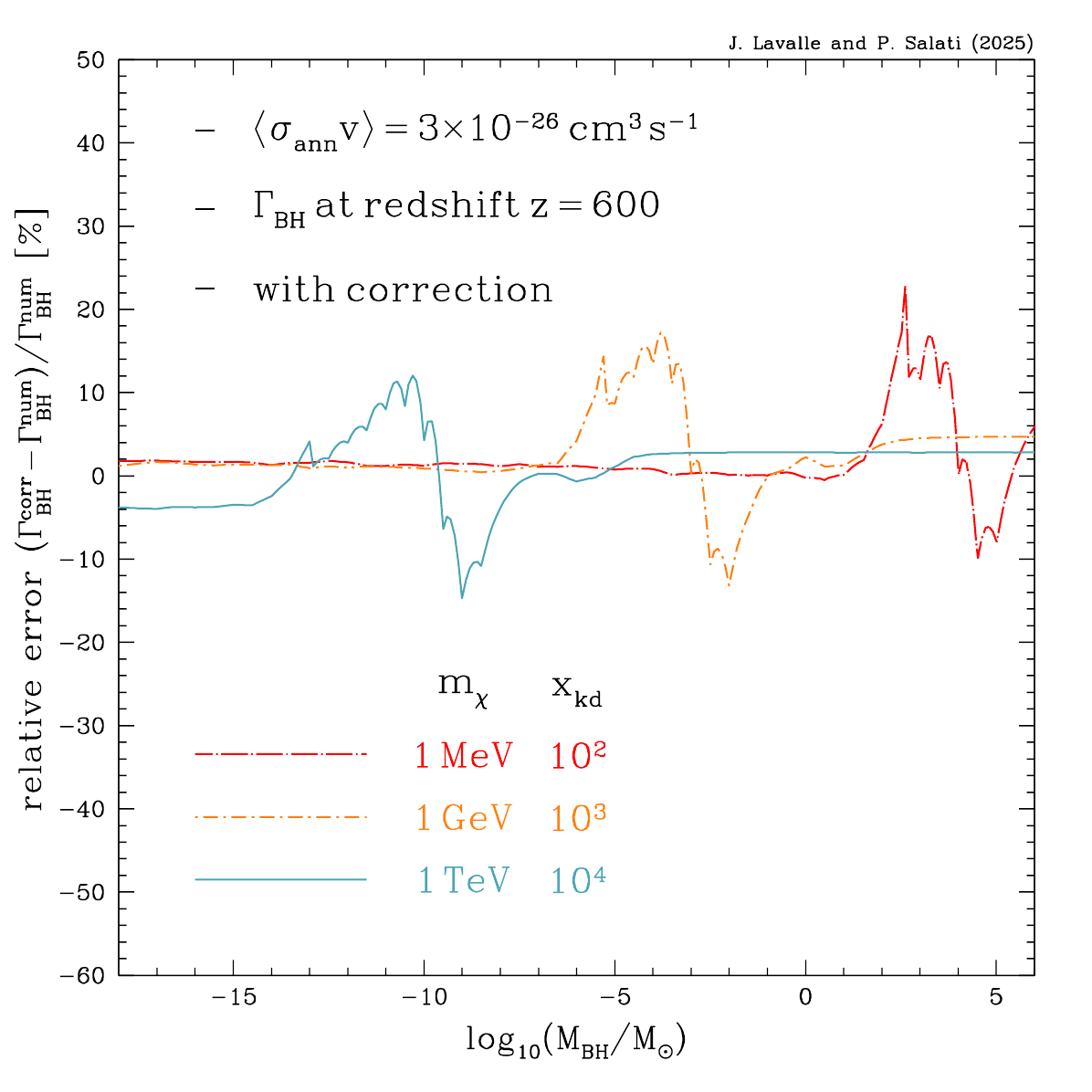}
\caption{The relative error induced on the minispike annihilation rate $\Gamma_{\rm BH}$ by using the soft approximation~(\ref{eq:soft_approximation_DM_rho}) instead of the numerical integral~(\ref{eq:post_collapse_density}) for the DM post-collapse density is presented in the {\bf left panel} as a function of PBH mass. Three representative configurations have been selected for the WIMP mass $m_{\chi}$ and the kinetic decoupling parameter $x_{\rm kd}$.
The annihilation rate is derived at a redshift $z$ of $600$, while a thermal annihilation cross-section is assumed.
In the {\bf right panel}, the soft approximation has been corrected close to the minispike surface by using expansions~(\ref{eq:correction_soft}). The agreement with the numerical result improves significantly.
\label{fig:error_Gamma_BH_various_models}}
\end{figure}

To close this section, we explore the impact on the minispike annihilation rate $\Gamma_{\rm BH}$ of using the soft approximation~(\ref{eq:soft_approximation_DM_rho}) instead of the result of integral~(\ref{eq:post_collapse_density}) to derive the post-collapse DM density $\rho(\tilde{r})$.
In the left panel of Fig.~\ref{fig:error_Gamma_BH_various_models}, we plot the relative difference between both rates as a function of PBH mass for three different models and redshift $z=600$.
%
%
The agreement between the full numerical result, and the calculation based on the soft approximation, is fairly good for a 1~TeV DM particle (solid blue). Over most of the PBH mass range, the soft approximation yields the numerical result up to a relative error of $\pm 10 \%$. The error reaches however $+  16 \%$ and $- 14 \%$ at PBH mass $10^{-13} \; {\rm M_{\odot}}$ and $10^{-9} \; {\rm M_{\odot}}$. This is already an excellent result given that $\Gamma_{\rm BH}$ varies logarithmically over more than 39 decades.

\vskip 0.1cm
The situation deteriorates though for smaller DM masses.
For a 1~GeV WIMP, the error is acceptably bounded by $\pm 10 \%$, except for PBH masses between $10^{-11}$ and $10^{-3} \; {\rm M_{\odot}}$. The dotted short-dashed orange curve exhibits a peak at $5 \times 10^{-6} \; {\rm M_{\odot}}$ where it reaches +35\%.
%
%
%
The situation worsens even further in the 1~MeV case (dotted long-dashed red) for which the error exceeds +14\% between $3 \times 10^{-3}$ and $10^{4} \; {\rm M_{\odot}}$, peaking at +70\% for a mass of $4.5 \times 10^{2} \; {\rm M_{\odot}}$.
%
%
%
As already noticed at the end of section~\ref{subsec:num_vs_approx}, the post-collapse density $\rho(\tilde{r})$ is overestimated by the soft approximation~(\ref{eq:soft_approximation_DM_rho}) close to the surfaces of minispikes, and so is the annihilation rate. This effect is most important for PBH masses close to $M_{1}$. The peaks, which the dotted short-dashed orange and dotted long-dashed red curves exhibit, can be actually associated to the critical mass $M_{1}$, respectively equal to $5.282 \times 10^{-6}$ and $4.512 \times 10^{2} \; {\rm M_{\odot}}$.

\vskip 0.1cm
The larger the volume of the surface layer where the density is overestimated, relative to the volume of the entire minispike, the larger the annihilation rate derived using the soft approximation compared to the full numerical result. This effect is most sensitive for light DM species.
Actually, at fixed kinetic decoupling parameter $x_{\rm kd}$, the smaller $m_{\chi}$, the smaller the decoupling temperature $T_{\rm kd}$. In  the phase diagram of Fig.~\ref{fig:phase_diagram}, the dotted short-dashed red line corresponding to $\tilde{r}_{\rm kd}$ is shifted upward as $T_{\rm kd}$ decreases, hence a larger critical mass $M_{1}$ and a smaller extension of the corresponding minispike in reduced coordinates. The overestimation of $\rho(\tilde{r})$ becomes more acute in this situation than for heavier DM species.

\vskip 0.1cm
In the next section, we will concentrate on DM particles heavier than a few tens of GeV. In that mass range, the soft approximation can perfectly be used to calculate decent values of $\Gamma_{\rm BH}$.
One of the goals of this work is nevertheless to derive an acceptable approximation for the post-collapse density $\rho(\tilde{r})$,  which ought to be valid over the entire parameter space. That is why we have slightly corrected the soft approximation by respectively replacing, in the asymptotic expressions~(\ref{eq:rho_asymptotic_slope_3_2}) and (\ref{eq:rho_slope_9_4}), the integrals ${\cal I}_{3/2}^{\rm asy}$ and ${\cal I}_{9/4}^{\rm asy}$ by
\beq
{\cal I}_{3/2} = {\cal I}_{3/2}^{\rm asy} \, - \, \frac{3 \pi}{8} \frac{\tilde{r}}{\tilde{r}_{\rm eq}}
\;\;\;\text{and}\;\;\;
{\cal I}_{9/4} = {\cal I}_{9/4}^{\rm asy} \, - \, \frac{4}{3} \sqrt{2 \pi} \, \left\{ \frac{\tilde{r}}{\tilde{r}_{\rm eq}} \right\}^{\! 3/4}.
\label{eq:correction_soft}
\eeq
These relations have been derived in our previous study\footnote{More precisely, we have borrowed from~\cite{BoudaudEtAl2021} expansions~(5.22) for $\rho_{3/2}$ and (5.50) for $\rho_{9/4}$.}\cite{BoudaudEtAl2021}. The profile close to the surface is now better described, and so is the annihilation rate. As is clear in the right panel of Fig.~\ref{fig:error_Gamma_BH_various_models}, the tension between the approximated and fully numerical results for $\Gamma_{\rm BH}$ is alleviated by taking into account expansions~(\ref{eq:correction_soft}).
Although there is not much of an effect at 1~TeV, the situation dramatically improves for the other cases, with a discrepancy lying almost always in the band $\pm$15\%. The peaks of the left panel have been substantially erased. In particular, while $\Gamma_{\rm BH}$ spans more than 66 decades in the 1~MeV case, the error is now confined between -10\% and +22\%.

\vskip 0.1cm
The same analysis has been performed at redshift $z = 0$ (today). Results do not change much with respect to the case $z = 600$.
When corrections~(\ref{eq:correction_soft}) are not used, the agreement between the annihilation rates derived using the soft approximation~(\ref{eq:soft_approximation_DM_rho}) and integral~(\ref{eq:post_collapse_density}) slightly deteriorates, especially for a 1~MeV particle.
Including these corrections noticeably improves the agreement, as is the case at redshift~600.
For a 1~TeV species, the discrepancy lies in the band $\pm$14\%.
For a 1~GeV particle, it is confined between -8\% and peaks at +32\%.
For a 1~MeV WIMP, it lies between -10\% and +2\% below $1 \; {\rm M_{\odot}}$. Above that value, as PBH mass increases, the relative difference peaks down at -30\% and increases up to +24\% for very heavy objects.

\vskip 0.1cm
All these results are remarkable and make us confident that the soft approximation~(\ref{eq:soft_approximation_DM_rho}), once corrected by~(\ref{eq:correction_soft}), can be safely used for fast and robust calculations of the post-collapse density $\rho(\tilde{r})$ and minispike annihilation rate $\Gamma_{\rm BH}$. That is why we have already used it in Fig.~\ref{fig:Gamma_BH_vs_DM_parameters}. We will also use it in the next section.

\section{Observational constraints}
\label{sec:constraints}

In the mixed scenario explored in this article, primordial black holes and thermal species contribute respectively a fraction $f_{\rm BH}$ and $f_{\rm DM} \equiv 1 - f_{\rm BH}$ to the astronomical DM. In the previous sections, results were derived assuming a fraction $f_{\rm DM}$ of $1$. From now on, that fraction is left free to vary in the range from $0$ (no WIMPs) to $1$ (no PBHs).
Whatever the value of $f_{\rm DM}$, the total amount of DM has been fixed once and for all by observations of the CMB~\cite{AghanimEtAl2020}. All the relations previously derived between temperature and energy density of the primordial plasma, cosmological time and radius of influence for a given PBH mass can still be used.

\vskip 0.1cm
The only modification bears on the cosmological density $\rho_{i}$ of DM thermal particles, which is to be multiplied by a factor $f_{\rm DM} \leq 1$. According to its definition~(\ref{eq:post_collapse_density}), so is the post-collapse density $\rho(\tilde{r})$.
On the other hand, the saturation density $\rho_{\rm sat}$ is only determined by the WIMP mass $m_{\chi}$ and annihilation cross-section $\asv$, as well as by the duration $\Delta t$. Once WIMP annihilations are taken into account, the DM density of minispikes is given by relation~(\ref{eq:rho_after_DM_annihilation_a}) where the post-collapse density $\rho(\tilde{r})$ is rescaled by a factor $f_{\rm DM}$. This yields
\beq
\rho(t , \tilde{r}) = f_{\rm DM} \times \frac{\rho'_{\rm sat} \, \rho(\tilde{r})}{\rho'_{\rm sat} + \rho(\tilde{r})}
\;\;\;\text{where}\;\;\;
\rho'_{\rm sat} \equiv {\rho_{\rm sat}}/{f_{\rm DM}} \,.
\label{eq:rho_after_DM_annihilation_b}
\eeq
Up to a global rescaling by a factor $f_{\rm DM}$, the entire minispike structure is the same as that considered in the previous section, but now with a saturation density $\rho'_{\rm sat}$ instead of $\rho_{\rm sat}$. The smaller the fraction $f_{\rm DM}$, the larger $\rho'_{\rm sat}$. This leads us to the scaling relation for the annihilation rate
\beq
\Gamma_{\rm BH} \left\{ \rho_{\rm sat} , f_{\rm DM} \rho_{i} \right\} = f_{\rm DM}^{2} \times
\Gamma_{\rm BH} \left\{ \rho'_{\rm sat} , \rho_{i} \right\}.
\label{eq:scaling_G_BH_f_DM}
\eeq
This relation has immediate implications for the asymptotic expressions~(\ref{eq:scaling_G_BH_3_2_a}), (\ref{eq:scaling_G_BH_9_4_a}) and (\ref{eq:scaling_G_BH_0_a}) derived in section~\ref{subsec:Gamma_BH_comparison}.

\vskip 0.1cm
Using relation~(\ref{eq:r_tilde_T_a}), we first remark that the radius of the saturation plateau shrinks with $f_{\rm DM}$ like
\beq
\tilde{r}_{\rm T} \propto {1}/{{\rho'}_{\rm sat}^{2/3}} \propto f_{\rm DM}^{2/3} \,.
\eeq
The coordinates of the transition point T in the $(\tilde{r} , M_{\rm BH})$ plane change. In the phase diagram of Fig.~\ref{fig:saturated_PD_and_sat_profiles} for instance, T is shifted toward the upper-left direction along the line connecting points B$_{1}$ and B$_{2}$. The coefficient ${\cal C}_{2}$ of Eq.~(\ref{eq:scaling_G_BH_9_4_a}) corresponds now to $\rho'_{\rm sat}$ and grows slightly.
Depending on whether the saturation density is larger or smaller than $\rho_{\rm B_{1}}$, the transition mass $M_{\rm T}$ increases like
\beq
m_{\rm T} \propto f_{\rm DM}^{-1/3}
\text{ if $\rho_{\rm sat} \geq \rho_{\rm B_{1}}$   or   } \propto f_{\rm DM}^{-1}
\text{ if $\rho'_{\rm sat} \leq \rho_{\rm B_{1}}$}\,.
\label{eq:definition_m_T_c}
\eeq
Making use of the rescaling relation~(\ref{eq:scaling_G_BH_f_DM}), we realize that the asymptotic expressions~(\ref{eq:scaling_G_BH_3_2_a}) and (\ref{eq:scaling_G_BH_9_4_a}) are still valid, with same normalization coefficients, but include now the fraction of DM species with
\beq
\Gamma_{\rm BH}^{(3/2)} \propto f_{\rm DM}^{2}
\;\;\;\text{while}\;\;\;
\Gamma_{\rm BH}^{(9/4)} \propto f_{\rm DM}^{4/3} \,.
\label{eq:G_BH_scaling_f_DM}
\eeq
The coefficients ${\cal K}_{3/2}$ and ${\cal K}_{9/4}$ also vary as they depend on $\tilde{r}_{\rm T}$.
Quite the contrary, $\Gamma_{\rm BH}^{(0)}$ is not sensitive to $f_{\rm DM}$. In that configuration, annihilations have completely erased the minispike structure and transformed it into a homogeneous distribution with density $\rho_{\rm sat}$.

\vskip 0.1cm
Having understood how the annihilation rate $\Gamma_{\rm BH}$ depends on the WIMP fraction $f_{\rm DM}$, we are ready to explore two kinds of constraints on our mixed DM scenario.
\renewcommand{\labelitemi}{$-$}
\begin{itemize}
\item{First, we can set limits on the abundance $f_{\rm BH}$ of PBHs assuming that most of the DM is made of thermal species. The goal is to close the asteroid window between $10^{-16}$ and $10^{-11} \; {\rm M_{\odot}}$, supplementing the bounds already derived in the literature~\cite{CarrEtAl2021,GreenEtAl2021}.
}
\item{Then, if GW observatories were to discover PBHs and measure their mass and contribution to the DM, stringent limits could be derived on the properties of WIMPs. This point has been disregarded in the literature and is one of the novelties of this work.
}
\end{itemize}
The constraints discussed hereafter arise from the signals produced by a population of annihilating minispikes during the post-recombination era. We will concentrate on the $\gamma$-ray extragalactic background, and on the injection of energy in the primordial plasma in connection with CMB angular distortions.
As a word of caution, we wish to warn the reader that what follows is mostly about setting up a detailed methodology to extract and interpret constraints. A detailed data analysis and statistical inference of the bounds is left to future and dedicated work.

\subsection{Bounds on the fraction $f_{\rm BH}$ of primordial black holes}
\label{subsec:f_PBH}

\subsubsection{Cosmological $\gamma$-ray background}
\label{subsubsec:f_BH_gamma_ray}

If a fraction $f_{\rm BH}$ of the DM is made of PBHs with monochromatic mass $M_{\rm BH}$, we expect the number density of these objects at redshift $z$ to be
\beq
n_{\rm BH}(z) = \frac{f_{\rm BH \,} \rho_{\rm dm}^{0}}{M_{\rm BH}} \, (1+z)^{3} \,,
\label{eq:n_BH_of_z_a}
\eeq
where the present DM density $\rho_{\rm dm}^{0}$ has been measured by the Planck collaboration~\cite{AghanimEtAl2020} to be equal to $2.25 \times 10^{-30} \; {\rm g \, cm^{-3}}$.
The associated DM minispikes continuously annihilate, producing standard model particles and eventually high-energy photons whose energy distribution, per annihilation event, is described by the primary spectrum
\beq
{\cal Q}_{\gamma}^{\rm prim}(E_{\gamma}) = \left. \frac{dN_{\gamma}}{dE_{\gamma}} \right|_{E_{\gamma}}.
\eeq
A cosmological background of $\gamma$-rays is produced, with flux at the Earth (e.g.~\cite{UllioEtAl2002})
\beq
\Phi_{\gamma}^{\rm BH}(E_{\gamma}) =
\frac{c}{4 \pi} \, \frac{f_{\rm BH \,} \rho_{\rm dm}^{0}}{M_{\rm BH}}
\int \! \frac{dz}{H(z)} \, \Gamma_{\rm BH}(z) \,
e^{-\tau_{\rm opt}(E_{\gamma} , z)} \,
{\cal Q}_{\gamma}^{\rm prim}(E'_{\gamma}) \,.
\label{eq:definition_Phi_gamma_a}
\eeq
The integral is performed along the line of sight, with $z$ running from $0$ up to $({m_{\chi}}/{E_{\gamma}})~{-}~1$. Since high-energy photons interact with the CMB or the intergalactic medium as they propagate to the Earth, they cannot come from far away. Their flux undergoes extinction with optical depth $\tau_{\rm opt}$.
The photon energy at the source is $E'_{\gamma} = (1+z) E_{\gamma}$.
The annihilation rate of minispikes $\Gamma_{\rm BH}$ also depends on redshift, as showed in Fig.~\ref{fig:Gamma_BH_vs_DM_parameters}.
The expansion rate of the universe at redshift $z$ may be expressed as
\beq
H(z) \simeq {H_{0}} \, \sqrt{\Omega_{\Lambda} + \Omega_{\rm m} (1+z)^{3}} \,,
\eeq
where $H_{0}$ is the Hubble constant, while $\Omega_{\rm m} = \Omega_{\rm dm} + \Omega_{\rm b}$ and $\Omega_{\Lambda}$ denote the matter and dark energy abundances today.
Observations can set limits on the extragalactic $\gamma$-ray background, providing an upper bound on $\Phi_{\gamma}^{\rm BH}$ and eventually on the fraction $f_{\rm BH}$.
The problem is quite involved though for at least two reasons
\renewcommand{\labelitemi}{$-$}
\begin{itemize}
\item{Annihilating DM particles inside minispikes also produce electrons and positrons. These subsequently scatter on the CMB, which they convert into inverse Compton (IC) photons. This secondary component must be added to the primary source ${\cal Q}_{\gamma}^{\rm prim}$.
}
\item{As regards $\gamma$-ray observations, several backgrounds must be thoroughly modeled. Interactions of high-energy cosmic ray protons and helium nuclei on the interstellar medium of the Milky Way, for instance, produce a Galactic foreground which needs to be substracted from data to get the extragalactic background. Another component arises from IC photons produced by Galactic electrons and positrons scattering on the CMB and stellar light.
Such an analysis has been performed by the Fermi-LAT collaboration who derived~\cite{Fermi-LAT:2014ryh} the extragalactic background, which is the sum of the isotropic background with the cumulative intensity from resolved Fermi-LAT sources. The latter  needs to be modeled to extract the former, which is the only useful component for our analysis.
}
\end{itemize}

This onerous task has been first performed by Ando and Ishiwata~\cite{AndoEtAl2015} (see also Blanco and Hooper \cite{Blanco:2018esa}) who considered various classes of extragalactic sources, namely blazars, star-forming galaxies and misaligned active galactic nuclei. Using the Fermi-LAT result~\cite{Fermi-LAT:2014ryh}, they performed a Bayesian statistical analysis to derive 95\% credible lower limits on the lifetime $\tau_{\rm dm}$ of decaying DM species, for a large variety of DM masses and annihilation channels.

\vskip 0.1cm
As originally proposed by Boucenna et al.~\cite{BoucennaEtAl2018}, the bounds derived by Ando and Ishiwata can be recast into constraints on $f_{\rm BH}$. Actually, in our mixed scenario, each PBH with its minispike can be seen as some sort of huge particle which continuously decays. In that respect, genuine decaying DM species and annihilating minispikes around primordial black holes behave in a similar way. How these objects are clustered does not come into play. This is not quite the case with annihilation, for which concentration plays a major role.

\vskip 0.1cm
Recasting the Ando and Ishiwata results~\cite{AndoEtAl2015} requires to take a decaying particle twice as massive as in the minispike scenario. We note that this little subtlety was actually missed in most of previous works \cite{BoucennaEtAl2018,AdamekEtAl2019,GinesEtAl2022}. As an illustration, let us consider the production of $b \bar{b}$ quark pairs. With that prescription, each $b$ jet is always produced with the energy $m_{\chi}$. Because of extinction, most of the photons detected at the Earth originate from the local universe. The rate of production of quark pairs, which in principle should be integrated along the line of sight, can be roughly taken at redshift $z \simeq 0$ (today) and is respectively given by
\beq
\frac{{\rho_{\rm dm}^{0}}{\tau_{\rm dm}^{-1}}}{2 m_{\chi}}
\;\text{(decaying DM)}
\;\;\;\text{and}\;\;\;
\frac{f_{\rm BH \,} \rho_{\rm dm \,}^{0} \Gamma_{\rm BH}}{M_{\rm BH}} \, 
\;\text{(minispike scenario).}
\label{eq:two_models_a}
\eeq
Because of the correspondence between decaying DM and annihilating minispikes, the expressions in Eq.~(\ref{eq:two_models_a}) can be equated. The lower limit $\tau_{\rm \, dm}^{\rm min}$ on the lifetime of a decaying DM species can be converted into a bound on PBHs, which we express as
\beq
f_{\rm BH} \Gamma_{\rm BH} \leq
\left\{ \frac{M_{\rm BH}}{2 m_{\chi}} \right\}
\left\{ \frac{1}{\tau_{\rm \, dm}^{\rm min}} \right\}.
\label{eq:maximum_f_BH_G_BH_a}
\eeq
An upper limit applies on the product $f_{\rm BH} \Gamma_{\rm BH}$. Notice that $\Gamma_{\rm BH}$ also depends on $f_{\rm BH}$ through the WIMP fraction $f_{\rm DM} = 1 - f_{\rm BH}$. We remark that $f_{\rm BH} \Gamma_{\rm BH}$ vanishes for $f_{\rm BH} = 0$ or $1$, and that it reaches a maximum while $f_{\rm BH}$ runs from $0$ to $1$.
If that maximum is smaller than the right-hand side term of Eq.~(\ref{eq:maximum_f_BH_G_BH_a}), no limit applies on the fraction of PBHs.
In the opposite situation, $f_{\rm BH}$ must be either small or close to $1$. An entire range of values is excluded, over which $f_{\rm BH} \Gamma_{\rm BH}$ exceeds the Ando and Ishiwata constraint.
Before proceeding any further, we have to emphasize that formally, since minispike annihilation has a time dependence that differs from standard dilution, it is very likely that the bound on decaying DM can actually not be recast as straightforwardly in terms of a bound on our mixed PBH-WIMP scenario. A more involved analysis is left to future work. Note that other complementary approaches to extract bounds from the extragalactic diffuse gamma-ray background have been followed, like for instance comparing the signal integrated at high redshift before galaxy formation with the current Fermi-LAT sensitivity~\cite{CarrEtAl2021a}.
%
\begin{figure}[h!]
\centering
\includegraphics[width=0.495\textwidth]{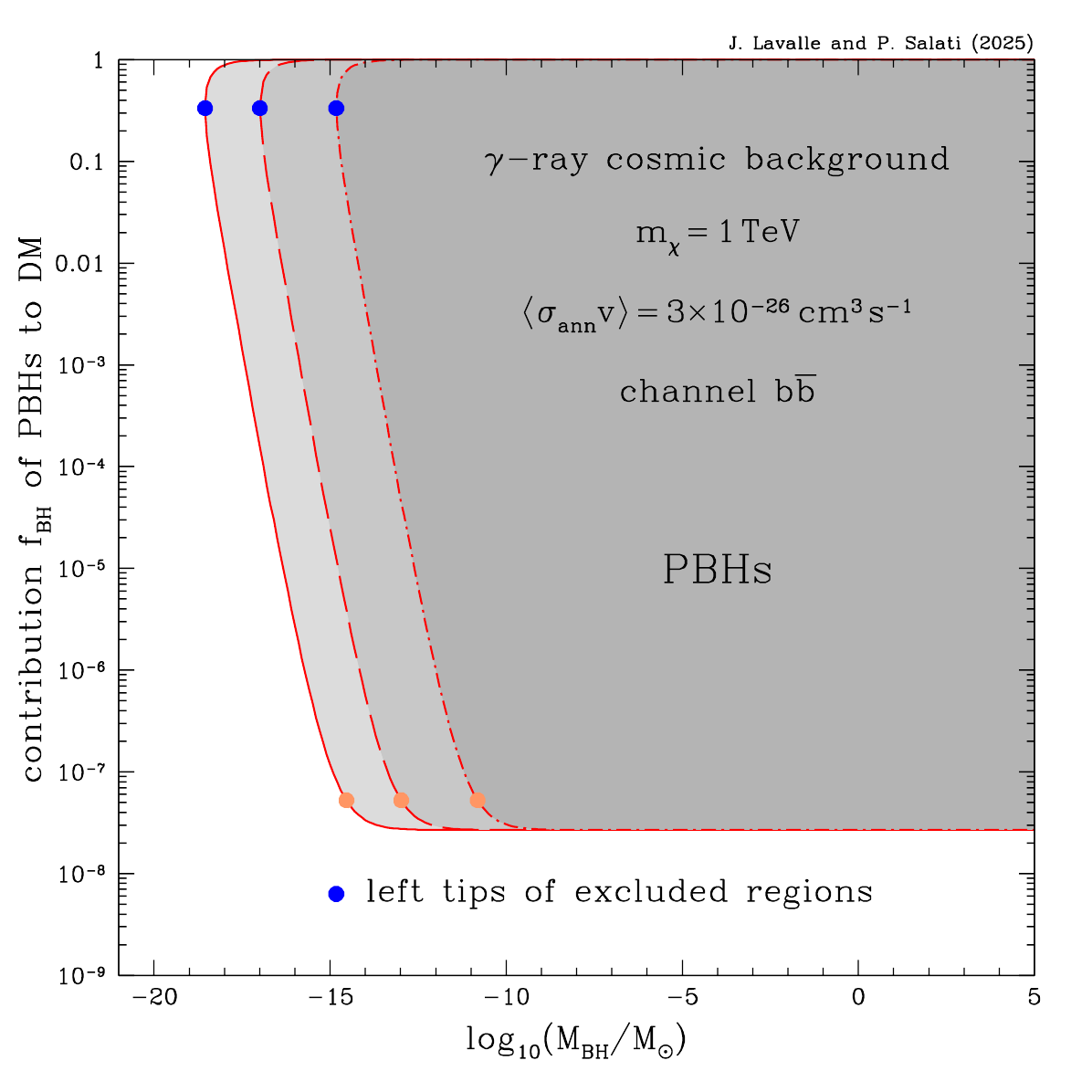}
\includegraphics[width=0.495\textwidth]{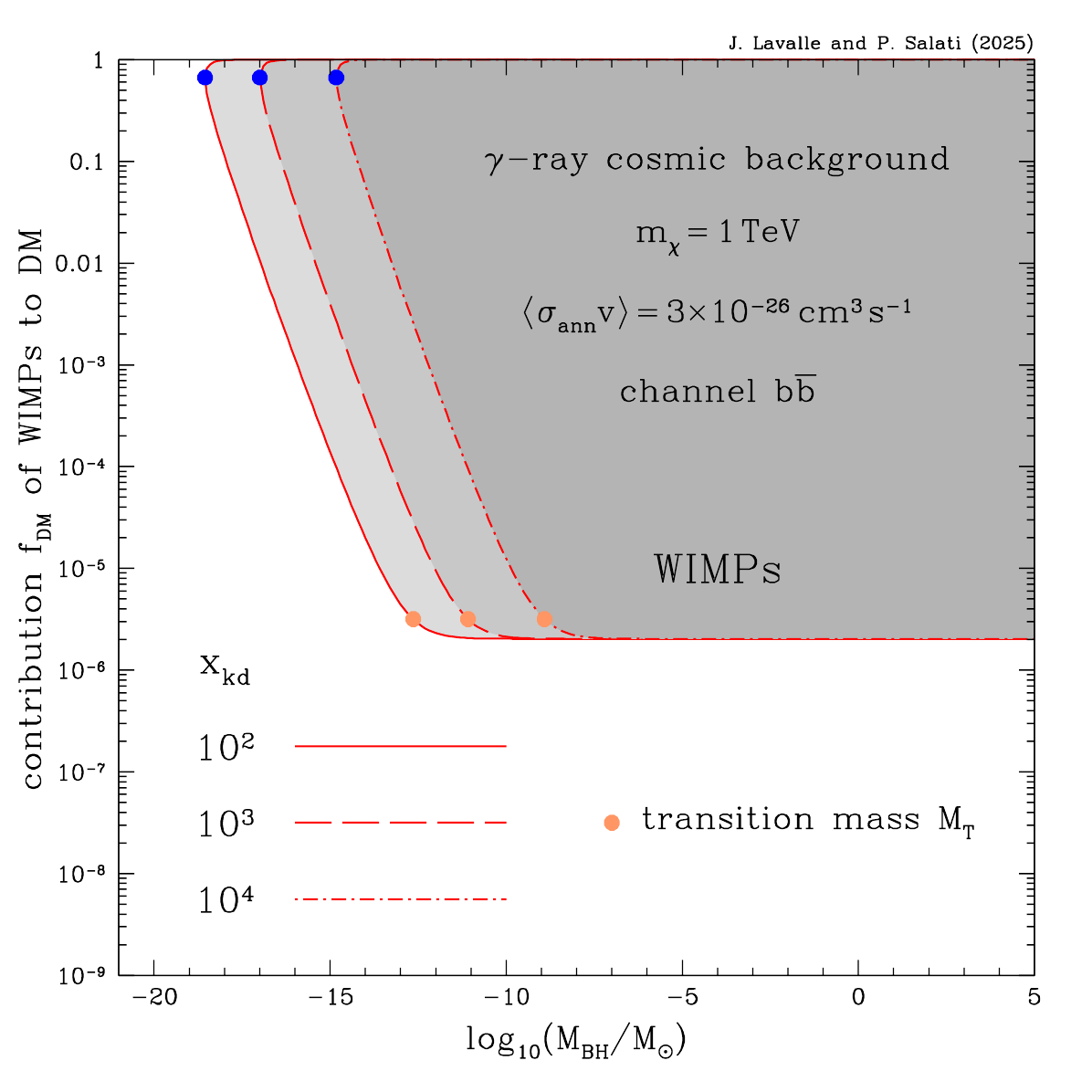}
\caption{The Fermi-LAT measurements~\cite{Fermi-LAT:2014ryh} of the extragalactic $\gamma$-ray background yield constraints on the contributions of PBHs and thermal species to the overall DM, respectively displayed in the {\bf left} and {\bf right panels}.
Ando and Ishiwata~\cite{AndoEtAl2015} have used the Fermi-LAT data to set bounds on decaying DM particles, which have been recast into the forbidden gray regions in the case of a 1~TeV WIMP with thermal annihilation cross-section.
The various shades of gray and contour types correspond to different values of the kinetic decoupling parameter $x_{\rm kd}$, as indicated.
Each panel features the same excluded domains observed from a different perspective, depending on whether $f_{\rm BH}$ or $f_{\rm DM}$ is used.
For heavy black holes, the PBH fraction $f_{\rm BH}$ is constrained to be either very small ({\bf left panel}) or very close to $1$ ({\bf right panel}). As already noticed by~\cite{LackiEtAl2010}, PBHs contribute all or almost nothing to the astronomical DM.
Light PBHs evade the $\gamma$-ray constraint to the left of the blue dots.
\label{fig:gamma_bounds_on_f_BH_and_f_DM}}
\end{figure}
%

\vskip 0.1cm
In Fig.~\ref{fig:gamma_bounds_on_f_BH_and_f_DM}, we have used the results that Ando and Ishiwata have displayed in the Fig.~3 of their analysis~\cite{AndoEtAl2015}. The lower limit $\tau_{\rm \, dm}^{\rm min}$ on the lifetime of a 2~TeV DM particle decaying into $b \bar{b}$ pairs is found to be $3.02 \times 10^{27} \; {\rm s}$.
%
%
In the left panel, the regions excluded by the cosmological $\gamma$-ray background are plotted in the $(M_{\rm BH} , f_{\rm BH})$ plane for a 1~TeV annihilating DM species, and three different values of the kinetic decoupling parameter $x_{\rm kd}$, each corresponding to a particular shade of gray. The forbidden domains are delineated by a red line whose type depends on $x_{\rm kd}$ as specified in the right panel. We emphasize that these bounds have to be appreciated for their qualitative and pedagogical values only, as we did not conduct any serious statistical analysis and further neglected the time dependence in the spike decay rate when recasting bounds derived for DM decay. Comparisons with other bounds from different probes are given in appendix section~\ref{append:comparison} for the sake of order-of-magnitude discussion.
For heavy black holes, the fraction $f_{\rm BH}$ cannot exceed $2.69 \times 10^{-8}$ whatever $M_{\rm BH}$ and $x_{\rm kd}$.
%
%
On the contrary, for light PBHs, the upper limit on $f_{\rm BH}$ becomes less stringent. It even disappears below some critical mass, respectively equal to $2.81 \times 10^{-19}$, $1.01 \times 10^{-17}$ and $1.50 \times 10^{-15} \; {\rm M_{\odot}}$ as $x_{\rm kd}$ is increased from $10^{2}$ to $10^{4}$. At the left tips of the gray excluded areas, corresponding to the blue dots in both panels, the PBH fraction is equal to ${1}/{3}$. Above that value, the red contours turn back rightward, while a lower limit, very close to $1$, applies this time on $f_{\rm BH}$.
%
%

\vskip 0.1cm
To observe it, the gray forbidden zones of the left panel have been plotted in the right panel as a function of WIMP fraction $f_{\rm DM}$. The lower limit on $f_{\rm BH}$ translates into an upper bound on $f_{\rm DM}$ at $2.03 \times 10^{-6}$, which is also independent of PBH mass and kinetic decoupling parameter. Once again, the constraint on $f_{\rm DM}$ weakens for light black holes and disappears at the same PBH masses as in the left panel, with a fraction $f_{\rm DM}$ equal to ${2}/{3}$ as could have been guessed.
Each panel of Fig.~\ref{fig:gamma_bounds_on_f_BH_and_f_DM} features the same excluded regions observed from a different perspective, depending on whether $f_{\rm BH}$ or $f_{\rm DM}$ is used.

\vskip 0.1cm
From the upper limits found on $f_{\rm BH}$ (left panel) and $f_{\rm DM}$ (right panel), we conclude that PBHs contribute either all or almost nothing to the astronomical DM, if the partner DM species is made of $s$-wave annihilating WIMPs, and if PBHs are typically heavier than the asteroid mass range. This curious property has been discovered more than 15~years ago by~Lacki and Beacom~\cite{LackiEtAl2010}, though on more qualitative grounds insofar as no exact solutions to the minispike profiles were known at the time. It has since then been confirmed by several studies~\cite{AdamekEtAl2019,CarrEtAl2021a}, which we improve over thanks to our careful analytical understanding, further correcting for the overlooked factor of $2$ in WIMP mass necessary to convert DM decay bounds into minispike annihilation bounds.
Primordial black holes and such $s$-wave annihilating particles seem to be incompatible. The former attract the latter to create dense minispikes which should have already been observed. The absence of a $\gamma$-ray excess points toward the scarcity or the overabundance of PBHs with respect to WIMPs.
Several remarks are now in order to analyze the structure of the forbidden regions.
\renewcommand{\labelitemi}{$-$}
\begin{itemize}
\item{In the right parts of the gray domains, the red cuves are horizontal. The upper bounds on $f_{\rm BH}$ (left panel) and $f_{\rm DM}$ (right panel) are insensitive to $M_{\rm BH}$ and $x_{\rm kd}$. In the regime of heavy black holes, the minispike annihilation rate is well approximated by the asymptotic expression~(\ref{eq:scaling_G_BH_9_4_a}) supplemented by~(\ref{eq:G_BH_scaling_f_DM}), hence
\beq
\Gamma_{\rm BH}^{(9/4)} \propto {\asv}^{1/3} \, {\mu}^{-4/3} \, m \; f_{\rm DM}^{4/3} \,,
\eeq
with no dependence on $x_{\rm kd}$. We recall that $m = {M_{\rm BH}}/{\rm M_{\odot}}$ and $\mu = {m_{\chi}}/{\rm 1 \; TeV}$. The PBH mass disappears from the inequality~(\ref{eq:maximum_f_BH_G_BH_a}), together with the kinetic decoupling parameter.}
\item{In this high PBH mass regime, the product $f_{\rm BH} \Gamma_{\rm BH}$ scales like $f_{\rm DM}^{4/3}$ when $f_{\rm BH} \simeq 1$, or like $f_{\rm BH}$ when $f_{\rm DM} \simeq 1$. It is no surprise if the upper bounds on the fractions of WIMPs and PBHs are related through
\beq
f_{\rm DM}^{\rm max} \simeq \left( f_{\rm \, BH}^{\rm max} \right)^{3/4} \,,
\eeq
as can be checked directly with the numerical values given above.
}
\item{For light PBHs, relations~(\ref{eq:scaling_G_BH_3_2_a}) and (\ref{eq:G_BH_scaling_f_DM}) should be used  for the minispike annihilation rate, which now scales like
\beq
\Gamma_{\rm BH}^{(3/2)} \propto {\asv} \, \mu^{4} \, x_{\rm kd}^{-3} \, m^{3} \, f_{\rm DM}^{2} \,.
\eeq
The upper bound on the product $f_{\rm BH} f_{\rm DM}^{2}$ is proportional to ${x_{\rm kd}^{3}}/{M_{\rm BH}^{2}}$. So are the left boundaries of the excluded domains in each panel of Fig.~\ref{fig:gamma_bounds_on_f_BH_and_f_DM}. Actually, the upper limits on the fractions $f_{\rm BH}$ and $f_{\rm DM}$ scale respectively like
\beq
f_{\rm \, BH}^{\rm max} \propto x_{\rm kd}^{3} \, m^{-2} \; \text{(PBHs)}
\;\;\;\text{while}\;\;\;
f_{\rm DM}^{\rm max} \propto x_{\rm kd}^{3/2} \, m^{-1} \; \text{(WIMPs)} \,.
\eeq
In the left and right panels, we do observe that the left borders of the gray regions drop respectively like ${1}/{M_{\rm BH}^{2}}$ and ${1}/{M_{\rm BH}}$.
At fixed fraction $f_{\rm BH}$, or $f_{\rm DM}$, the PBH masses at the left boundaries are expected to scale with kinetic decoupling parameter like
\beq
M_{\rm \, BH}^{\rm max} \propto x_{\rm kd}^{3/2} \,.
\eeq
As $x_{\rm kd}$ is increased from $10^{2}$ to $10^{3}$, the gray regions shift righward in agreement with that relation. But between $10^{3}$ and $10^{4}$, the observed shift is more important than expected. To explain this discrepancy, we note that the effective number of entropic degrees of freedom of the primordial plasma at kinetic decoupling $h_{\rm kd} \equiv h_{\rm eff}(T_{\rm kd})$ comes actually into play in relation~(\ref{eq:scaling_G_BH_3_2_a}). Taking it into account yields the new scaling
\beq
M_{\rm \, BH}^{\rm max} \propto x_{\rm kd}^{3/2} \, h_{\rm kd}^{-1} \,.
\eeq
As $x_{\rm kd}$ is increased from $10^{3}$ to $10^{4}$, the kinetic decoupling temperature $T_{\rm kd}$ drops from 1~GeV down to 100~MeV, i.e. below the quark/hadron phase transition, with a sensible reduction of $h_{\rm kd}$ by a factor $\sim 4$ which cannot be disregarded anymore.
%
%
}
\item{The transition between light and heavy PBH regimes occurs at the critical mass $M_{\rm T}$ defined in section~\ref{subsec:Gamma_BH_behavior}. In the configuration under scrutiny, the saturation density $\rho_{\rm sat}$ is 6 orders of magnitude larger than $\rho_{\rm B_{1}}$ (see table~\ref{tab:rho_sat_vs_rho_B1_B2}). We can apply relations~(\ref{eq:r_tilde_T_a}), (\ref{eq:definition_m_T_a}) and (\ref{eq:definition_m_T_c}) to derive $M_{\rm T}$, whose positions in both panels of Fig.~\ref{fig:gamma_bounds_on_f_BH_and_f_DM} are indicated by orange dots.
}
\item{The left tips of the excluded zones are indicated by blue dots. At these positions, the maximal value of $f_{\rm BH} \Gamma_{\rm BH}$ is barely equal to the right-hand side of inequality~(\ref{eq:maximum_f_BH_G_BH_a}). The former scales like
\beq
f_{\rm BH} \Gamma_{\rm BH} \simeq f_{\rm BH} \Gamma_{\rm BH}^{(3/2)} \propto f_{\rm BH} \left( 1 - f_{\rm BH} \right)^{2} \,,
\eeq
since we are deeply in the low mass regime. The maximum, where the blue dots sit, is reached at $f_{\rm BH} = {1}/{3}$, hence $f_{\rm DM} = {2}/{3}$, as already mentioned.
}
\end{itemize}

Before moving to the CMB distortion signal, we stress that for the extragalactic gamma-ray bounds discussed above, we have disregarded the contribution from DM annihilation outside PBH spikes. Such a contribution should in principle be taken into account in a global analysis. We have disregarded it because it would require integrating over all DM structures that have formed along the line of sight, owing to domination of the clumpy DM component over the homogeneous one in the total annihilation yield \cite{UllioEtAl2002,FornasaEtAl2015}. This goes beyond the scope of this paper. In contrast, for CMB distortions, as we shall see, the diffuse signal can be easily estimated for the homogeneous DM component because density fluctuations are still in the fully linear regime and actually  vanishingly small. Therefore, clumpiness can essentially be neglected at the relevant redshifts, except precisely for the contribution of PBH spikes.

\subsubsection{Injection of energy in the post-recombination plasma}
\label{subsubsec:f_BH_CMB}

In the conventional WIMP scenario, thermal DM species still annihilate after recombination and inject energy into the primordial plasma with rate
\beq
\dEOwimpsinj \!\! = \frac{1}{2} \asv n_{\chi}^{2} \times 2 m_{\chi} \times f_{\rm eff} \,.
\eeq
The three terms in the right-hand side of this relation respectively refer to the number of annihilations per unit volume and time, the amount of energy released by each annihilation, and the fraction of that energy that is eventually deposited into the plasma.
Just after recombination, DM particles are quasi-homogeneously distributed, and their average cosmological number density, at redshift $z$, is given by
\beq
n_{\chi}(z) = \frac{\rho_{\rm dm}^{0}}{m_{\chi}} \, (1+z)^{3} \,.
\eeq
The rate of energy injection inside the plasma becomes
\beq
\dEOwimpsinj \!\! = p_{\rm ann} \times \left\{ \rho_{\rm dm}^{0} (1+z)^{3} \right\}^{2}
\;\;\;\text{where}\;\;\;
p_{\rm ann} \equiv f_{\rm eff} \frac{\asv}{m_{\chi}} \,.
\eeq
The parameter $p_{\rm ann}$ naturally appears in the previous expression. The energy transferred to the intergalactic medium (IGM) reheats and reionizes it. This modifies the optical depth against Thomson scattering of the photons produced on the last scattering surface, with the consequence of blurring CMB angular anisotropies. The effect is most important at redshift $z_{\rm CMB}$ of order $600$~\cite{Finkbeiner_2012}, hence the limit\footnote{More precisely, we have borrowed relation~(89c) from the Planck analysis of the CMB, and corrected for the misprint $10^{-28}$ instead of $10^{28}$.}~\cite{AghanimEtAl2020}
\beq
p_{\rm ann} \leq p_{\rm \, ann}^{\rm max} = 3.2 \times 10^{-28} \; {\rm cm^{3} \, s^{-1} \, GeV^{-1}} \,.
\eeq

\vskip 0.1cm
In our mixed scenario with both a cosmological diffuse component of WIMPs, and very dense minispikes around PBHs, DM annihilation is also associated to energy injection in the IGM with rate
\beq
\dEtotinj \!\! = \dEwimpsinj \!\! + \dEspikesinj \,.
\eeq
Disregarding for the moment the contribution from the diffuse WIMP component, we concentrate on the effect of DM annihilation inside minispikes, whose rate of energy deposition at redshift~$z$ may be expressed as
\beq
\dEspikesinj \!\! = \Gamma_{\rm BH}(z) n_{\rm BH}(z) \times 2 m_{\chi} \times f_{\rm eff} \,.
\eeq
The first term in the right-hand side of this relation is the number of annihilations per unit volume and time. The number density $n_{\rm BH}(z)$ of PBHs is given by Eq.~(\ref{eq:n_BH_of_z_a}). Requiring that at redshift $z_{\rm CMB}$, where CMB anisotropy data are most sensitive, minispikes do not inject in the IGM more energy than what the Planck limit on $p_{\rm ann}$ allows, we get the condition
\beq
\dEspikesinj \!\! \simeq \dEtotinj \!\! \leq p_{\rm \, ann}^{\rm max} \times \left\{ \rho_{\rm dm}^{0} (1+z_{\rm CMB})^{3} \right\}^{2} \,,
\eeq
which translates into the bound
\beq
f_{\rm BH} \Gamma_{\rm BH} \leq
\left\{ \frac{M_{\rm BH}}{2 m_{\chi}} \right\}
\left\{ \frac{1}{\tau_{\rm CMB}^{\rm \, min}} \right\}
\;\;\;\text{where}\;\;\;
\frac{1}{\tau_{\rm CMB}^{\rm \, min}} \equiv \frac{p_{\rm \, ann}^{\rm max}}{f_{\rm eff}} \, \rho_{\rm dm}^{0} \, (1+z_{\rm CMB})^{3} \,.
\label{eq:maximum_f_BH_G_BH_b}
\eeq
Introducing the effective lifetime $\tau_{\rm CMB}^{\rm \, min}$, the constraint arising from CMB distortions takes the same form as the limit~(\ref{eq:maximum_f_BH_G_BH_a}) derived from the extragalactic $\gamma$-ray background. Although not physically motivated, $\tau_{\rm CMB}^{\rm \, min}$ plays the same role as $\tau_{\rm \, dm}^{\rm min}$. For particles annihilating into $b \bar{b}$ pairs, the fraction $f_{\rm eff}$ of the energy released and effectively injected in the post-recombination plasma is of order $0.1$ \cite{SlatyerEtAl2009,Slatyer2016}, hence a value of $1.139 \times 10^{24} \; {\rm s}$ for $\tau_{\rm CMB}^{\rm \, min}$, independent of $m_{\chi}$.
%
%
The minispike annihilation rate $\Gamma_{\rm BH}$ must be effectively derived at redshift $z_{\rm CMB}$, and not today.

\begin{figure}[h!]
\centering
\includegraphics[width=0.495\textwidth]{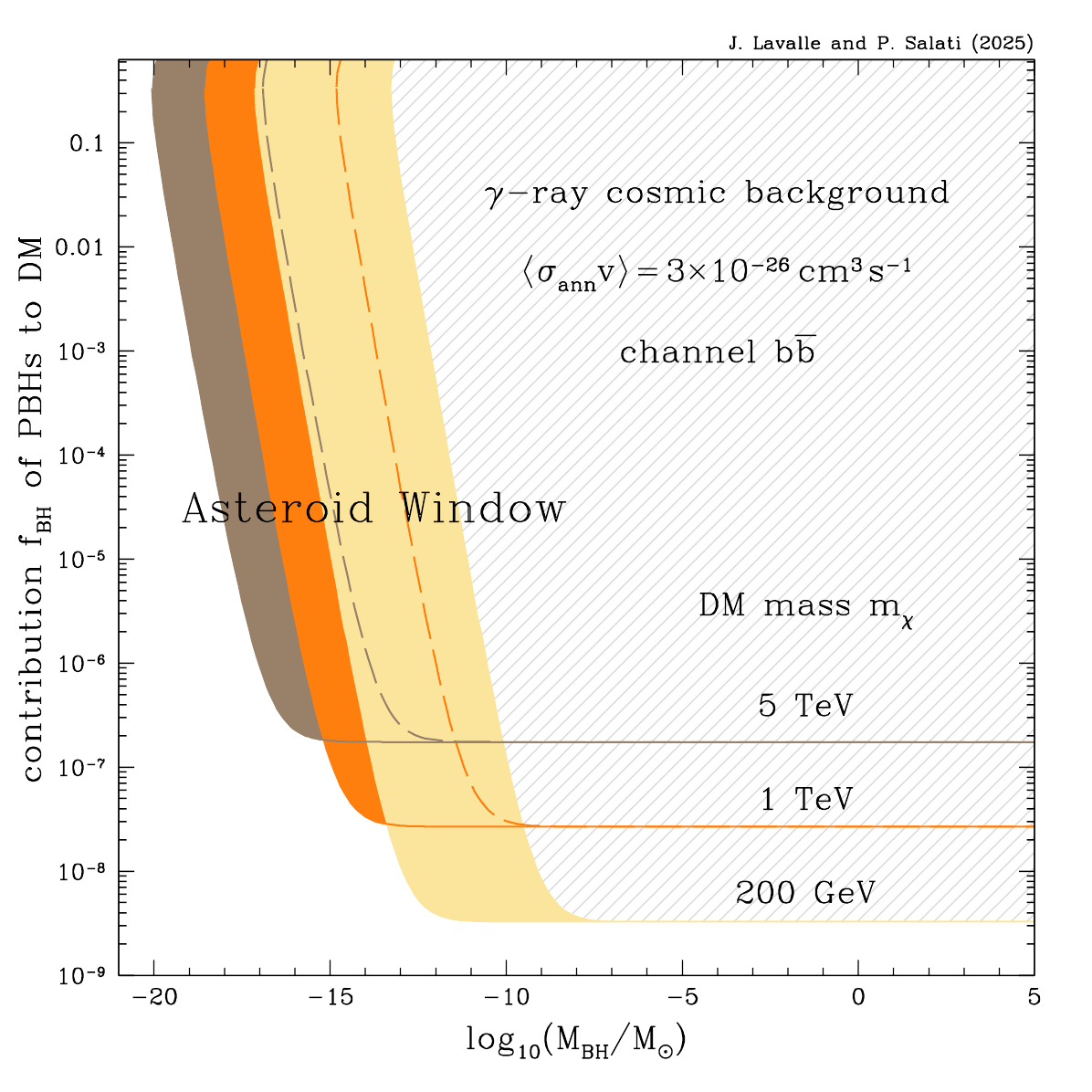}
\includegraphics[width=0.495\textwidth]{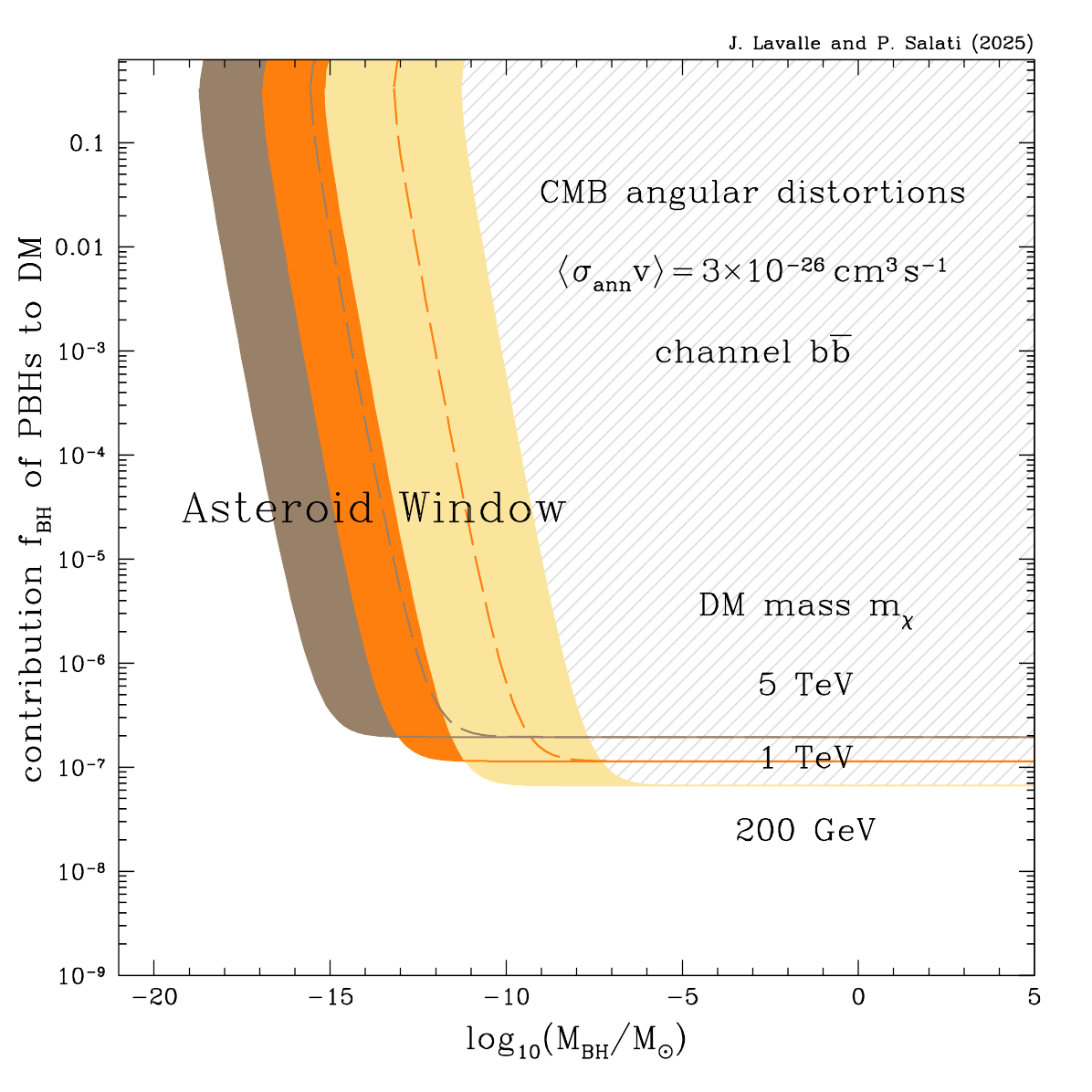}
\caption{Measurements of the extragalactic $\gamma$-ray background~\cite{Fermi-LAT:2014ryh,AndoEtAl2015} and of the angular distortions of the CMB~\cite{AghanimEtAl2020} set constrains on the fraction $f_{\rm BH}$ respectively displayed in the {\bf left} and {\bf right panels}.
Three different WIMP masses have been selected, with thermal annihilation cross-section and $b \bar{b}$ final state.
The hatched gray domain, which corresponds to a 200~GeV particle and a kinetic decoupling parameter of $10^{4}$, is excluded by observations. As $x_{\rm kd}$ is decreased down to $10^{2}$, this forbidden region expands to encompass the light-brown vertical band.
The orange and dark-brown bands correspond respectively to 1 and 5~TeV DM species. 
All excluded areas encroach on the asteroid window, which is closed over a large range of $f_{\rm BH}$ values.
\label{fig:gamma_CMB_bounds_on_f_BH}}
\end{figure}

\vskip 0.1cm
In Fig.~\ref{fig:gamma_CMB_bounds_on_f_BH}, the upper bounds on $f_{\rm BH}$ extracted from the extragalactic $\gamma$-ray background and CMB distortions are respectively displayed in the left and right panels. These plots are remarkably similar. The annihilation cross-section $\asv$ has been set equal to its thermal value, with $b \bar{b}$ final state. Three different WIMP masses are considered.
In each panel, the hatched region corresponds to a 200~GeV particle, with kinetic decoupling parameter $10^{4}$. This domain, which is excluded by observations, behaves like the gray areas in the left panel of Fig.~\ref{fig:gamma_bounds_on_f_BH_and_f_DM}, the topologies of which have already been thoroughly discussed.
The hatched area expands leftward to enclose the light-brown vertical band when $x_{\rm kd}$ is decreased down to $10^{2}$.
%
%

\vskip 0.1cm
We observe the same behavior for the 1 and 5~TeV cases, respectivelely color coded in orange and dark-brown.
The lower borders of the excluded zones are delineated by horizontal lines. Acually, for massive black holes, i.e. above the transition mass $M_{\rm T}$, the upper limit on the fraction $f_{\rm BH}$ neither depends on PBH mass nor on $x_{\rm kd}$. It scales like
\beq
f_{\rm \, BH}^{\rm max} \propto m_{\chi}^{1/3} \, {\Delta t}^{2/3} \left\{ \frac{1}{\tau_{\rm \, obs}^{\rm min}} \right\},
\label{eq:scaling_f_BH_a}
\eeq 
where $\Delta t = t_{\rm U}(z) - t_{\rm eq}$, with $t_{\rm U}(z)$ and $t_{\rm eq}$ respectively the age of the universe at redshift $z$ and at matter-radiation equality. The lifetime $\tau_{\rm \, obs}^{\rm min}$ denotes alternatively $\tau_{\rm \, dm}^{\rm min}$ (extragalactic $\gamma$-ray background) or $\tau_{\rm CMB}^{\rm \, min}$ (CMB distortions).
The former lifetime is extracted from Fig.~3 of~\cite{AndoEtAl2015} for the $b \bar{b}$ channel. Taking a decaying DM species twice as massive as our annihilating WIMP, $\tau_{\rm \, dm}^{\rm min}$ is found respectively equal to $1.49 \times 10^{28}$, $3.02 \times 10^{27}$ and $7.79 \times 10^{26} \; {\rm s}$ at 400~GeV, 2~TeV and 10~TeV.
We emphasize that in contrast with the CMB bounds, the extragalactic bounds on the decaying DM lifetime does not exhibit any particular relation with the DM particle mass. So the hierarchy between the bounds at different decaying DM particle mass is purely accidental here, and could have actually been inverted with other choices of mass.

\vskip 0.1cm
Although the extragalactic $\gamma$-ray background has little to do with plasma reionization and CMB distortions, the upper limits which these observations set on $f_{\rm BH}$ are almost identical at 5~TeV, with $f_{\rm \, BH}^{\rm max} \simeq 2 \times 10^{-7}$. With $\tau_{\rm \, dm}^{\rm min}$ almost three orders of magnitude larger than $\tau_{\rm CMB}^{\rm \, min}$, we would have expected $\gamma$-rays to be much more constraining than CMB distortions. This difference in lifetime is actually compensated by the increase of the duration $\Delta t$ from $2.98 \times 10^{13} \; {\rm s}$ at $z_{\rm CMB} = 600$ up to $4.36 \times 10^{17} \; {\rm s}$ today.
%
%
%
%
%
%
As the DM mass $m_{\chi}$ is decreased from 5~TeV down to 200~GeV, the constraint on $f_{\rm BH}$ becomes stronger, in agreement with scaling relation~(\ref{eq:scaling_f_BH_a}).
In the right panel of Fig.~\ref{fig:gamma_CMB_bounds_on_f_BH}, $f_{\rm \, BH}^{\rm max}$ is shifted downward by a factor $5^{1/3} \simeq 1.71$ between adjacent lines.
%
%
%
%
%
In the left panel, the lower limit $\tau_{\rm \, dm}^{\rm min}$ comes into play, which varies approximately like ${m_{\chi}^{-1}}$ in the range of WIMP masses considered. There is this time a shift by a factor $5^{4/3} \sim 8$ between lines.

\vskip 0.1cm
As the WIMP mass is decreased, the excluded domains in the plane $(M_{\rm BH} , f_{\rm BH})$ are not only shifted downward, as already discussed. They also move righward. The explanation is given in the right panel of Fig.~\ref{fig:Gamma_BH_vs_DM_parameters}, in section~\ref{subsec:Gamma_BH_behavior}. The long-dashed curve exhibits the effect on the minispike annihilation rate $\Gamma_{\rm BH}$ of reducing $m_{\chi}$ from 1~TeV to 10~GeV. The annihilation rate decreases at low PBH masses, while it increases for heavy objects. The long-dashed curve is shifted rightward and upward with respect to the solid line of the fiducial configuration. Decreasing the WIMP mass makes the transition mass $M_{\rm T}$ increase and the colored bands of Fig.~\ref{fig:gamma_CMB_bounds_on_f_BH} move rightward to higher masses.

\vskip 0.1cm
The colored bands, and associated excluded areas, extend over the asteroid window, i.e. roughly between $10^{-16}$ and $10^{-11} \; {\rm M_{\odot}}$. Until recently, this region was not constrained by observations~\cite{GreenEtAl2021,CarrEtAl2021}. A new analysis~\cite{Esser:2025pnt}, based on the effect of asteroid-mass PBHs on the heaviest stars of ultra-faint dwarf galaxies, has set constraints on $f_{\rm BH}$. In Triangulum~II, objects around $10^{19} \; {\rm g}$ are excluded at the $2\sigma$ ($3\sigma$) level from contributing more than $\sim 55\%$ ($\sim 78\%$) to the ambient DM, while the possibility that $f_{\rm BH}$ may reach $1$ is excluded at the $3.7\sigma$ level.
In that context, the bounds featured in Fig.~\ref{fig:gamma_CMB_bounds_on_f_BH} are a useful complement. In our mixed scenario, the fraction of the DM in the form of PBHs is much more severely constrained than in~\cite{Esser:2025pnt}, with an upper bound as low as a few billionths in the case of a 200~GeV particle.
However, as pointed out by Lacki and Beacom~\cite{LackiEtAl2010}, nothing prevents PBHs from constituting all the astronomical DM. In the plots of Fig.~\ref{fig:gamma_CMB_bounds_on_f_BH}, a very narrow strip close to $f_{\rm BH} \simeq 1$ is still allowed. As already mentioned above, we remind that qualitative comparisons with other bounds from different probes are given in appendix section~\ref{append:comparison}.
%
\begin{figure}[h!]
\centering
\includegraphics[width=0.495\textwidth]{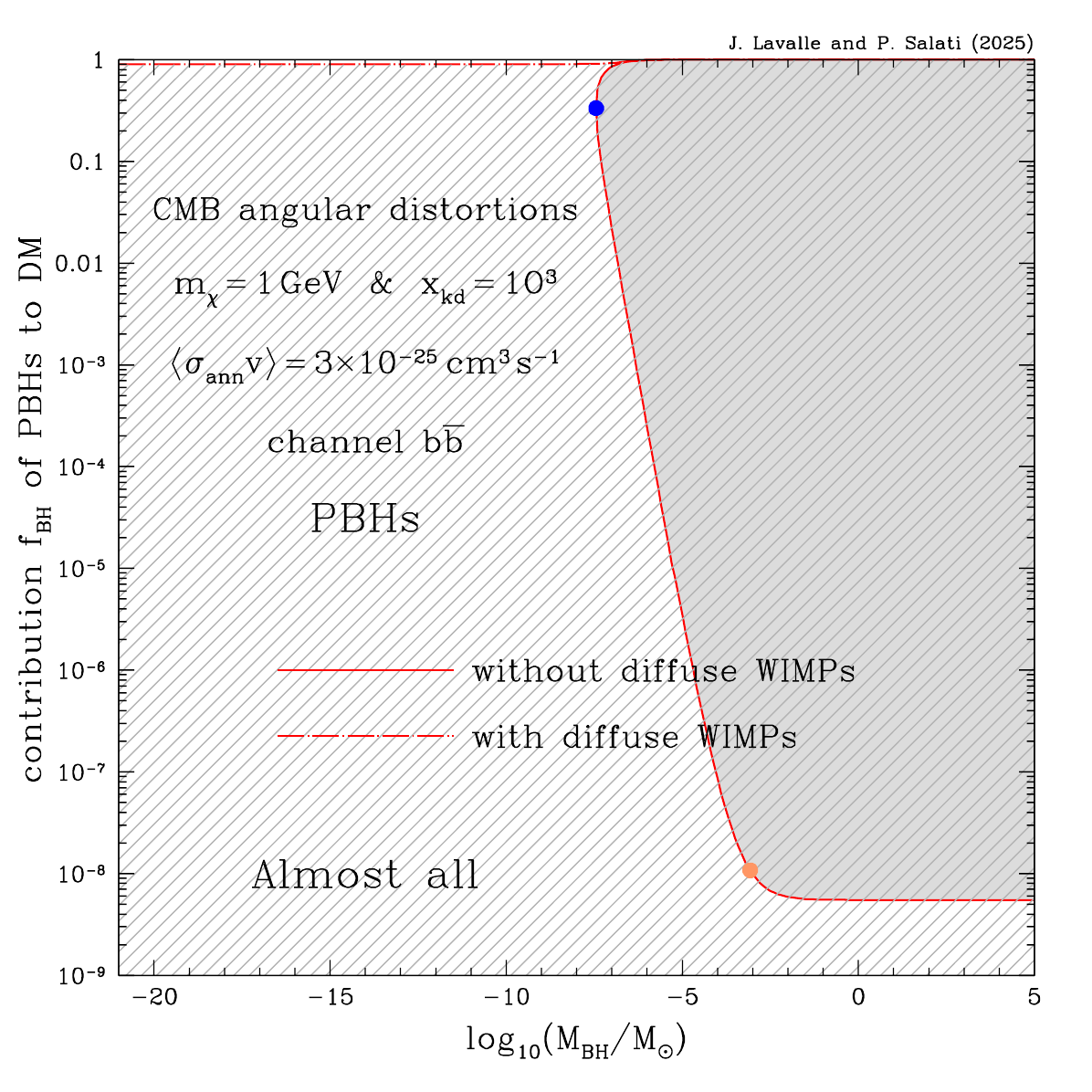}
\includegraphics[width=0.495\textwidth]{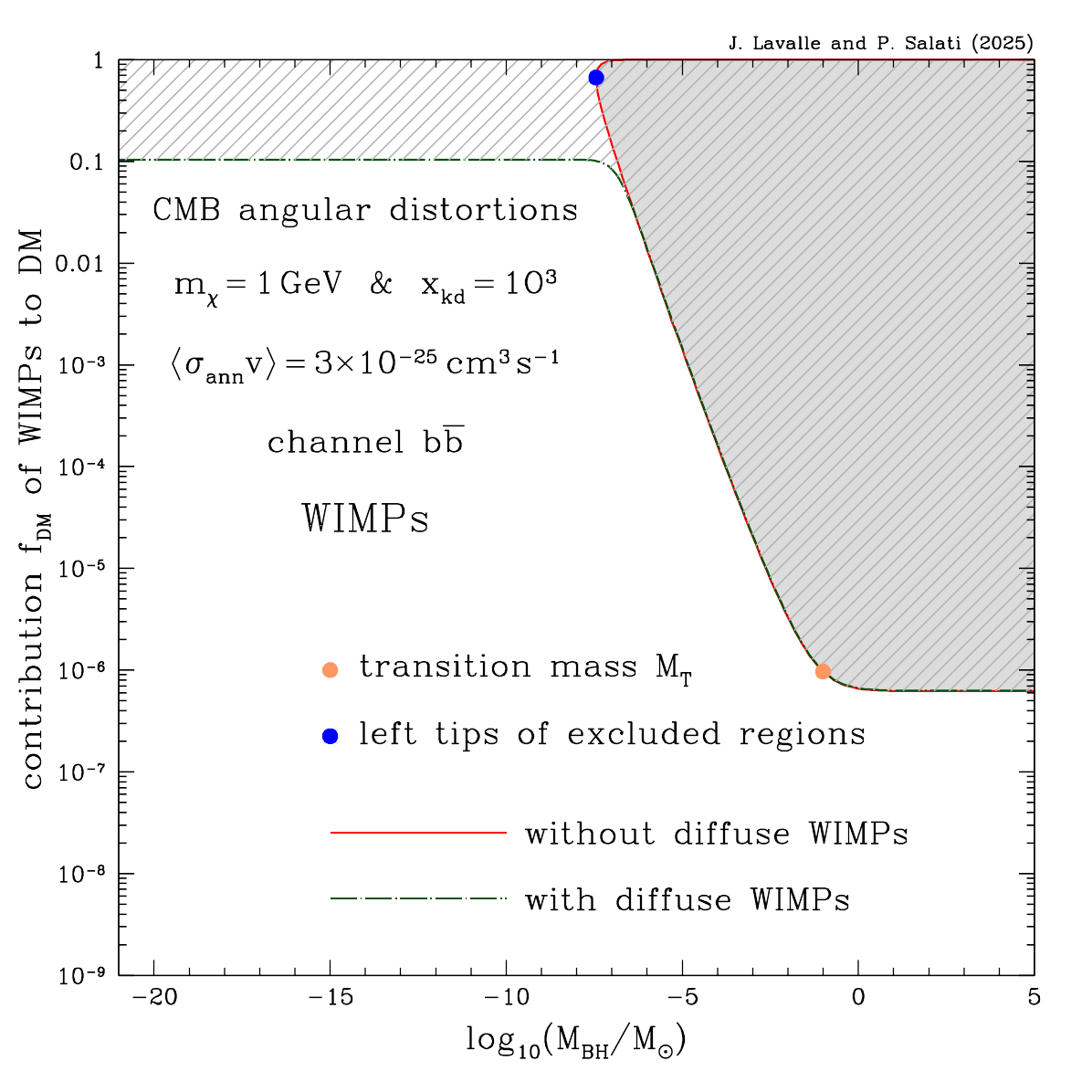}
\caption{Same as in Fig.~\ref{fig:gamma_bounds_on_f_BH_and_f_DM} with constraints from CMB distortions instead of extragalactic $\gamma$-ray background. For simplicity, one value only has been assumed for the kinetic decoupling parameter~$x_{\rm kd}$. The contributions of PBHs and WIMPs to the DM are respectively displayed in the {\bf left} and {\bf right panels}.
Considering energy injection in the post-recombination plasma induced solely by minispike annihilation yields the light-gray regions. These excluded domains expand to encompass the hatched areas when the contribution from the diffuse WIMPs is added.
To make this contribution significant, we have used an unusually large WIMP annihilation cross-section $\asv$ (ten times the thermal value).
Although no bound applies in the former case below PBH mass of $3.6 \times 10^{-8} \; {\rm M_{\odot}}$, featured by the blue dots, the diffuse WIMP component is constrained to contribute less than 10\% to the DM once it is included in the plasma reheating and reionization.
The fraction $f_{\rm BH}$ is constrained to be at least larger than 90\%. For a WIMP of 1 TeV though, the hatched region would be completely removed.
\label{fig:CMB_bounds_on_f_BH_and_f_DM}}
\end{figure}
%

\vskip 0.1cm
As a final point, we would like to examine how the CMB bounds are modified when the diffuse component of WIMPs is now taken into account. It contributes an energy injection
\beq
\dEwimpsinj = f_{\rm DM}^{2} \! \dEOwimpsinj \equiv
\left( f_{\rm DM}^{2} \, p_{\rm ann} \right) \times \left\{ \rho_{\rm dm}^{0} (1+z)^{3} \right\}^{2} \,.
\eeq
The constraint on the contribution from minispikes becomes
\beq
\dEspikesinj \!\! = \dEtotinj \!\! - \dEwimpsinj \!\! \leq \left( p_{\rm \, ann}^{\rm max} - f_{\rm DM}^{2} \, p_{\rm ann} \right)
\times \left\{ \rho_{\rm dm}^{0} (1+z_{\rm CMB})^{3} \right\}^{2} \,,
\eeq
which eventually translates into
\beq
f_{\rm BH} \Gamma_{\rm BH} \leq
\left\{ \frac{M_{\rm BH}}{2 m_{\chi}} \right\}
\left\{ \frac{1}{\tau_{\rm CMB}^{\rm \, min}} \right\}
\left\{ 1 - f_{\rm DM}^{2} \! \left( \frac{p_{\rm ann}}{p_{\rm \, ann}^{\rm max}} \right) \! \right\}.
\label{eq:maximum_f_BH_G_BH_c}
\eeq
As long as the ratio ${p_{\rm an}}/{p_{\rm \, ann}^{\rm max}}$ is small compared to $1$, the bound on $f_{\rm BH}$ does not change much. In Fig.~\ref{fig:gamma_CMB_bounds_on_f_BH} for instance, $p_{\rm ann}$ is equal to $3 \times 10^{-30} \; {\rm cm^{3} \, s^{-1} \, GeV^{-1}}$ for a 1~TeV species, to be compared to a CMB limit $p_{\rm \, ann}^{\rm max}$ two orders of magnitude larger.
Including the diffuse WIMP component strenghtens the constraint on the fraction of PBHs. The larger ${p_{\rm ann}}$, the lower the upper bound $f_{\rm \, BH}^{\rm max}$. If ${p_{\rm ann}}$ turns out to be larger than the limit $p_{\rm \, ann}^{\rm max}$, a small admixture of PBHs is completely ruled out insofar as the astronomical DM would be essentially made of diffuse WIMPs, a possibility forbidden by CMB data.

\vskip 0.1cm
On the other hand, PBHs evade observational constraints when their contribution to DM is close to $1$. Changing perspective, we can examine whether or not WIMPs are allowed even though ${p_{\rm ann}}$ is larger than $p_{\rm \, ann}^{\rm max}$. Such a possibility is excluded in the pure WIMP scenario. But the presence of PBHs could alleviate CMB constraints.
In Fig.~\ref{fig:CMB_bounds_on_f_BH_and_f_DM}, the case of a 1~GeV DM species is featured, with annihilation cross-section $\asv$ an order of magnitude larger than the thermal value. With such parameters, and assuming that $f_{\rm eff}$ is still equal to $0.1$, $p_{\rm ann}$ reaches $3 \times 10^{-26} \; {\rm cm^{3} \, s^{-1} \, GeV^{-1}}$, i.e. two orders of magnitude larger than the CMB limit.

\vskip 0.1cm
The fractions $f_{\rm BH}$ and $f_{\rm DM}$ are respectively presented in the left and right panels. The gray regions are derived from the conservative limit~(\ref{eq:maximum_f_BH_G_BH_b}). These domains are similar to the excluded areas of Fig.~\ref{fig:gamma_bounds_on_f_BH_and_f_DM}, although the latter was drawn assuming a 1~TeV species, and translate now the CMB data. No constraint applies below a PBH mass of $3.6 \times 10^{-8} \; {\rm M_{\odot}}$, featured by the blue dots.
On the contrary, the hatched regions are obtained by taking into account the diffuse WIMP contribution and by applying the more restrictive condition~(\ref{eq:maximum_f_BH_G_BH_c}). PBHs are essentially excluded, apart from a very small fringe located in the upper part of the left panel. The right panel features the WIMP fraction $f_{\rm DM}$, which is now constrained to be less than 10\% even for light black holes. In this region, PBHs must contribute more than 90\% to the DM.
While minispike annihilation provides most of the energy injected into the post-recombination plasma for PBH masses above $3.6 \times 10^{-8} \; {\rm M_{\odot}}$, the diffuse WIMP component takes over below that value. Both regimes clearly show up in the structure of the hatched region of the right panel.
Note finally that, had we assumed a WIMP mass of 1 TeV instead of 1 GeV, the hatched area would have completely disappeared since the diffuse contribution of such heavy particles is currently not constrained by CMB data.

\subsection{Bounds on WIMP properties}
\label{subsec:sav_WIMPs}

Changing perspective, we assume now that a population of primordial black holes has been discovered, contributing a fraction $f_{\rm BH}$ to the astronomical DM. Searches for such objects are underway, and it is conceivable that GW observatories will find them in the near future, should they exist.
Such a discovery would also be of great interest for thermal DM species. Even a minute admixture of PBHs to DM implies stringent limits on the annihilation cross-section of WIMPs, sometimes orders of magnitude below the thermal value of $3 \times 10^{-26} \; {\rm cm^{3} \, s^{-1}}$.

\subsubsection{Structure of the excluded region in the $(m_{\chi} , \asv)$ plane}
\label{subsubsec:structure_excluded_region_sig_an_v_vs_m_chi}

From now on, we conservatively concentrate on the limits set by the contribution of minispikes alone to either the cosmological $\gamma$-ray background, or the injection of energy in the post-recombination plasma. These constraints are respectively described by relations~(\ref{eq:maximum_f_BH_G_BH_a}) and (\ref{eq:maximum_f_BH_G_BH_b}). They can be summarized by the inequality
\beq
f_{\rm BH} \Gamma_{\rm BH} \leq
\left\{ \frac{M_{\rm BH}}{2 m_{\chi}} \right\}
\left\{ \frac{1}{\tau_{\rm \, obs}^{\rm min}} \right\},
\label{eq:maximum_f_BH_G_BH_d}
\eeq
with $\tau_{\rm \, obs}^{\rm min}$ being either equal to $\tau_{\rm \, dm}^{\rm min}$ or $\tau_{\rm CMB}^{\rm \, min}$.
For pedagogical purposes, we choose the CMB constraints. They are the least stringent, with $\tau_{\rm CMB}^{\rm \, min}$ of order $10^{24} \; {\rm s}$ while $\tau_{\rm \, dm}^{\rm min}$ lies between $10^{26}$ and $10^{28} \; {\rm s}$. Furthermore, as long as $f_{\rm eff}$ does not depend on WIMP mass, the effective lifetime $\tau_{\rm CMB}^{\rm \, min}$ is constant. This will make the discussion clearer.
%

\begin{figure}[h!]
\centering
\includegraphics[width=0.495\textwidth]{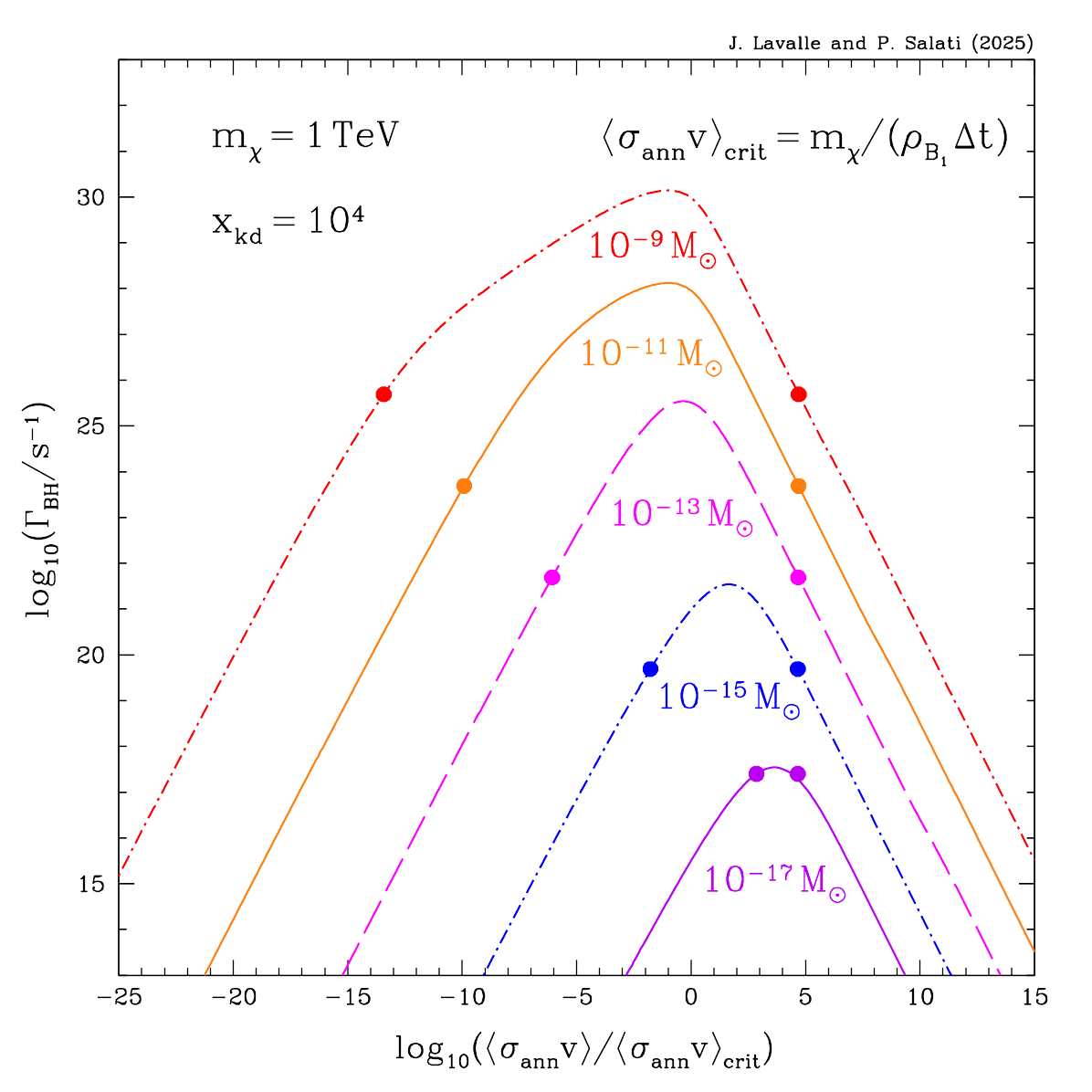}
\includegraphics[width=0.495\textwidth]{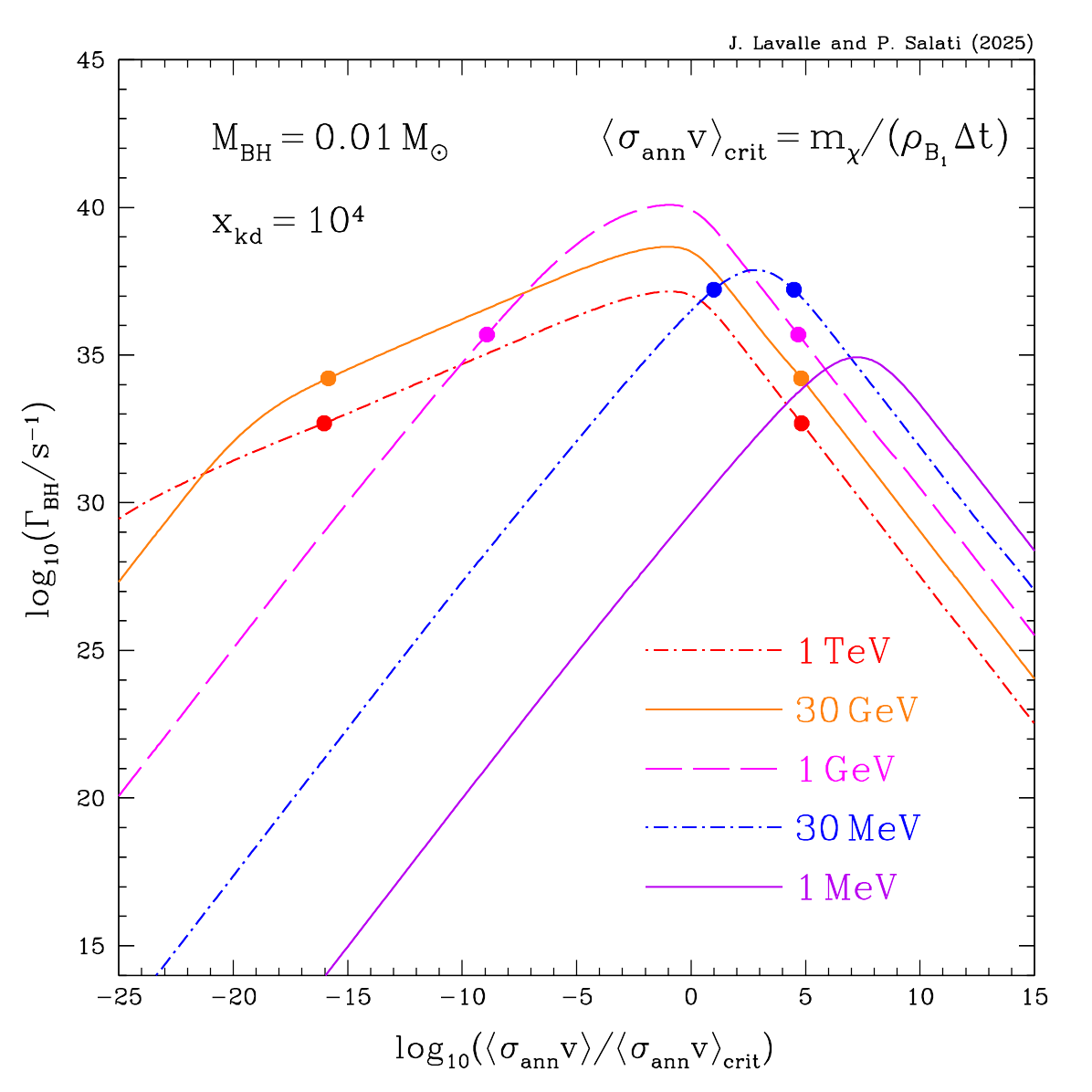}
\caption{The minispike annihilation rate $\Gamma_{\rm BH}$ is plotted as a function of the WIMP annihilation cross-section $\asv$.
In the {\bf left panel}, the properties of the thermal DM particle are specified, with a mass $m_{\chi}$ of 1~TeV and a kinetic decoupling parameter $x_{\rm kd}$ of $10^{4}$. The PBH mass is varied from $10^{-17}$ to $10^{-9} \; {\rm M_{\odot}}$. A power law behavior with slope ${1}/{3}$ appears for $M_{\rm BH}$ larger than $10^{-13} \; {\rm M_{\odot}}$, a value close to the critical mass $M_{1}$.
In the {\bf right panel}, the PBH mass is set equal to $0.01 \; {\rm M_{\odot}}$ while the WIMP mass $m_{\chi}$ is increased from 1~MeV up to 1~TeV. The slope ${1}/{3}$ also appears for the heaviest particles.
For each curve, the colored dots indicate the critical value of $\Gamma_{\rm BH}$ above which the constraint~(\ref{eq:maximum_f_BH_G_BH_b}) is violated, with $f_{\rm eff}$ and $f_{\rm BH}$ respectively set equal to $0.1$ and $10^{-5}$. In the case of the solid purple curve of the {\bf right panel}, this condition is always met.
\label{fig:G_BH_vs_sigav_various_cases}}
\end{figure}

\vskip 0.1cm
To understand how relation~(\ref{eq:maximum_f_BH_G_BH_d}) constrains the annihilation cross-section, we have plotted in Fig.~\ref{fig:G_BH_vs_sigav_various_cases} the minispike annihilation rate $\Gamma_{\rm BH}$ as a function of $\asv$.
We have normalized the cross-section with the critical cross-section ${\asv}_{\rm crit}$ as given in Eq.~(\ref{eq:sig_v_max_universal}).
In the left panel, the WIMP properties are specified while the PBH mass is varied. All curves have a maximum and exhibit a power law behavior, with slope $+1$ as $\Gamma_{\rm BH}$ increases with $\asv$, and slope~$-1$ as it decreases. We also notice that above $10^{-13} \; {\rm M_{\odot}}$, a portion with slope ${1}/{3}$ appears. The dotted short-dashed red and solid orange curves clearly flatten before reaching their maxima.

\vskip 0.1cm
The explanation of how $\Gamma_{\rm BH}$ evolves with $\asv$ has been partially given in section~\ref{subsec:Gamma_BH_comparison}. The minispike annihilation rate follows three distinct regimes depending on how the PBH mass $M_{\rm BH}$, the critical value $M_{1}$ and the transition mass $M_{\rm T}$ compare to each other. Once the WIMP mass $m_{\chi}$ and kinetic decoupling parameter $x_{\rm kd}$ are defined, as is the case in the left panel of Fig.~\ref{fig:G_BH_vs_sigav_various_cases}, the phase diagram of Fig.~\ref{fig:phase_diagram} is completely determined. So is the critical mass $M_{1}$, which we find here equal to $1.097 \times 10^{-13} \; {\rm M_{\odot}}$, a value corresponding to the long-dashed magenta curve.
%
%
We also should keep in mind that the saturation density $\rho_{\rm sat}$ and the corresponding transition mass $M_{\rm T}$ decrease as $\asv$ increases, while all the other parameters are kept fixed.
Two cases can essentially be distinguished.
\begin{enumerate}
\item{$M_{1} \leq M_{\rm BH}$ -- The annihilation rate is well approximated by $\Gamma_{\rm BH}^{(3/2)}$ and its definition~(\ref{eq:scaling_G_BH_3_2_a}) as long as the transition mass $M_{\rm T}$ exceeds $M_{\rm BH}$, i.e. for small values of the annihilation cross-section. In this regime, $\Gamma_{\rm BH}$ scales like $\asv$, hence the slope~$+1$.
A transition is expected when $M_{\rm T}$ and $M_{\rm BH}$ become equal. In the case of the dotted short-dashed red curve, this takes place when $M_{\rm T}$ reaches down $10^{-9} \; {\rm M_{\odot}}$, at the reduced annihilation cross-section of $3.16 \times 10^{-12}$. A close inspection of the plot indicates that the transition is very smooth and starts already an order of magnitude below that value.
%
%

Then, for $M_{1} \leq M_{\rm T} \leq M_{\rm BH}$, the annihilation rate is given by $\Gamma_{\rm BH}^{(9/4)}$ and relation~(\ref{eq:scaling_G_BH_9_4_a}). In this intermediate regime, $\Gamma_{\rm BH}$ scales like ${\asv}^{1/3}$, hence the slope~${1}/{3}$ which the dotted short-dashed red curve of the left panel clearly exhibits.
The maximal annihilation rate is expected to be reached when the saturation density $\rho_{\rm sat}$ becomes equal to $\rho_{\rm B_{1}}$, i.e. when $M_{\rm T}$ has decreased down to $M_{1}$. This occurs when the annihilation cross-section is equal to ${\asv}_{\rm crit}$ as defined by relation~(\ref{eq:sig_v_max_universal}) that yields in our case a value of $4.4 \times 10^{-16} \; {\rm cm^{3} \, s^{-1}}$ independent of PBH mass.
%
%
Once again, the maximum is very smooth and is reached numerically slightly below ${\asv}_{\rm crit}$.

Finally, for $M_{\rm T} \leq M_{1}$,  the approximation $\Gamma_{\rm BH}^{(0)}$ as defined by~(\ref{eq:scaling_G_BH_0_a}) comes into play, hence a minispike annihilation rate decreasing like ${1}/{\asv}$ and the slope~$-1$ which the curves exhibit when the annihilation cross-section is large.}
\item{$M_{\rm BH} \leq M_{1}$ -- Once again, for $M_{\rm T}$ exceeding $M_{\rm BH}$, $\Gamma_{\rm BH}^{(3/2)}$ is a good approximation for the annihilation rate. In the left panel of Fig.~\ref{fig:G_BH_vs_sigav_various_cases}, the dotted short-dashed blue and solid purple curves do feature a power law increase with slope~$+1$ when $\asv$ is small. This regime extends up to the critical cross-section ${\asv}_{\rm crit}$ as defined now by the more involved expression~(\ref{eq:sig_v_max_normalization}). At $10^{-15}$ and $10^{-17} \; {\rm M_{\odot}}$, the maxima are respectively expected at the reduced cross-sections of $10^{2}$ and $10^{4}$, in good agreement with the numerical results of Fig.~\ref{fig:G_BH_vs_sigav_various_cases}.

Finally, for $M_{\rm T}$ smaller than $M_{\rm BH}$, $\Gamma_{\rm BH}^{(0)}$ and its expression~(\ref{eq:scaling_G_BH_0_a}) can be used. The annihilation rate $\Gamma_{\rm BH}$ is inversely proportional to the annihilation cross-section, hence the slope~$-1$ which the curves exhibit above ${\asv}_{\rm crit}$.
}
\end{enumerate}
To summarize, the minispike annihilation rate increases with slope $+1$ as long as the annihilation cross-section is smaller than the critical value ${\asv}_{\rm crit}$. Above that value, $\Gamma_{\rm BH}$ decreases with slope $-1$.
If $M_{\rm BH}$ is larger than the critical mass $M_{1}$, ${\asv}_{\rm crit}$ is defined by~(\ref{eq:sig_v_max_universal}) and a regime with slope ${1}/{3}$ appears before the maximum is reached.
In the opposite situation, there is no such regime, and ${\asv}_{\rm crit}$ is defined by~(\ref{eq:sig_v_max_normalization}).
%
%

\vskip 0.1cm
In the right panel of Fig.~\ref{fig:G_BH_vs_sigav_various_cases}, the PBH mass is set equal to $0.01 \; {\rm M_{\odot}}$, while the WIMP mass $m_{\chi}$ is decreased from 1~TeV down to 1~MeV. We observe the same trend as in the left panel, with the slope~${1}/{3}$ appearing for the heaviest DM species.
At fixed kinetic decoupling parameter $x_{\rm kd}$, $T_{\rm kd}$ is proportional to $m_{\chi}$. In the phase diagram of Fig.~\ref{fig:phase_diagram}, decreasing the WIMP mass translates into shifting upward the dotted short-dashed red line (radius of influence $\tilde{r}_{\rm kd}$ at kinetic decoupling), moving it closer to the solid red line (radius of influence $\tilde{r}_{\rm eq}$ at matter-radiation equality). The intersection between the former line and the dotted long-dashed gray vertical also moves upward. So do points A and B$_{2}$, hence an increase of $M_{2}$ and an even larger increase of $M_{1}$ insofar as the radial infall domain with slope 9/4 has shrinked. As $m_{\chi}$ decreases, $M_{1}$ increases.
For heavy DM particles, as long as $M_{1}$ is smaller than $M_{\rm BH}$, the curves exhibit a maximum when $\rho_{\rm sat}$ is equal to $\rho_{\rm B_{1}}$. They also contain a portion with slope ${1}/{3}$, as is particularly clear in the 30~GeV and 1~TeV cases.
At 1~GeV, the critical mass $M_{1}$ is still equal to $3.7 \times 10^{-4} \; {\rm M_{\odot}}$, to be compared to a value of $16.7 \; {\rm M_{\odot}}$ at 30~MeV.
%
%
%
The transition occurs in between. The curves corresponding to a WIMP mass of 1 and 30~MeV exhibit the same kind of behavior as the smallest PBH masses of the left panel. The maximum is reached for a saturation density $\rho_{\rm sat}$ smaller than $\rho_{\rm B_{1}}$, i.e. for a reduced cross-section larger than $1$. The regime with slope~${1}/{3}$ has also disappeared.

\vskip 0.1cm
In both panels of Fig.~\ref{fig:G_BH_vs_sigav_various_cases}, the colored dots indicate the critical value of $\Gamma_{\rm BH}$ over which constraint~(\ref{eq:maximum_f_BH_G_BH_d}) is broken. For clarity, we have used the CMB bound~(\ref{eq:maximum_f_BH_G_BH_b}) and set $f_{\rm eff}$ and $f_{\rm BH}$ respectively equal to $0.1$ and $10^{-5}$.
In the right panel, where the PBH mass has been set equal to $0.01 \; {\rm M_{\odot}}$, the solid purple curve does not have any dot.
Whatever the annihilation cross-section, a 1~MeV DM species fulfills the CMB observational constraint on minispikes (and not on the diffuse contribution which is strongly constrained in this light mass range).
This is not the case for the other configurations. On each curve of these, the portion lying between dots does not meet condition~(\ref{eq:maximum_f_BH_G_BH_b}) and is excluded. The left and right dots respectively yield the upper and lower limits $\asv_{\rm L}$ and $\asv_{\rm R}$ set by CMB measurements on WIMP annihilation cross-section. The range extending between these values violates observations, hence the existence of an excluded domain in the $(m_{\chi} , \asv)$ plane.

\vskip 0.1cm
At fixed $f_{\rm BH}$ and $\tau_{\rm \, obs}^{\rm min}$, the minispike annihilation rate, at the dots, scales like
\beq
\Gamma_{\rm BH} \propto \frac{M_{\rm BH}}{m_{\chi}} \,.
\eeq
This scaling clearly appears in Fig.~\ref{fig:G_BH_vs_sigav_various_cases}. In the left panel, dots shift downward by two orders of magnitude each time the PBH mass is decreased while in the right panel, they shift upward by a factor $\sim 30$ each time the WIMP mass is decreased.
We also notice the peculiar alignment of the right dots. These are located on the decreasing parts of curves. In this regime, $\Gamma_{\rm BH}$ is given by relation~(\ref{eq:scaling_G_BH_0_a}) which, combined with condition~(\ref{eq:maximum_f_BH_G_BH_d}), yields the scaling
\beq
\asv_{\rm R} \propto f_{\rm BH} \times m_{\chi} \times {{\Delta}t}^{-2} \times \tau_{\rm \, obs}^{\rm min}
\;\;\;\text{(upper boundary).}
\label{eq:scaling_sigav_upper_contour}
\eeq
In the left panel, the WIMP mass is set equal to 1~TeV and does not vary. In the right panel, the annihilation cross-section is expressed in units of the critical value~(\ref{eq:sig_v_max_universal}). The WIMP mass cancels out in the ratio ${\asv_{\rm R}}/{{\asv}_{\rm crit}}$, hence the vertical alignment which we also observe in that panel.
%
\begin{figure}[h!]
\centering
\includegraphics[width=0.70\textwidth]{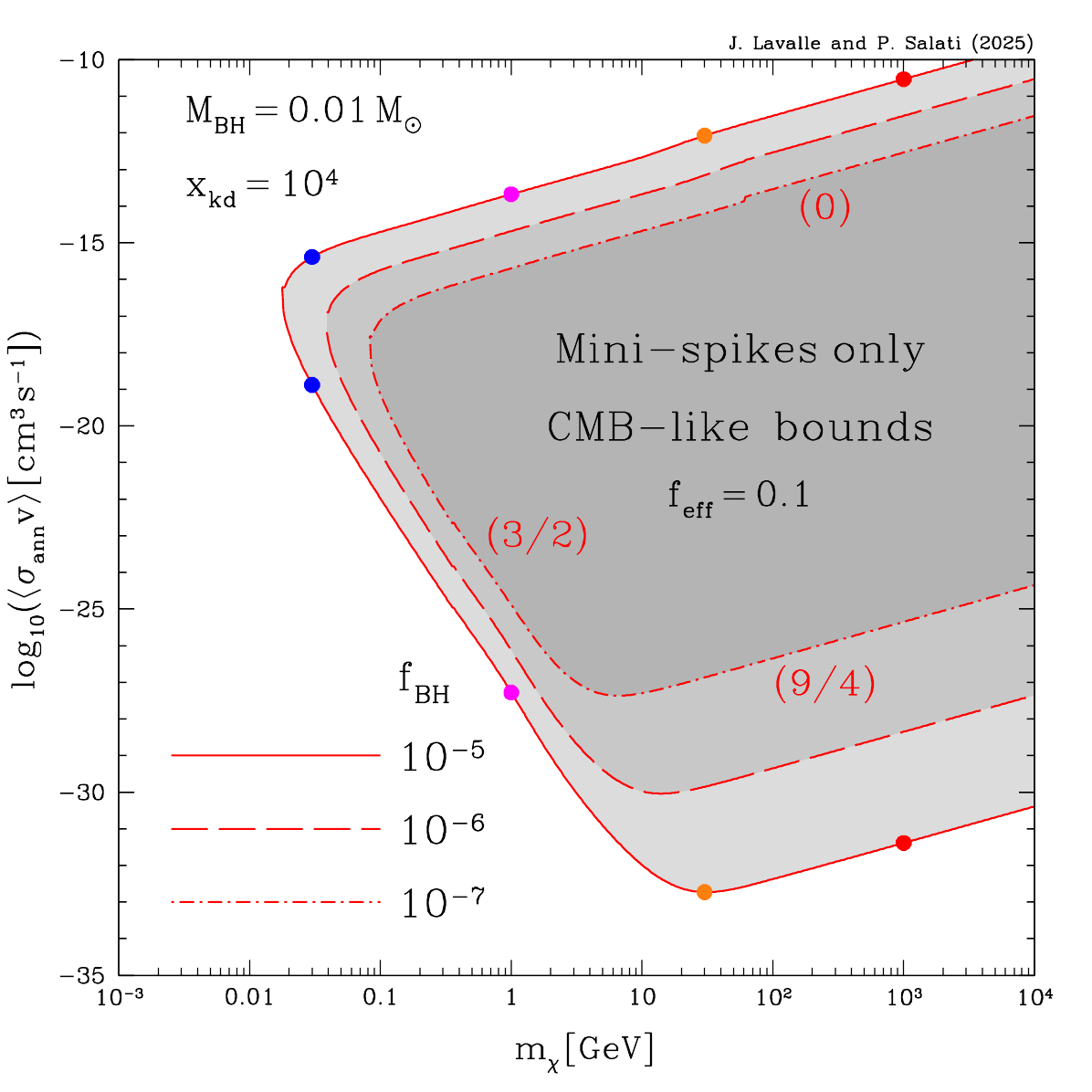}
\caption{In this pedagogical plot, the bounds set by cosmological observations on DM annihilation inside minispikes only are featured in the $(m_{\chi} , \asv)$ plane. The CMB constraint~(\ref{eq:maximum_f_BH_G_BH_b}), based on Planck determination of $p_{\rm \, ann}^{\rm max}$~\cite{AghanimEtAl2020}, has been chosen. The fraction $f_{\rm eff}$ has been set equal to $0.1$ for simplicity.
We assume the existence of a population of PBHs with mass $0.01 \; {\rm M_{\odot}}$, contributing a fraction $f_{\rm BH}$ to the astronomical DM.
The various shades of gray and contour types correspond to different values of this contribution, as indicated.
The colored dots extracted from the right panel of Fig.~\ref{fig:G_BH_vs_sigav_various_cases} have been put on this diagram. They lie, as they should, on the boundary of the light-gray domain.
We clearly distinguish, in the structure of the excluded regions, three different regimes for the minispike annihilation rate. The lower-right, lower-left and upper edges correspond respectively to the asymptotic behaviors~(\ref{eq:scaling_G_BH_9_4_a}), (\ref{eq:scaling_G_BH_3_2_a}) and (\ref{eq:scaling_G_BH_0_a}) of $\Gamma_{\rm BH}$.
\label{fig:excluded_zone_m_chi_sigav_plane}}
\end{figure}
%

\vskip 0.1cm
Inspired by our analysis of Fig.~\ref{fig:G_BH_vs_sigav_various_cases}, we are ready to construct in the $(m_{\chi} , \asv)$ plane the region excluded by Planck observations~\cite{AghanimEtAl2020} of the CMB and determination of $p_{\rm \, ann}^{\rm max}$.
We start scanning over WIMP mass. For each selected $m_{\chi}$, we first check if the maximal minispike annihilation rate, i.e. the value of $\Gamma_{\rm BH}$ taken at ${\asv}_{\rm crit}$, fulfills condition~(\ref{eq:maximum_f_BH_G_BH_b}).
If it does not, we use dichotomy to derive the upper and lower bounds $\asv_{\rm L}$ and $\asv_{\rm R}$ on the annihilation cross-section. Values of $\asv$ lying in between are excluded.
In Fig.~\ref{fig:excluded_zone_m_chi_sigav_plane}, such a construction is presented for a PBH mass of $0.01 \; {\rm M_{\odot}}$ and three different values of the fraction $f_{\rm BH}$. Each of these corresponds to a specific shade of gray and contour type, as indicated.
The light-gray domain, drawn for a PBH fraction of $10^{-5}$, is most extended. We have put the colored dots of the right panel of Fig.~\ref{fig:G_BH_vs_sigav_various_cases} in the $(m_{\chi} , \asv)$ diagram. They all lie along the solid red contour which delineates that forbidden area, as they should.
We observe that the smaller $f_{\rm BH}$, the less extended the excluded zones.
Each contour clearly exhibits three portions over which $\Gamma_{\rm BH}$ has a specific behavior.
\renewcommand{\labelitemi}{$-$}
\begin{itemize}
\item{The upper segment corresponds to the lower bound $\asv_{\rm R}$, and to the declining parts of the curves in Fig.~\ref{fig:G_BH_vs_sigav_various_cases}. The annihilation rate is well described by the asymptotic expression~(\ref{eq:scaling_G_BH_0_a}) which, combined with the generic constraint~(\ref{eq:maximum_f_BH_G_BH_d}), yields the scaling~(\ref{eq:scaling_sigav_upper_contour}).
The upper borders of the forbidden regions increase with $m_{\chi}$ with slope $+1$, and extend upward by an order of magnitude each time the fraction $f_{\rm BH}$ is increased.
}
\item{In the lower-left section of contours, the WIMP mass is small. Inspired by the right panel of Fig.~\ref{fig:G_BH_vs_sigav_various_cases}, we guess that the annihilation rate behaves asymptotically according to expression~(\ref{eq:scaling_G_BH_3_2_a}) which, once combined with the generic bound~(\ref{eq:maximum_f_BH_G_BH_d}), leads to the scaling
\beq
\asv_{\rm L} \propto {f_{\rm BH}^{-1}} \, {M_{\rm BH}^{-2}} \times {m_{\chi}^{-5}} \, x_{\rm kd}^{3}\times \frac{1}{\tau_{\rm \, obs}^{\rm min}}
\;\;\;\text{(lower-left boundary).}
\label{eq:scaling_sigav_lower_left_contour}
\eeq
In the plot, we do observe the contours dropping with $m_{\chi}$ while following a characteristic slope of $-5$. They also shift downward by an order of magnitude each time $f_{\rm BH}$ is increased.

For a PBH fraction of $10^{-5}$, we remark that the lower-left boundary extends approximately from the left tip of the light-gray area at 18~MeV up to the orange dot at 30~GeV. At the latter location, the transition mass $M_{\rm T}$ associated to the upper bound~$\asv_{\rm L}$ is equal to $1.7 \times 10^{-3}$, well below $M_{\rm BH}$\footnote{The transition between the lower-left and lower-right boundaries can be more accurately determined by finding the intersection of the corresponding tangent lines. A closer inspection shows that the transition takes place actually at 13~GeV, a value at which the transition mass $M_{\rm T}$ is equal to $1.1 \times 10^{-2} \; {\rm M_{\odot}}$.}.
Above 30~GeV, we enter in a new regime.
%
%
%
}
\item{Along the lower-right edges of the exclusion regions, the WIMP mass is large. We can approximate $\Gamma_{\rm BH}$ by its asymptotic expression~(\ref{eq:scaling_G_BH_9_4_a}). Actually, the transition mass $M_{\rm T}$ corresponding to the upper limit~$\asv_{\rm L}$ is well below $M_{\rm BH}$. Combining relation~(\ref{eq:scaling_G_BH_9_4_a}) with the generic cosmological constraint~(\ref{eq:maximum_f_BH_G_BH_d}) yields the scaling
\beq
\asv_{\rm L} \propto {f_{\rm BH}^{-3}} \times {m_{\chi}} \times {{\Delta t}}^{2} \times
\left\{ \frac{1}{\tau_{\rm \, obs}^{\rm min}} \right\}^{\! 3}
\;\;\;\text{(lower-right boundary).}
\label{eq:asv_L_lower_right}
\eeq
The lower-right and upper frontiers of the excluded regions are parallel, moving upward as $m_{\chi}$ increases with slope~$+1$. The forbidden domains shift downward by three orders of magnitude each time $f_{\rm BH}$ is increased. The bound $\asv_{\rm L}$ is also very sensitive to the lower limit $\tau_{\rm \, obs}^{\rm min}$. We expect the $\gamma$-ray constraints to be significantly more stringent than those extracted from the CMB. This conclusion should be nevertheless mitigated by the much larger cosmic duration ${\Delta}t$ in the former case.
}
\end{itemize}

\subsubsection{Constraints on $\asv$ from CMB angular distortions}
\label{subsubsec:CMB_sig_an_v}

\begin{figure}[h!]
\centering
\includegraphics[width=0.495\textwidth]{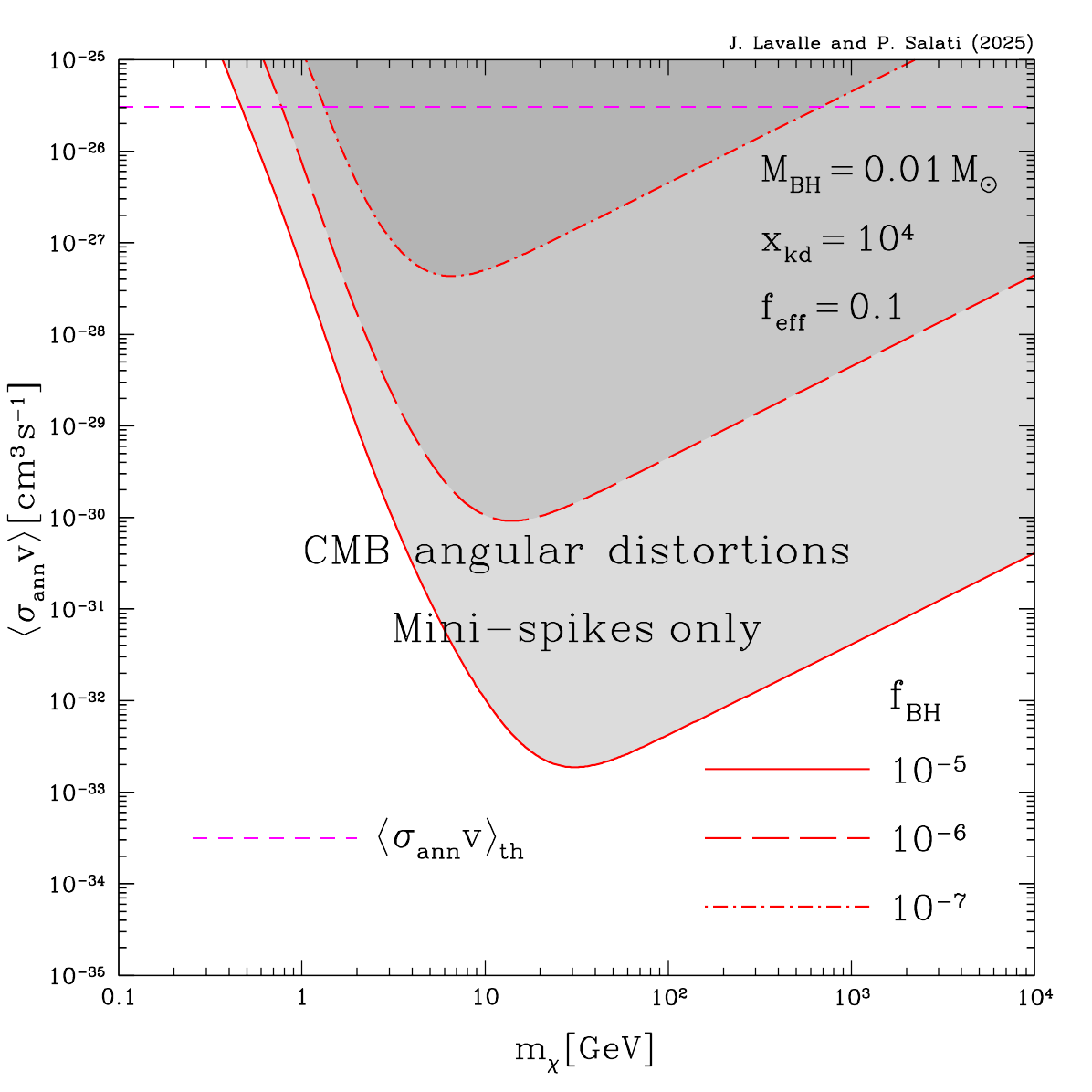}
\includegraphics[width=0.495\textwidth]{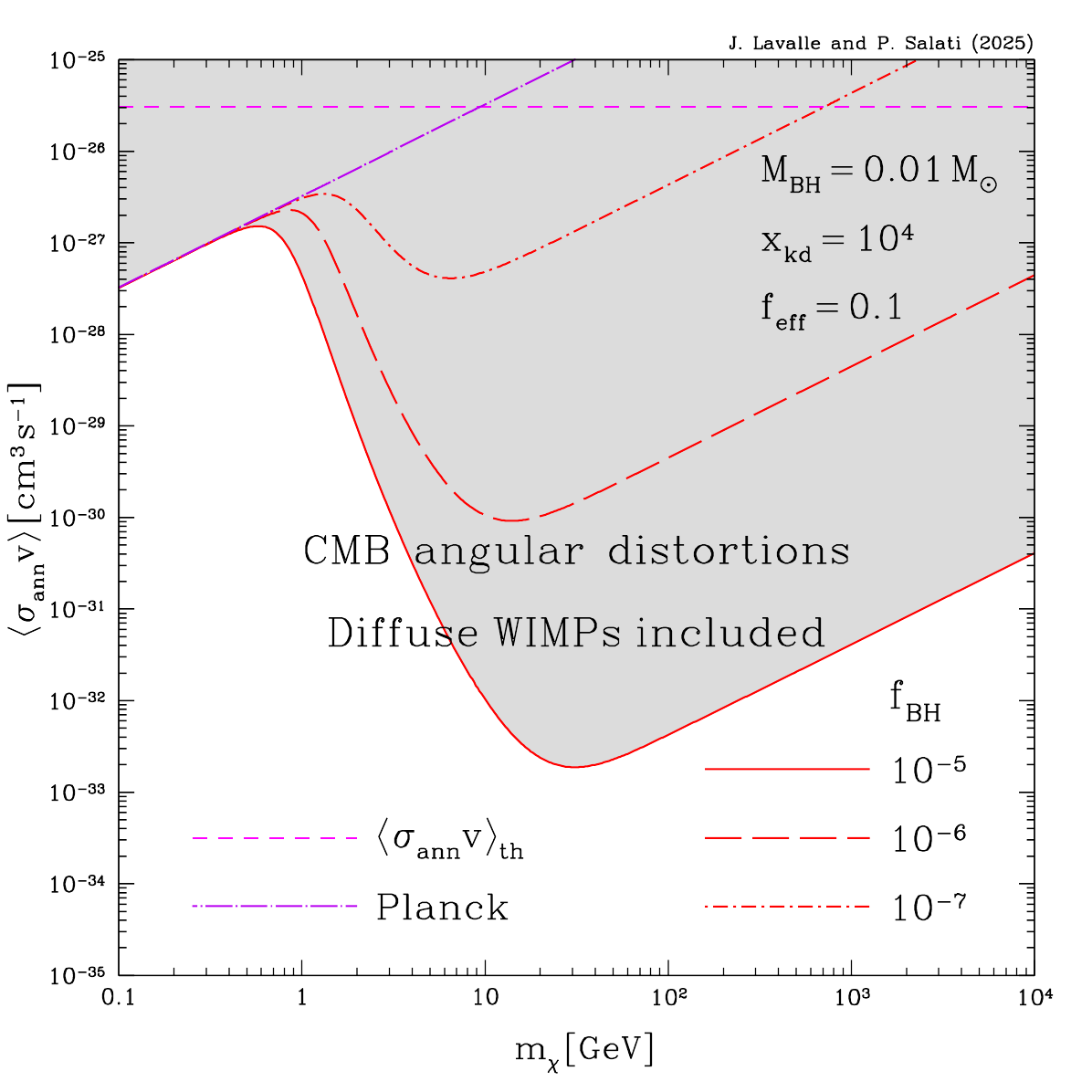}
\caption{Planck observations of the CMB angular distortions constrain the amount of energy that is released by DM annihilations and eventually injected in the intergalactic medium after recombination~\cite{AghanimEtAl2020}. In these plots, a population of PBHs with mass $0.01 \; {\rm M_{\odot}}$ is assumed, which contributes a fraction $f_{\rm BH}$ to the astronomical DM. The CMB limit on $p_{\rm \, ann}^{\rm max}$ translates into an upper bound~$\asv_{\rm L}$ on the annihilation cross-section.
In the {\bf left panel}, the exclusive contribution of minispikes is considered, while the diffuse WIMP component is also taken into account in the {\bf right panel}.
The short-dashed magenta horizontal line indicates the thermal value of $\asv$.
The dotted long-dashed purple line is the upper limit set by Planck in the conventional WIMP scenario without primordial black holes.
\label{fig:CMB_sigav_vs_m_chi_MSO_DWI}}
\end{figure}

The left panel of Fig.~\ref{fig:CMB_sigav_vs_m_chi_MSO_DWI} is an enlargement of Fig.~\ref{fig:excluded_zone_m_chi_sigav_plane} and focuses on the upper bound $\asv_{\rm L}$ set by CMB observations on the annihilation cross-section between 100~MeV and 10~TeV. In this panel, DM annihilation taking place only inside minispikes is considered, so that we have implemented constraint~(\ref{eq:maximum_f_BH_G_BH_b}).
The short-dashed magenta line corresponds to the thermal value ${\asv}_{\rm th}$ required in the canonical scenario to get the observed DM abundance from WIMP freeze-out.
We notice that even a small admixture of PBHs to the overall DM budget induces stringent constraints on $\asv$. For a fraction $f_{\rm BH}$ of $10^{-6}$, for instance, the upper bound stands more than four orders of magnitude below the thermal value at 10~GeV.
The CMB constraints weaken considerably though in the sub-GeV region where essentially minispikes do not provide any leverage on WIMPs.
%
%

\vskip 0.1cm
The gap left open at small masses can be closed if the contribution of the diffuse population of DM species is now taken into account, together with minispikes. In the right panel, constraint~(\ref{eq:maximum_f_BH_G_BH_c}) has been implemented. The dotted long-dashed purple line is the upper limit set by Planck~\cite{AghanimEtAl2020} on $\asv$ in the conventional WIMP scenario, i.e. without black holes, and is trivially expressed as
\beq
\asv \leq \frac{p_{\rm \, ann}^{\rm max}}{f_{\rm eff}} \, m_{\chi} \,.
\label{eq:Planck_asv_upper_limit}
\eeq
A comparison between the left and right panels shows that the upper bound $\asv_{\rm L}$ is either dominated by minispikes, or by the diffuse WIMP population, the most stringent constraint prevailing over the other. Relation~(\ref{eq:Planck_asv_upper_limit}) provides a conservative upper limit which $\asv$ cannot exceed.
Without minispikes, thermal DM species, at least in their simplest incarnation with $s$-wave annihilation, are ruled out below 10~GeV but allowed above.

\vskip 0.1cm
This possibility would be further jeopardized if a population of light black holes were discovered in the near future and its abundance $f_{\rm BH}$ measured. As showed in the right panel, DM particles with mass below 700~GeV are already excluded by a PBH fraction $f_{\rm BH}$ as small as $10^{-7}$. We can derive a lower limit on the WIMP mass $m_{\chi}$ by requiring that the upper bound~$\asv_{\rm L}$ should exceed the thermal value ${\asv}_{\rm th}$. Using the scaling~(\ref{eq:asv_L_lower_right}), we get
\beq
m_{\chi} \geq 7 \times 10^{5} \; {\rm GeV} \left\{ \frac{f_{\rm BH}}{10^{-6}} \right\}^{\! 3} \left\{ \frac{f_{\rm eff}}{0.1} \right\}^{\! 3}.
\eeq
This constraint is remarkably stringent and points toward heavy DM candidates.

\vskip 0.1cm
The efficiency factor $f_{\rm eff}$ has been heuristically set equal to $0.1$ insofar as our study aims primarily at clearing the field and paving the road for a forthcoming detailed investigation. We intend to investigate the CMB constrains on $\asv$ using the public code {\tt CLASS} (work in preparation). This would allow to get the correct variations of $f_{\rm eff}$ with cosmic time as well as its dependence on WIMP mass and annihilation channel.
The second benefit would be to follow in time the actual energy deposition in the post-recombination plasma. We have recycled here limits derived in the conventional scenario of a pure diffuse WIMP component, for which the rate of energy injection scales like $(1 + z)^{6}$ and peaks at $z_{\rm CMB}$ of order $600$. In the case of minispikes, this rate scales either like $(1 + z)^{3}$ or $(1 + z)^{4}$ depending on whether the annihilation rate $\Gamma_{\rm BH}$ is given by relation~(\ref{eq:scaling_G_BH_3_2_a}) or (\ref{eq:scaling_G_BH_9_4_a}). The redshift at which most of the energy deposition occurs should be different from the value of $600$ assumed here. We nevertheless expect the results derived using {\tt CLASS} to be qualitatively similar to those presented in this article.

\subsubsection{Constraints on $\asv$ from cosmological $\gamma$-ray background}
\label{subsubsec:gamma_ray_sig_an_v}

From their analysis of the cosmological $\gamma$-ray background, Ando and Ishiwata have derived bounds on decaying DM particles~\cite{AndoEtAl2015} which we have translated, in the framework of our mixed DM model, into condition~(\ref{eq:maximum_f_BH_G_BH_a}). As our focus is here on the WIMP annihilation cross-section, we apply the same method as in the previous sections, and implement that constraint to derive the upper bound $\asv_{\rm L}$.
%
\begin{figure}[h!]
\centering
\includegraphics[width=0.495\textwidth]{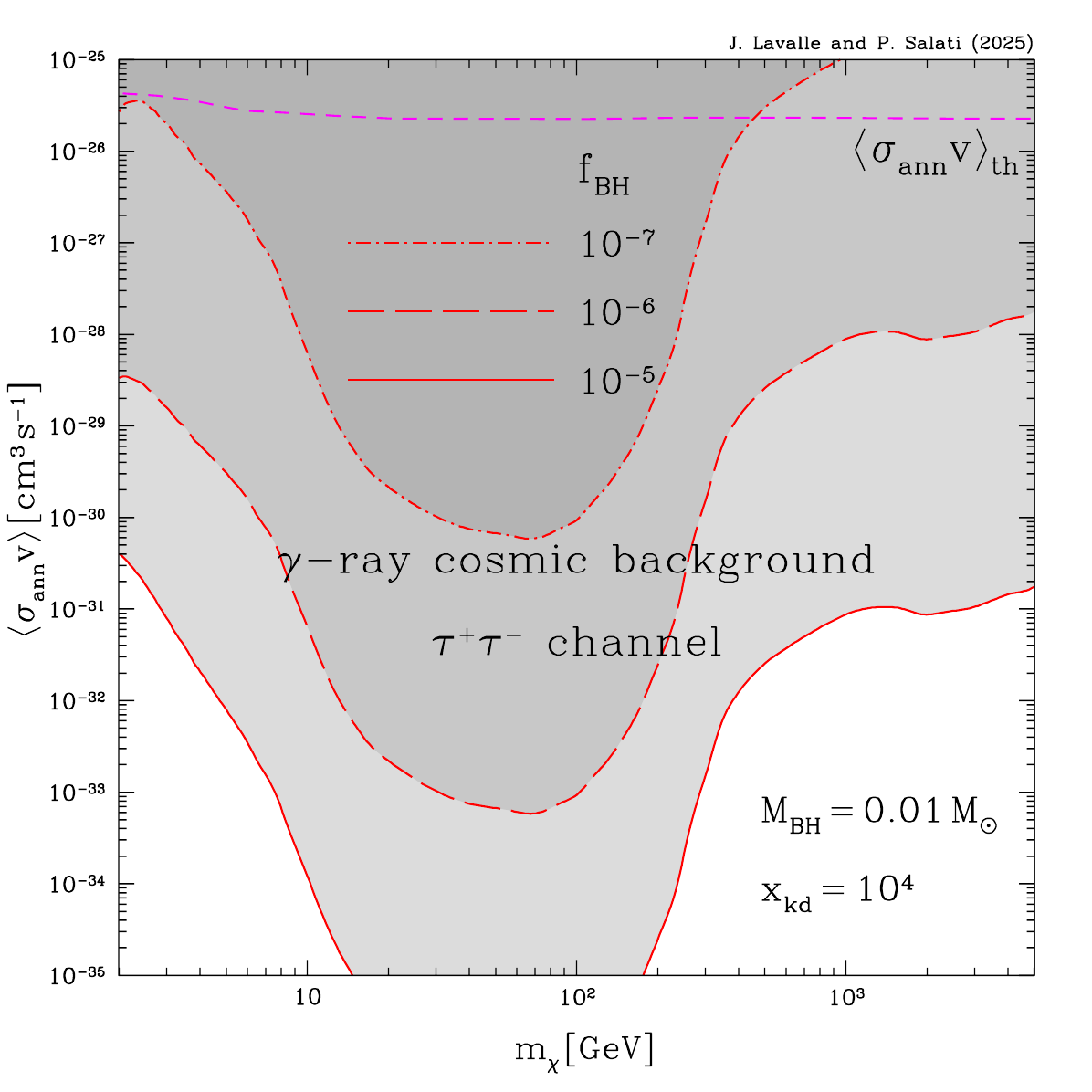}
\includegraphics[width=0.495\textwidth]{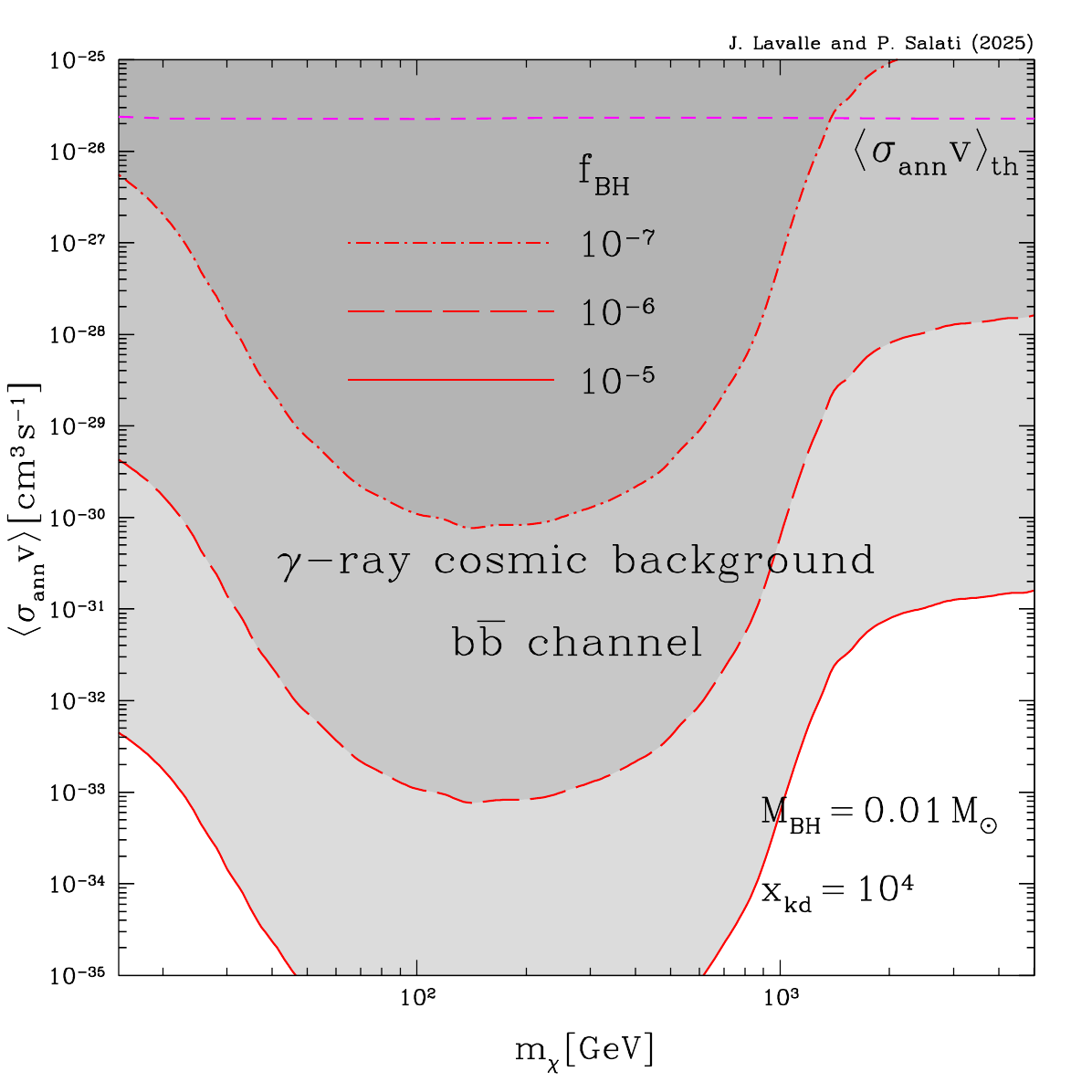}
\caption{The upper bound on the annihilation cross-section set by the cosmological $\gamma$-ray background is plotted as a function of WIMP mass. The ${\tau^{+}}{\tau^{-}}$ and $b \bar{b}$ annihilation channels are respectively displayed in the {\bf left} and {\bf right panels}.
A population of $0.01 \; {\rm M_{\odot}}$ primordial black holes is assumed, which contributes a fraction $f_{\rm BH}$ to the astronomical DM.
The $\gamma$-ray constraint~(\ref{eq:maximum_f_BH_G_BH_a}) has been implemented and translated into the excluded regions shaded in gray. Three values of $f_{\rm BH}$ are presented, each corresponding to a particular shade of gray and contour type as indicated.
The short-dashed magenta horizontal line refers to the thermal value of the annihilation cross-section.
The constraints on $\asv$ are even more stringent than those derived from the CMB.
\label{fig:gamma_sigav_vs_m_chi_tau_bb}}
\end{figure}
%

\vskip 0.1cm
Should the lower limit $\tau_{\rm \, dm}^{\rm min}$ on the lifetime of decaying DM species had been constant, the exclusion plots of Figs.~\ref{fig:excluded_zone_m_chi_sigav_plane} and \ref{fig:gamma_sigav_vs_m_chi_tau_bb} would have been identical, up to a rescaling of the gray shaded regions. However, as is clear in Fig.~3 of Ando and Ishiwata analysis~\cite{AndoEtAl2015}, the lifetime $\tau_{\rm \, dm}^{\rm min}$ varies strongly with $m_{\chi}$, evolving between $10^{26}$ and $10^{28} \; {\rm s}$. As a consequence, the contours of the excluded domains of Fig.~\ref{fig:gamma_sigav_vs_m_chi_tau_bb} exhibit festoon-like features. They also extend very deep downward.
For the  ${\tau^{+}}{\tau^{-}}$ annihilation channel, for instance, the most stringent upper limit on $\asv$ is reached at 70~GeV and is equal to $6 \times 10^{-34} \; {\rm cm^{3} \, s^{-1}}$ for a PBH fraction of a millionth. For the $b \bar{b}$ channel, we get $8 \times 10^{-34} \; {\rm cm^{3} \, s^{-1}}$ at 150~GeV.
We observe that the $\gamma$-ray bounds are even more stringent than those derived from the CMB. However, they are limited to WIMP masses below 5~TeV insofar as Ando and Ishiwata have considered DM decaying species lighter than 10~TeV. In that range, $s$-wave annihilating DM particles are forbidden, unless $f_{\rm BH}$ does not exceed 3.5 billionths and PBHs are less massive than the asteroid mass scale. We still note that our choice of annihilation channels restricts from below the range of WIMP masses that can be probed, typically above the GeV mass scale; in contrast CMB distortions would provide competitive limits for lighter WIMP masses if light leptonic annihilation channels are open ($e^+e^-$ or $\mu^+\mu^-$, see e.g.~\cite{Slatyer2016}).

\vskip 0.1cm
To summarize our investigation of the CMB and $\gamma$-ray bounds on $\asv$, the message which we would like to convey at this stage is the potentially strong perturbation which a tiny admixture of primordial black holes would bring on the phenomenology of thermal DM species. The simplest DM candidate, whose annihilation proceeds through $s$-wave, would be essentially ruled out if there are PBHs in the asteroid mass window or heavier, representing a DM fraction typically larger than a millionth.
We emphasize again that statistically meaningful bounds require a more involved data analysis, which goes beyond the scope of this paper. Here, though, we have provided a very detailed qualitative understanding of those bounds and associated subtleties.

\section{Conclusions}
\label{sec:conclusions}

In this work, we have explored a scenario where the astronomical DM is made of both PBHs and thermal particles processed through freeze-out. The latter collapse around the former during the radiation dominated era to form very dense minispikes whose large densities are the remnants of the primeval universe. We completely elucidated the structure of these ultra-compact haloes in a former publication~\cite{BoudaudEtAl2021}. The DM post-collapse profiles exhibit four asymptotic behaviours which we explained by the interplay between the orbital injection of particles and the allowed phase space for their capture around the central object.

\vskip 0.1cm
Using the results of this analysis, we have constructed in this article a simple and fairly accurate approximation, dubbed soft, for the DM density profile. Once the WIMP mass $m_{\chi}$ and kinetic decoupling temperature $T_{\rm kd}$ are specified, the phase space diagram in the radius vs PBH mass plane is completely set, together with critical masses $M_{1}$ and $M_{2}$. By comparing these values with $M_{\rm BH}$, it is straightforward to determine the asymptotic behaviors of the density $\rho$ as a function of radius $\tilde{r}$. The soft approximation combines these behaviors to yield a density that behaves smoothly and is satisfactorily close to the numerical result. Using it requires just a few lines of code and is extremely fast, in contrast with the numerical integration.

\vskip 0.1cm
We have then focused on the minispike annihilation rate $\Gamma_{\rm BH}$. The on-going annihilation of DM species constantly reshapes the structure of minispikes. We have heuristically described this remodeling by the existence of an annihilation plateau, which results from the beheading of the central parts of minispikes. Its density $\rho_{\rm sat}$ can be associated to a critical value $M_{\rm T}$ of PBH mass that delineates two specific asymptotic behaviors of the density $\rho$.
We have thoroughly analyzed how $\Gamma_{\rm BH}$ depends on WIMP parameters and redshift.
We have explained these dependencies by the interplay between PBH mass $M_{\rm BH}$, transition mass $M_{\rm T}$ and critical mass $M_{1}$, and identifed three asymptotic regimes for the minispike annihilation rate.
By correcting the soft approximation near the surface of minispikes, we have obtained values for $\Gamma_{\rm BH}$ close, almost always, within $\pm 15\%$ to those derived using the lengthy numerical calculation. In the worst case of a 1~MeV particle, the discrepancy between both results is confined in the band between -10\% and +22\%, while $\Gamma_{\rm BH}$ spans more than 66 orders of magnitude.

\vskip 0.1cm
We have finally analyzed how the extragalactic $\gamma$-ray background and CMB angular distortions constrain our mixed DM scenario.
We started by fixing WIMP parameters and concentrated on the fraction $f_{\rm BH}$. Our results confirm what Lacki and Beacom~\cite{LackiEtAl2010} found more than 15~years ago, i.e. primordial black holes contribute either all or almost nothing to the astronomical DM in this particular mixed scenario, and assuming these black holes are essentially more massive than the asteroid mass scale. The former possibility is excluded by a host of cosmological and astrophysical arguments~\cite{GreenEtAl2021,CarrEtAl2021}, except precisely in the asteroid window. Until recently, this region was not constrained by observations. A recent analysis~\cite{Esser:2025pnt} sets a weak upper limit of order $55\%$ on $f_{\rm BH}$. In that context, the qualitative bounds that we have derived are a useful complement---we leave more quantitative determinations of these limits to future work, including estimates of potential late-time spike disruption \cite{RaidalEtAl2024}. The fraction of DM in the form of PBHs is severely constrained, with an upper bound as low as a few billionths in the case of a 200~GeV species annihilating into $b \bar{b}$ pairs with thermal cross-section.

\vskip 0.1cm
One of the most important results of this article is the upper bound set by minispikes on the $s$-wave annihilation cross-section of thermal particles. To our knowledge, this is the first time that such an analysis is performed in the context of mixed DM. The limits set by cosmological observations on $\asv$ are astonishingly stringent, even if the contribution of PBHs to DM is vanishingly small.
This result is exciting insofar as GW observatories are currrently searching for sub-solar compact objects. Should these be discovered, thermal DM species would be partially ruled out, at least those annihilating through $s$-wave.

\vskip 0.1cm
This work is nevertheless a preliminary investigation of a rich and promising topic. It is meant to pave the road for further studies. It may be improved in various ways.
To begin with, PBHs are considered in our approach as point-like masses and treated in the framework of Newtonian mechanics. It would be valuable to re-examine the construction of DM spikes in the light of general relativity. This includes deriving relativistic corrections for the orbits close to the horizon, and disregarding in integral~(\ref{eq:post_collapse_density}) those crossing it.
We have also seen in section~\ref{sec:DM_profiles} that the velocity distribution of DM close to a PBH may be perturbed by its gravity. The vicinity of a black hole is actually heated by the on-going accretion of the ambient plasma onto that object. However, the region within the inner few Schwarzschild radii has little impact on the construction and structure of a DM spike.

\vskip 0.1cm
We may alternatively wonder if accretion modifies the PBH mass during the formation of the spike, i.e. after kinetic decoupling. We have so far assumed that $M_{\rm BH}$ was constant. Using the Bondi-Hoyle-Lyttleton theory~\cite{Edgar:2004mk,MackEtAl2007,Serpico2024}, we can show that this mass undergoes a relative increase equal, up to the numerical factor ${9 \sqrt{3}}/{8} \sim 2$, to the ratio ${r_{\rm S}}/{r_{\rm H}}$, with ${r_{\rm S}}$ the Schwarzschild radius and ${r_{\rm H}}$ the Hubble radius. For the vast majority of black holes formed in the early universe before kinetic decoupling, the former radius is vanishingly smaller than the latter, and $M_{\rm BH}$ is constant during the construction of the spike.

\vskip 0.1cm
The critical mass above which this no longer holds is given, in the phase diagram of Fig.~\ref{fig:phase_diagram}, by the intersection of the dotted short-dashed red line (radius of influence $\tilde{r}_{\rm kd}$) with the vertical line corresponding to the Schwarzschild radius ($\tilde{r}_{\rm S} = 1$).
This mass, of order $10 \; {\rm M_{\odot}}$ in this example, corresponds to the most massive BH that can form at kinetic decoupling. For such an object, the Schwarzschild radius $r_{\rm S}$, the Hubble radius $r_{\rm H}$ and the influence radius $r_{\rm kd}$ are approximately equal to each other at that time.
A 10 ${\rm M_{\odot}}$ black hole accretes matter just after its creation and grows. The formation of the DM spike around it should be considered starting slightly after kinetic decoupling, when $r_{\rm H}$ has sufficiently increased with respect to $r_{\rm S}$ so that the accretion rate ${\dot{M}_{\rm BH}}/{M_{\rm BH}}$ has become negligible. If we require that rate to be less than, say, 10\% for instance, the Hubble radius $r_{\rm H}$ must be at least ten times larger than the Schwarzschild radius $r_{\rm S}$. The plasma temperature corresponding to that value is a factor $\sqrt{10}$ smaller than $T_{\rm kd}$, while the influence radius of the PBH is a factor $10^{2/3}$ larger than the Schwarzschild radius. We recover the conventional situation developed in this article if we consider the DM spike above a reduced influence radius $\tilde{r}_{i} \sim 5$.
For even heavier PBHs, the starting point of the spike formation should also be taken after the creation of the black hole. The same reasoning applies and the DM spike should be considered above $\tilde{r}_{i} \sim 5$.

\vskip 0.1cm
To summarize, the accretion of plasma onto a PBH generates a minute variation of its mass $M_{\rm BH}$ after kinetic decoupling. This is not so for the heaviest objects for which a lower cut-off of order $\tilde{r}_{i} \sim 5$ should be applied on the post-collapse distribution. Given that DM annihilation subsequently erases the density profile, we do not expect accretion to modify the conclusions which we have derived in section~\ref{sec:constraints}.

\vskip 0.1cm
As already mentioned, we have set the efficiency factor $f_{\rm eff}$ equal to $0.1$ in our study on the energy injected in the post-recombination plasma by DM annihilation inside minispikes. We plan to reinvestigate the CMB constrains using the public code {\tt CLASS}. The factor $f_{\rm eff}$ and redshift $z_{\rm CMB}$ of injection will be correctly taken care of.

\vskip 0.1cm
Another project should focuse on the possibility that minispikes may appear today as bright point-like sources in the $\gamma$-ray sky. These ultra-compact haloes behave like the DM skirts potentially surrounding intermediate mass black holes. It is conceivable that the observational programs dedicated to the latter could be applied to search for the former.

\vskip 0.1cm
Finally, we have heuristically used the plateau approximation to describe how annihilations reshape the inner profiles of minispikes as the universe expands. This is correct if DM particles move along circular orbits around the central black hole, which is not quite the case for minispikes.
In that context, some analyses~\cite{Vasiliev:2007vh,Shapiro:2016ypb} indicate that the saturation plateau should be replaced by a weak cusp with slope lying between $0$ and $1/2$. But these studies explore the formation of black holes at the centers of already existing DM clumps during the matter dominated era. In our case, the context is different since DM collapses during the radiation era around already existing PBHs. The initial phase space density is also completely determined in our case, with a velocity distribution far from spherical.
An analysis dedicated to this problem should be undertaken. This would make our bounds on the annihilation cross-section firmer, and would also allow us to deal with $p$-wave annihilating species.

\acknowledgments
We are indebted to Thomas Lacroix and Martin Stref for their participation in the very early stages of this work. We would also like to thank Vivian Poulin, Pearl Sandick, and Tracy Slatyer for inspiring discussions on aspects complementary to this paper. P.S. would like to thank colleagues from LUPM for several visits during which part of this work has been completed.
We also would like to thank the referee for her/his constructive remarks that helped improve the quality of discussions in our paper.
This work has benefited financial support from the ANR project ANR-18-CE31-0006 ({\em GaDaMa}), and from European Union's Horizon 2020 research and innovation program under the Marie Sk\l{}odowska-Curie grant agreement N$^{\rm o}$ 101086085-ASYMMETRY---in addition to recurrent public funding from CNRS, the University of Montpellier, and the University of Savoie-Mont-Blanc.

\newpage
\appendix
\section{Accelerating the code}
\label{append:fast_code}

In order to accelerate the code, a few quantities of interest can be calculated beforehand once and for all, and stored in tables. Once the code runs, these quantities can be interpolated from these tables very quickly and with excellent precision. We give here a few examples of this procedure.

\subsection{DM cosmological density $\rho_{i}$ as a function of radius $\tilde{r}$}
\label{append:rho_i_vs_tilde_r_0}

A key issue in the calculation of the DM minispike density surrounding a PBH is the pre-collapse density profile $\rho_{i}$. DM species located at radius $\tilde{r}$ start to fall onto the black hole at cosmological time $t_{i}$, when they enter inside the gravitational horizon of the object. At that precise moment, the radius $\tilde{r}$ can be identified with the radius $\tilde{r}_{\rm inf}$ of the sphere of influence. The pre-collapse density $\rho_{i}$ is furthermore the DM cosmological density at time $t_{i}$, hence a unique relation between cosmological time $t_{i}$, plasma temperature $T_{i}$, DM cosmological density $\rho_{i}$, plasma density $\rho_{\rm tot}$ and radius of influence $\tilde{r} \equiv \tilde{r}_{\rm inf}$. We can tabulate in advance the relation between $\rho_{i}$ and the radius in the case of a $1 \; {\rm M_{\odot}}$ PBH. As the code runs, we can very easily derive for any given PBH mass the pre-collapse DM density from the radius, or proceed reversely.

\vskip 0.1cm
We first want to derive the pre-collapse DM density $\rho_{i}$ from the radius $\tilde{r}$ around a PBH with mass $m = {M_{\rm BH}}/{{\rm M_{\odot}}}$. Relation~(\ref{eq:definition_r_inf_b}) allows to rescale the radius $\tilde{r}$ into the radius $\tilde{r}_{0}$ corresponding to a $1 \; {\rm M_{\odot}}$ black hole. At time $t_{i}$, the radii of influence $\tilde{r}$ and $\tilde{r}_{0}$ are related to the plasma density $\rho_{\rm tot}$ through
\beq
{\rho_{\rm tot}(t_{i})} \, \tilde{r}_{0}^{3} = {\rho_{\rm tot}(t_{i})} \, m^{2} \, \tilde{r}^{3} = {\rho_{\rm S}^{0}} \,,
\label{eq:definition_r_inf_c}
\eeq
with $\rho_{\rm S}^{0}$ a constant defined in Eq.~(\ref{eq:rho_S_0}). Once the radius $\tilde{r}$ and the PBH mass $m$ are given, the rescaling of the radius for a $1 \; {\rm M_{\odot}}$ object takes the form
\beq
\tilde{r}_{0} = \tilde{r} \; m^{2/3} \,.
\label{eq:r_tilde_to_r0_tilde}
\eeq

\vskip 0.1cm
To proceed any further and calculate the cosmological DM density $\rho_{i}$ corresponding to $\tilde{r}_{0}$, i.e. to the epoch $t_{i}$ and temperature $T_{i}$ at which the radius of influence of a $1 \; {\rm M_{\odot}}$ PBH is actually equal to $\tilde{r}_{0}$, we need to use a few ingredients. To commence, the energy and entropy densities of the primordial plasma at temperature $T_{i}$ are given by
\beq
\rho_{\rm tot}(T_{i}) = \frac{\pi^{2}}{30} \, g_{\rm eff}(T_{i}) \, T_{i}^{4}
\;\;\;\text{and}\;\;\;
{\cal S}_{\rm tot}(T_{i}) = \frac{2 \pi^{2}}{45} \, h_{\rm eff}(T_{i}) \, T_{i}^{3} \,,
\label{eq:plasma_E_and_S_densities}
\eeq
where $g_{\rm eff}$ and $h_{\rm eff}$ denote respectively the effective number of degrees of freedom (spin states) relative to energy and entropy.
As the universe expands adiabatically, the product ${\cal S}_{\rm tot}(T_{i}) \, a^{3}$, where $a$ is the scale factor at time $t_{i}$, is furthermore constant.
Finally, the cosmological DM density $\rho_{i}$ at time $t_{i}$ is related to the present DM density $\rho_{\rm dm}^{0}$ through
\beq
\rho_{i}(t_{i}) = \frac{\rho_{\rm dm}^{0}}{a^{3}} \,,
\eeq
with $a$ the scale factor at time $t_{i}$. The Planck collaboration~\cite{AghanimEtAl2020} has measured $\rho_{\rm dm}^{0}$ to be equal to $2.25401 \times 10^{-30} \; {\rm g \, cm^{-3}}$.
At matter-radiation equality, this translates into a cosmological DM density of
$\rho_{i}^{\rm eq} \simeq 8.88262 \times 10^{-20} \; {\rm g \, cm^{-3}}$.

\vskip 0.1cm
Tossing these ingredients together, we can readily relate the cosmological DM density $\rho_{i}$ at time $t_{i}$ and temperature $T_{i}$ to its value $\rho_{i}^{\rm eq}$ at matter-radiation equality through
\beq
\rho_{i} \equiv \rho_{\rm dm}(t_{i}) = \rho_{i}^{\rm eq}
\left\{ \frac{\tilde{r}_{\rm eq}^{0}}{\tilde{r}_{0}} \right\}^{\! 9/4}
\left\{ \frac{g_{\rm eff}(T_{\rm eq})}{g_{\rm eff}(T_{i})} \right\}^{\! 3/4}
\left\{ \frac{h_{\rm eff}(T_{i})}{h_{\rm eff}(T_{\rm eq})} \right\}.
\label{eq:rho_i_vs_r_tilde_EQ}
\eeq
At matter-radiation equality, the reduced radius of influence of a $1 \; {\rm M}_{\odot}$ black hole is found to be $ \tilde{r}_{\rm eq}^{0} = 2.87404 \times 10^{11}$.
We recover the ${\tilde{r}}^{-9/4}$ scaling of the second relation of Eq.~(\ref{eq:pre_collapse}).
Notice that for any given PBH mass, the ratio ${\tilde{r}}/{\tilde{r}_{\rm eq}}$ is always equal to ${\tilde{r}_{0}}/{\tilde{r}_{\rm eq}^{0}}$.
In our code, we have tabulated the combination
\beq
\frac{\rho_{i}}{\rho_{i}^{\rm eq}} \times \left\{ \frac{\tilde{r}_{0}}{\tilde{r}_{\rm eq}^{0}} \right\}^{\! 9/4} \equiv
\left\{ \frac{g_{\rm eff}(T_{\rm eq})}{g_{\rm eff}(T_{i})} \right\}^{\! 3/4}
\left\{ \frac{h_{\rm eff}(T_{i})}{h_{\rm eff}(T_{\rm eq})} \right\} \equiv {\cal C}_{1}(T_{i}) \,,
\label{eq:def_cal_C}
\eeq
as a function of the ratio ${\tilde{r}_{\rm eq}^{0}}/{\tilde{r}_{0}}$. The factor ${\cal C}_{1}$ is a very slowly growing function of temperature. It actually increases from $1$ at matter-radiation equality up to $3.43074$ at $100 \; {\rm TeV}$.
%
The table can be built once and for all as follows.
\renewcommand{\labelitemi}{$-$}
\begin{itemize}
\item{Scan over $\tilde{r}_{0}$ from ${\tilde{r}_{\rm eq}^{0}}$ downward.}
\item{Use relation~(\ref{eq:definition_r_inf_c}) to derive the plasma energy density $\rho_{\rm tot}$ from $\tilde{r}_{0}$.}
\item{Compute the plasma temperature $T_{i}$ corresponding to this density $\rho_{\rm tot}$. Since $g_{\rm eff}$ is a slowly varying function of the temperature, relation~(\ref{eq:plasma_E_and_S_densities}) can be used together with dichotomy.}
\item{Compute the coefficient ${\cal C}_{1}$ from Eq.~(\ref{eq:def_cal_C}) and store it together with the ratio ${\tilde{r}_{\rm eq}^{0}}/{\tilde{r}_{0}}$.}
\end{itemize}
Once the table is built, our code uses it to derive quickly $\rho_{i}$ as a function of $\tilde{r}$. Here are the steps of the procedure.
\renewcommand{\labelitemi}{$-$}
\begin{itemize}
\item{From the input radius $\tilde{r}$ (corresponding to PBH mass $m$), use Eq.~(\ref{eq:r_tilde_to_r0_tilde}) to derive $\tilde{r}_{0}$ (corresponding to a $1 \; {\rm M_{\odot}}$ black hole).}
\item{Compute the ratio ${\tilde{r}_{\rm eq}^{0}}/{\tilde{r}_{0}}$.}
\item{Interpolate in the table to get the value of ${\cal C}_{1}$.}
\item{Use finally relation~(\ref{eq:def_cal_C}) to get the cosmological DM density $\rho_{i}$.}
\end{itemize}

\subsection{Influence radius $\tilde{r}$ as a function of DM cosmological density $\rho_{i}$}
\label{append:tilde_r_0_vs_rho_i}

Our calculation of the annihilation rate $\Gamma_{\rm BH}$ requires to know the influence radius $\tilde{r}$ at which the DM cosmological density is $\rho_{i}$. The procedure of appendix~\ref{append:rho_i_vs_tilde_r_0} must be inverted. The line of reasoning is still the same: at cosmological time $t_{i}$ and temperature $T_{i}$, the radius of influence of a PBH with mass $m$ is $\tilde{r}$, while the DM cosmological density is $\rho_{i}$. All these quantities are related through expressions given in appendix~\ref{append:rho_i_vs_tilde_r_0}.
We can slightly modify Eq.~(\ref{eq:def_cal_C}) to get
\beq
\frac{\tilde{r}_{0}}{\tilde{r}_{\rm eq}^{0}} \times
\left\{ \frac{\rho_{i}}{\rho_{i}^{\rm eq}} \right\}^{\! 4/9} \equiv
\left\{ \frac{g_{\rm eff}(T_{\rm eq})}{g_{\rm eff}(T_{i})} \right\}^{\! 1/3}
\left\{ \frac{h_{\rm eff}(T_{i})}{h_{\rm eff}(T_{\rm eq})} \right\}^{\! 4/9} \equiv {\cal C}_{2}(T_{i}) \,,
\label{eq:def_cal_C_prime}
\eeq
We can store beforehand the ratio ${\rho_{i}}/{\rho_{i}^{\rm eq}}$ together with coefficient ${\cal C}_{2} \equiv {\cal C}_{1}^{\, 4/9}$ inside a table which we construct through several steps.
\renewcommand{\labelitemi}{$-$}
\begin{itemize}
\item{Scan over $\rho_{i}$ from $\rho_{i}^{\rm eq}$ upward.}
\item{The coefficient ${\cal C}_{2}$ is also a slowly varying function of plasma temperature $T_{i}$. It increases from $1$ at matter-radiation equality up to $1.72962$ at $100 \; {\rm TeV}$.
Since ${\cal C}_{2}$ is bounded by $1$ and $2$, we can use these values to derive the extreme radii $\tilde{r}_{0}^{\rm m}$ and $\tilde{r}_{0}^{\rm M}$ within which lies the true radius $\tilde{r}_{0}$.}
\item{Dichotomy can be used between the radii $\tilde{r}_{0}^{\rm m}$ and $\tilde{r}_{0}^{\rm M}$ to get the solution of Eq.~(\ref{eq:def_cal_C_prime}). Once a value of $\tilde{r}_{0}$ is chosen, derive the corresponding plasma temperature $T_{i}$ as explained in appendix~\ref{append:rho_i_vs_tilde_r_0} and compute ${\cal C}_{2}$. This coefficient increases with increasing $T_{i}$ and decreasing $\tilde{r}_{0}$. At each step of the dichotomy, compare the left-hand side term of relation~(\ref{eq:def_cal_C_prime}) with the value obtained for ${\cal C}_{2}$. If the former is larger than the latter, decrease $\tilde{r}_{0}$. Otherwise, increase it. After a few tens of iterations, the solution of Eq.~(\ref{eq:def_cal_C_prime}) is obtained.}
\item{Store in a table the ratio ${\rho_{i}}/{\rho_{i}^{\rm eq}}$ and the coefficient ${\cal C}_{2}$.}
\end{itemize}
When the code is running, it can quickly derive the radius $\tilde{r}_{0}$ from the cosmological DM density $\rho_{i}$ as follows.
\renewcommand{\labelitemi}{$-$}
\begin{itemize}
\item{From $\rho_{i}$ compute the ratio ${\rho_{i}}/{\rho_{i}^{\rm eq}}$.}
\item{Use the table to interpolate the correct value of ${\cal C}_{2}$.}
\item{Relation~(\ref{eq:def_cal_C_prime}) readily yields the radius $\tilde{r}_{0}$ of the sphere of influence of a $1 \; {\rm M_{\odot}}$ PBH at the exact moment $t_{i}$ for which the cosmological DM density is equal to $\rho_{i}$.}
\item{Invert Eq.~(\ref{eq:r_tilde_to_r0_tilde}) to derive the radius $\tilde{r}$ corresponding to the PBH mass $m$.}
\end{itemize}

\subsection{Calculation of $\upsilon^{\prime}$}
\label{append:upsilon_prime}

To derive the solution $\tilde{r}_{\rm t}$ of Eq.~(\ref{eq:transition_3_2_p_to_9_4_a}), we need to compute the cosmological DM density $\rho_{i}^{\rm t}$ at the exact time $t_{\rm t}$ and corresponding plasma temperature $T_{\rm t}$ at which the radius of influence of the PBH is equal to $\tilde{r}_{\rm t}$.
Inspired by relation~(\ref{eq:rho_i_vs_r_tilde_EQ}), we can express $\rho_{i}^{\rm t}$ as a function of the cosmological DM density $\rho_{i}^{\rm kd}$ at kinetic decoupling. Keeping in mind that the ratio ${\tilde{r}_{\rm t}}/{\tilde{r}_{\rm kd}}$ does not depend on PBH mass $m$, this yields
\beq
\rho_{i}^{\rm t} \equiv \rho_{\rm dm}(t_{\rm t}) = \rho_{i}^{\rm kd}
\left\{ \frac{\tilde{r}_{\rm kd}}{\tilde{r}_{\rm t}} \right\}^{\! 9/4}
\left\{ \frac{g_{\rm eff}(T_{\rm kd})}{g_{\rm eff}(T_{\rm t})} \right\}^{\! 3/4}
\left\{ \frac{h_{\rm eff}(T_{\rm t})}{h_{\rm eff}(T_{\rm kd})} \right\}.
\eeq
The radius $\tilde{r}_{\rm t}$ of transition between the asymptotic regimes of radial infall fulfills the equation
\beq
\frac{\tilde{r}_{\rm t}}{\tilde{r}_{\rm kd}} = \upsilon
\left\{ \frac{g_{\rm eff}(T_{\rm kd})}{g_{\rm eff}(T_{\rm t})} \right\} \!
\left\{ \frac{h_{\rm eff}(T_{\rm t})}{h_{\rm eff}(T_{\rm kd})} \right\}^{\! 4/3} \equiv \upsilon \, {\cal C}_{3}(T_{\rm t}) \,,
\label{eq:transition_3_2_p_to_9_4_c}
\eeq
where $\tilde{r}_{\rm t} = \tilde{r}_{\rm inf}(T_{\rm t})$ and $\tilde{r}_{\rm kd} = \tilde{r}_{\rm inf}(T_{\rm kd})$. The coefficient ${\cal C}_{3}$ is a very slowly increasing function of temperature insofar as
\beq
{\cal C}_{3}(T_{\rm t}) =
\left\{ {{\cal C}_{1}(T_{\rm t})}/{{\cal C}_{1}(T_{\rm kd})} \right\}^{4/3}.
\eeq
The solution of Eq.~(\ref{eq:transition_3_2_p_to_9_4_c}) can be derived with the help of a dichotomy starting from the bounding values $\tilde{r}_{\rm t}^{\rm m} = \tilde{r}_{\rm kd}$ and $\tilde{r}_{\rm t}^{\rm M} = \upsilon \, \tilde{r}_{\rm kd}$.
At the former (latter) bound, the left-hand side term of Eq.~(\ref{eq:transition_3_2_p_to_9_4_c}) is smaller (larger) than its right-hand side term. The plasma temperature is actually equal to $T_{\rm kd}$ at the influence radius $\tilde{r}_{\rm t}^{\rm m}$. Notice also that function ${\cal C}_{3}$ increases with $T_{\rm t}$ and decreases with influence radius $\tilde{r}_{\rm t}$.
Before running the code, we can store in a table the values of $\upsilon^{\prime} \equiv {\tilde{r}_{\rm t}}/{\tilde{r}_{\rm kd}}$ as a function of kinetic decoupling temperature $T_{\rm kd}$. The procedure is as follows.
\renewcommand{\labelitemi}{$-$}
\begin{itemize}
\item{Scan over $T_{\rm kd}$ from $T_{\rm eq} = 7.99240 \times 10^{-10} \; {\rm GeV}$ upward.}
%
\item{At fixed $T_{\rm kd}$, derive the solution $\tilde{r}_{\rm t}$ of Eq.~(\ref{eq:transition_3_2_p_to_9_4_c}) using the above mentioned dichotomy. If the left-hand side term of Eq.~(\ref{eq:transition_3_2_p_to_9_4_c}) is larger than its right-hand side term, decrease $\tilde{r}_{\rm t}$. Otherwise increase it. The solution is obtained after a few tens of iterations.}
\item{Store in a table the pre-calculated values of $T_{\rm kd}$ and $\upsilon^{\prime}$.}
\end{itemize}
Later on, as the code is running, $\upsilon^{\prime}$ can be interpolated whenever $T_{\rm kd}$ is defined. The radius of transition $\tilde{r}_{\rm t}$ between the radial infall asymptotic regimes with slopes $3/2$ and $9/4$ is given by relation~(\ref{eq:transition_3_2_p_to_9_4_b}).

\subsection{Transition between Keplerian and radial infall regimes}
\label{append:Lambda}

The radius $\tilde{r}_{\rm t}$ where the post-collapse DM profile transitions from the Keplerian regime to the radial infall regime with slope $9/4$ is defined by condition~(\ref{eq:transition_3_3_to_9_4_a}). It can be derived as the solution of Eq.~(\ref{eq:transition_3_3_to_9_4_b}) where the parameters of the model, i.e. the DM mass $m_{\chi}$ and kinetic decoupling temperature $T_{\rm kd}$ as well as the PBH mass $m$, come into play in the definition of parameter $\Lambda$.
To save time while the code is running, we can calculate in advance $\tilde{r}_{\rm t}$ as a function of $\Lambda$ by solving
\beq
\frac{\tilde{r}_{\rm t}}{\tilde{r}_{\rm eq}} =
\Lambda \,
\left\{ \frac{g_{\rm eff}(T_{\rm eq})}{g_{\rm eff}(T_{\rm t})} \right\} \!
\left\{ \frac{h_{\rm eff}(T_{\rm t})}{h_{\rm eff}(T_{\rm eq})} \right\}^{\! 4/3} \equiv
\Lambda \, {\cal C}_{4}(T_{\rm t}) \,.
\label{eq:transition_3_3_to_9_4_c}
\eeq
The function ${\cal C}_{4} \equiv {\cal C}_{1}^{\, 4/3}$ is an increasing function of $T_{\rm t}$.

\vskip 0.1cm
In Eq.~(\ref{eq:transition_3_3_to_9_4_c}), the radius $\tilde{r}_{\rm t}$ and temperature $T_{\rm t}$ are not independent. Actually the temperature of the plasma is $T_{\rm t}$ at the cosmological time $t_{\rm t}$ when the radius of influence of the PBH is precisey equal to~$\tilde{r}_{\rm t}$. As a consequence of Eq.~(\ref{eq:definition_r_inf_a}), the larger the temperature $T_{\rm t}$, the larger the plasma density $\rho_{\rm tot}$ and the smaller the radius $\tilde{r}_{\rm t}$. In Eq.~(\ref{eq:transition_3_3_to_9_4_c}), increasing the left-hand side term yields a decrease of the right-hand side term through $T_{\rm t}$. The solution of~(\ref{eq:transition_3_3_to_9_4_c}) can therefore be reached through a dichotomy. We notice that the coefficient ${\cal C}_{4}$ increases from $1$ at $T_{\rm eq}$, up to $5.17429$ at $100 \; {\rm TeV}$.
At fixed $\Lambda$, the solution $\tilde{r}_{\rm t}$ lies between the bounds
$\tilde{r}_{\rm t}^{\rm m} \equiv \Lambda \, \tilde{r}_{\rm eq}$ and, say,
$\tilde{r}_{\rm t}^{\rm M} \equiv 6 \Lambda \, \tilde{r}_{\rm eq}$.
We store the results as follows
\renewcommand{\labelitemi}{$-$}
\begin{itemize}
\item{Scan over $\Lambda$ from $1$ downward to $10^{-20}$ to reach a plasma temperature of $\sim 100 \; {\rm TeV}$.}
\item{For each value of $\Lambda$, use a dichotomy starting from the bounds $\tilde{r}_{\rm t}^{\rm m}$ and $\tilde{r}_{\rm t}^{\rm M}$. If the left-hand side term of Eq.~(\ref{eq:transition_3_3_to_9_4_c}) is larger than its right-hand side term, decrease $\tilde{r}_{\rm t}$. Otherwise increase it. The solution $\tilde{r}_{\rm t}$ is obtained after a few tens of iterations.}
\item{In a table, store in advance the values of $\Lambda$ and ${\cal C}_{4}$.}
\end{itemize}
As the code is running, the mass $m_{1}$ can be calculated as soon as $T_{\rm kd}$ and $x_{\rm kd}$ are defined. For each PBH mass $m$, $\Lambda$ can be readily computed. The radius $\tilde{r}_{\rm t}$ of transition between the Keplerian and radial infall (slope $9/4$) regimes can be derived as follows.
\renewcommand{\labelitemi}{$-$}
\begin{itemize}
\item{Interpolate from the table the value of ${\cal C}_{4}$ corresponding to $\Lambda$.}
\item{For a PBH with mass $m$, the radius $\tilde{r}_{\rm t}$ is given by $\Lambda \, {\cal C}_{4} \, \tilde{r}_{\rm eq}^{0} \, m^{-2/3}$.}
\end{itemize}

\section{Mass of the DM spike $M_{\rm halo}$}
\label{append:M_halo_vs_M_BH}

\begin{figure}[h!]
\centerline{
\includegraphics[width=0.70\columnwidth]{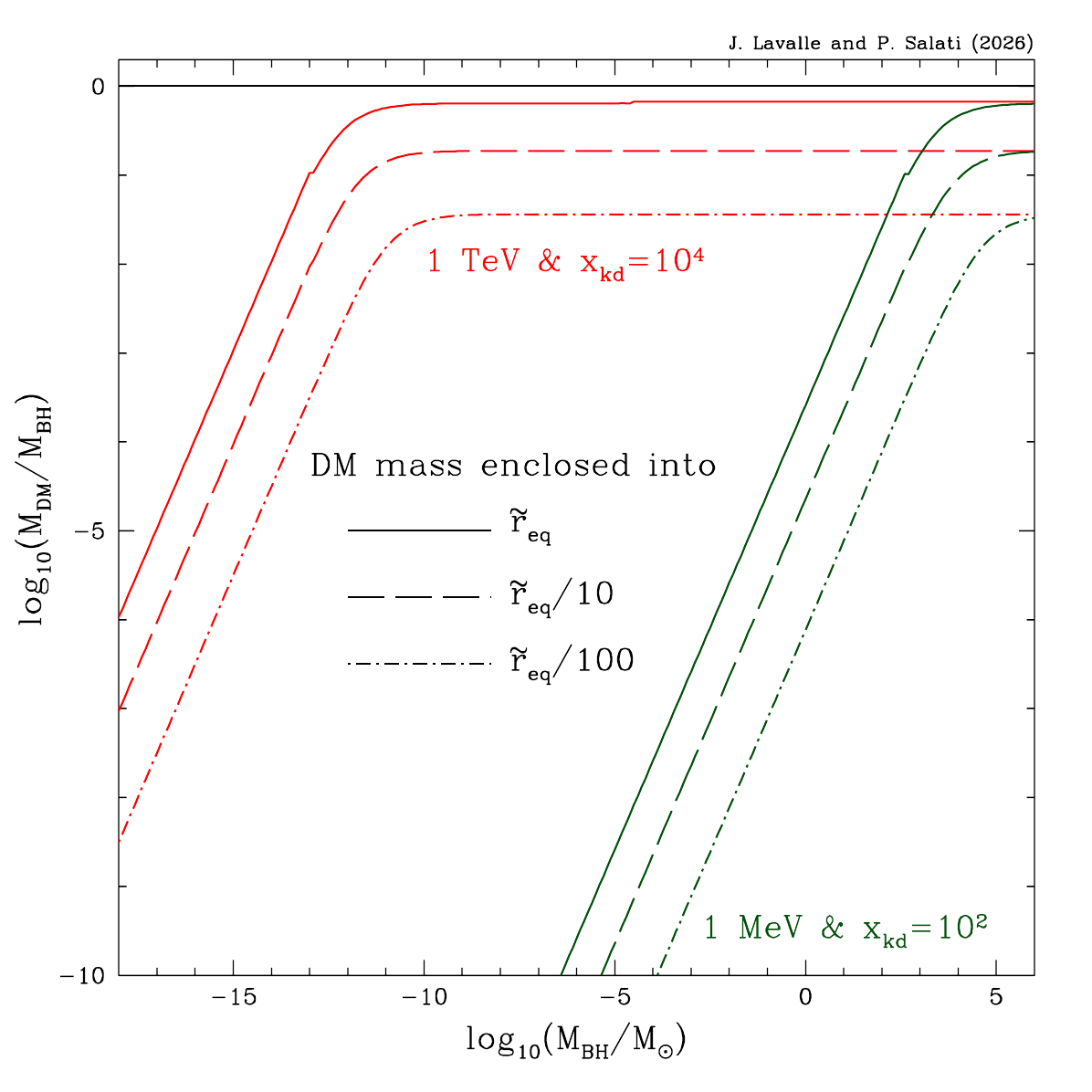}}
\vskip -0.5cm
\caption{\small
DM halo to PBH mass ratio as a function of PBH mass $M_{\rm BH}$. Two extreme configurations are considered. The case of a 1~TeV species with $x_{kd} = 10^{4}$ is plotted in red, while the case of a 1~MeV species with $x_{kd} = 10^{2}$ is plotted in green.
The DM halo mass is respectively integrated up to radius $\tilde{r}_{\rm eq}$ (solid), i.e. up to the surface of the spike, and also up to $\tilde{r}_{\rm eq}/10$ (long-dashed) and $\tilde{r}_{\rm eq}/100$ (dotted short-dashed).
For the post-collapse density, we use the soft approximation with corrections~(\ref{eq:correction_soft}).
}
\label{fig:DM_on_BH}
\end{figure}

As recalled in section~\ref{sec:intro}, the post-collapse DM density profiles are derived assuming that the local dynamics close to the BH is driven by the BH mass only. To check if this condition holds, we have plotted in Fig.~\ref{fig:DM_on_BH} the ratio of halo mass to PBH mass $M_{\rm halo}/M_{\rm BH}$ as a function of PBH mass $M_{\rm BH}$. We have considered two extreme configurations for WIMP mass $m_{\chi}$ and kinetic decoupling parameter $x_{\rm kd}$. The DM halo mass $M_{\rm halo}$ has been integrated up to different radii.
We first remark that all curves feature the same behavior and clearly exhibit two regimes, depending on $M_{\rm BH}$ being smaller or larger than the critical value $M_{1}$.
The halo mass is obtained by integrating the post-collapse DM density, at matter-radiation equality, from the Schwarzschild radius $r_{\rm S}$ up to the surface of the spike at $r_{\rm eq}$. DM annihilations have been switched off. We get
\beq
M_{\rm halo} = 4 \pi r_{\rm S}^{3} \, {\displaystyle \int_{1}^{\tilde{r}_{\rm eq}}} \tilde{r}^{3} \, {\rm d}{\ln}\tilde{r} \, \rho(\tilde{r}) \,.
\eeq
This integral is dominated by large radii. Most of the mass lies on the outskirts of the spike, the density slope being either $3/2$ or $9/4$.
%
%
Several remarks are in order.

\vskip 0.1cm
\noindent
$\bullet$
For $M_{\rm BH} \leq M_{1}$, most of the DM distribution is in the Keplerian regime as showed in Fig.~\ref{fig:phase_diagram}. The density profile is universal, with slope $3/2$. The halo mass scales like
\beq
M_{\rm halo} \propto  r_{\rm S}^{3} \,  \tilde{r}_{\rm eq}^{3/2} \propto m^{3} m^{-1} \propto M_{\rm BH}^{2} \,.
\eeq
The ratio $M_{\rm halo}/M_{\rm BH}$ is proportional to $M_{\rm BH}$. It is always smaller than $0.1$ and vanishes completely for light PBHs. Assumption (ii) is plainly justified. Notice that the asteroid window falls mostly in this regime.

\vskip 0.1cm
\noindent
$\bullet$
Such is not completely the case for $M_{\rm BH} \geq M_{1}$. A new regime sets in, where the ratio $M_{\rm halo}/M_{\rm BH}$ converges asymptotically toward a constant value, reached for $M_{\rm BH} \geq 10^{3} \, M_{1}$.
In the outer regions of the spike, the post-collapse density results from the radial infall of layers taking place between kinetic decoupling and matter-radiation equality. The density scales like $({\tilde{r}_{\rm eq}}/{\tilde{r}})^{9/4}$. The halo mass is given by
\beq
M_{\rm halo} \propto  r_{\rm S}^{3} \,  \tilde{r}_{\rm eq}^{3} \propto m^{3} m^{-2} \propto M_{\rm BH} \,,
\eeq
and the ratio $M_{\rm halo}/M_{\rm BH}$ is constant.
We can go a step further and integrate the halo mass up to a fraction $\alpha$ of the spike radius $\tilde{r}_{\rm eq}$. We expect $M_{\rm halo}$ to scale like $\alpha^{3/4}$ and get numerically a DM halo to PBH mass ratio respectively equal to $0.0364$, $0.188$ and $0.67$ as $\alpha$ is increased from $0.01$ to $1$ in Fig.~\ref{fig:DM_on_BH}. For heavy PBHs, the total DM halo mass $M_{\rm halo}$ is then a substantial fraction of PBH mass.
\renewcommand{\labelitemi}{$-$}
\begin{itemize}
\item{
We first note that the former does not exceed the latter.
}
\item{
We then observe that a WIMP orbiting the central object feels the mass $M_{\rm BH}$ as soon as it dives below the surface of the spike. At radius $\tilde{r}$, the halo mass has actually decreased by a factor $({\tilde{r}}/{\tilde{r}_{\rm eq}})^{3/4}$.
In this regime, the construction of the spike results from the radial infall of DM. As cosmic time goes on, more and more outer layers collapse onto the central object. During most of its construction, the halo mass is negligible with respect to $M_{\rm BH}$.
}
\item{The situation changes during the last stages of the spike formation, just before matter-radiation equality.
Species injected above ${\tilde{r}_{\rm eq}}/{10}$ see a larger mass (by a factor varying between $1.19$ and $1.67$) then those captured previously, i.e. closer to the PBH. These latecomers spend less time orbiting the black hole and crossing its spike than if they felt only the central mass $M_{\rm BH}$. At the apocenters of theirs trajectories, the gravitational pull is actually enhanced by the attraction from the halo. Insofar as their orbital period $T_{\rm orb}$ decreases, their contributions $\propto {2 \mathrm{d}t}/{T_{\rm orb}}$ to the post-collapse density inside the layer of radius $\tilde{r}$ and thickness $\mathrm{d} \tilde{r}$ change. Applying crudely Kepler's third law of planetary motion, we infer a relative increase at most equal to
\beq
{\displaystyle \frac{T_{\rm orb}(M_{\rm BH})}{T_{\rm orb}(M_{\rm BH} + M_{\rm halo})}} \simeq
\left\{ {\displaystyle \frac{M_{\rm BH} + M_{\rm halo}}{M_{\rm BH}}} \right\}^{\! 1/2} = 1.3 \,.
\eeq
This effect is mitigated by the decreasing crossing time $\mathrm{d}t$ since the radial velocity also increases. In particular, these latecomers contribute the same post-collapse density close to their injection radii, i.e. at the apocenters of their trajectories.
}
\item{In the depths of the spike, below its surface, contributions from the latecomers are already very small, since these species remain most of the time far from the PBH. We then expect minute variations of the post-collapse density in these regions. We conclude that assumption (ii) is also justified here.
}
\end{itemize}

\vskip 0.1cm
\noindent
All in all, we conclude that the local dynamics close to the BH is driven by the BH mass only. The mass $M_{\rm halo}$ is negligible with respect to $M_{\rm BH}$ for PBHs lighter than $M_{1}$. For heavier black holes, the mass of the spike amounts to a significant fraction (67\%) of $M_{\rm BH}$ only close to the surface, an inessential region for DM annihilations. Besides, the post-collapse density in this case undergoes everywhere a negligible variation when $M_{\rm halo}$ is taken into account on top of $M_{\rm BH}$.

\section{Effect of the Schwarzschild radius on the annihilation rate $\Gamma_{\rm BH}$}
\label{append:more_correct_Gamma_BH}

In this appendix, we refine the results derived in section~\ref{subsec:Gamma_BH_behavior}. The integral ${\cal J}_{\rm BH}$ defined in Eq.~(\ref{eq:Gamma_BH_c}) is now perfomed from the Schwarzschild radius of the central black hole by setting the lower bound $\tilde{r}_{\rm min}$ equal to $1$. The new expressions for ${\cal K}_{\rm BH}$ are collected in section~\ref{append:more_correct_Gamma_BH_slope_0} if a saturation plateau is assumed in the inner parts of minispikes, and in section~\ref{append:more_correct_Gamma_BH_slope_1_on_2} if a weak cusp with slope $1/2$ is assumed instead, as some numerical simulations suggest~\cite{Vasiliev:2007vh,Shapiro:2016ypb}.

\subsection{Slope $0$ and saturation plateau}
\label{append:more_correct_Gamma_BH_slope_0}

We use the saturation density and the PBH mass to classify the various configurations.

\vskip 0.1cm
\noindent $\bullet$ {\bf $\rho_{\rm sat} \leq \rho_{\rm B_{1}}$}
\renewcommand{\labelitemi}{$-$}
\begin{itemize}
\item{$M_{\rm BH} \leq M_{\rm T}$
\beq
{\cal K}_{\rm BH} = 1 - \frac{1}{\tilde{r}_{\rm T}^{3}} + 2 \ln \! \left\{ \frac{M_{\rm T}}{M_{\rm BH}} \right\}.
\label{eq:K_BH_small_mass_b_rS_on}
\eeq
}
\item{$M_{\rm T} \leq M_{\rm BH}$
\beq
{\cal K}_{\rm BH} = \left\{ \frac{M_{\rm T}}{M_{\rm BH}} \right\}^{\! 2}
\left\{ 1 - \frac{1}{\tilde{r}_{\rm eq}^{3}} \right\}
\;\text{if}\;
\tilde{r}_{\rm eq} \geq 1
\;\text{and $0$ otherwise.}
\label{eq:K_BH_large_mass_b_rS_on}
\eeq
For super-heavy black holes, the radius of the minispike may be less than $1$. The entire DM distribution vanishes inside the black hole horizon. This occurs for $\tilde{r}_{\rm eq} \leq 1$ which, by virtue of Eq.~(\ref{eq:definition_r_tilde_eq}), translates into
\beq
m \geq \left( \tilde{r}_{\rm eq}^{0} \right)^{3/2}
\;\;\;\text{or}\;\;\;
M_{\rm BH} \geq 1.54 \times 10^{17} \; {\rm M_{\odot}} \,.
\label{eq:M_BH_maximum}
\eeq
This critical value corresponds appproximately to the heaviest black hole that can form at matter-radiation equality. The mass $M_{\rm c}$ enclosed inside the causal horizon at that epoch is actually given by
\beq
M_{\rm c} = \frac{4}{3} \pi r_{\rm c}^{3 \,} \rho_{\rm tot}(t_{\rm eq})
\;\;\;\text{where}\;\;\;
r_{\rm c} =  c_{\,} t_{\rm eq} \,.
\eeq
We get $4.98 \times 10^{16} \; {\rm M_{\odot}}$, i.e. a factor 3 below the PBH mass previously derived.
We conclude that objects for which $\tilde{r}_{\rm eq} \leq 1$ cannot exist at matter-radiation equality. Their formation would violate causality.
}
\end{itemize}

\vskip 0.1cm
\noindent $\bullet$ {\bf $\rho_{\rm B_{1}} \leq \rho_{\rm sat} \leq \rho_{\rm B_{2}}$}
\renewcommand{\labelitemi}{$-$}
\begin{itemize}
\item{$M_{\rm BH} \leq M_{1}$
\beq
{\cal K}_{\rm BH} = 1 - \frac{1}{\tilde{r}_{\rm T}^{3}} + 3 \ln \! \left\{ \frac{\tilde{r}_{\rm eq}}{\tilde{r}_{\rm T}} \right\}.
\label{eq:K_BH_small_mass_a_rS_on}
\eeq
}
\item{$M_{1} \leq M_{\rm BH} \leq M_{\rm T}$
\beq
{\cal K}_{\rm BH} = 3 - \frac{1}{\tilde{r}_{\rm T}^{3}} + 3 \ln \! \left\{ \frac{\tilde{r}_{\rm J}}{\tilde{r}_{\rm T}} \right\} -
2 \left\{ \frac{\tilde{r}_{\rm J}}{\tilde{r}_{\rm eq}} \right\}^{\! 3/2}.
\label{eq:K_BH_medium_mass_rS_on}
\eeq
}
\item{$M_{\rm T} \leq M_{\rm BH}$\\
Depending on the values of $\tilde{r}_{\rm H}$ and $\tilde{r}_{\rm E} \equiv \tilde{r}_{\rm eq}$ with respect to $1$ (Schwarzschild radius), three possibilities can be distinguished. As long as $1 \leq \tilde{r}_{\rm H}$, we get
\beq
{\cal K}_{\rm BH} = \left\{ \frac{M_{\rm T}}{M_{\rm BH}} \right\}^{\! 2}
\left\{ 3 - \frac{1}{\tilde{r}_{\rm H}^{3}} - 2 \left( {\tilde{r}_{\rm H}}/{\tilde{r}_{\rm E}} \right)^{3/2} \right\}.
\label{eq:K_BH_large_mass_a_rS_on}
\eeq
For heavier black holes, $\tilde{r}_{\rm H}$ becomes smaller than $1$ while the surface of the minispike lies outside the horizon. This yields
\beq
{\cal K}_{\rm BH} = 2 \left\{ \frac{M_{\rm T}}{M_{\rm BH}} \right\}^{\! 2}
\tilde{r}_{\rm H}^{3/2}
\left\{ 1 - \frac{1}{\tilde{r}_{\rm eq}^{3/2}} \right\}.
\label{eq:K_BH_very_large_mass_a_rS_on}
\eeq
The integral ${\cal K}_{\rm BH}$ vanishes whenever $\tilde{r}_{\rm eq} \leq 1$. This extreme case corresponds to the super-heavy objects discussed above.
}
\end{itemize}

\subsection{Slope $1/2$}
\label{append:more_correct_Gamma_BH_slope_1_on_2}

If the saturation plateau is replaced by a weak cusp with slope $1/2$, the previous expressions for ${\cal K}_{\rm BH}$ are slightly modified as follows.

\vskip 0.1cm
\noindent $\bullet$ {\bf $\rho_{\rm sat} \leq \rho_{\rm B_{1}}$}
\renewcommand{\labelitemi}{$-$}
\begin{itemize}
\item{$M_{\rm BH} \leq M_{\rm T}$
\beq
{\cal K}_{\rm BH} =
\frac{3}{2} \left\{ 1 - \frac{1}{\tilde{r}_{\rm T}^{2}} \right\} +
2 \ln \! \left\{ \frac{M_{\rm T}}{M_{\rm BH}} \right\}.
\label{eq:K_BH_small_mass_b_rS_on_WC}
\eeq
}
\item{$M_{\rm T} \leq M_{\rm BH}$
\beq
{\cal K}_{\rm BH} = \frac{3}{2} \left\{ \frac{M_{\rm T}}{M_{\rm BH}} \right\}^{\! 2}
\left\{ 1 - \frac{1}{\tilde{r}_{\rm eq}^{2}} \right\}
\;\text{if}\;
\tilde{r}_{\rm eq} \geq 1
\;\text{and $0$ otherwise.}
\label{eq:K_BH_large_mass_b_rS_on_WC}
\eeq
}
\end{itemize}

\vskip 0.1cm
\noindent $\bullet$ {\bf $\rho_{\rm B_{1}} \leq \rho_{\rm sat} \leq \rho_{\rm B_{2}}$}
\renewcommand{\labelitemi}{$-$}
\begin{itemize}
\item{$M_{\rm BH} \leq M_{1}$
\beq
{\cal K}_{\rm BH} =
\frac{3}{2} \left\{ 1 - \frac{1}{\tilde{r}_{\rm T}^{2}} \right\} +
3 \ln \! \left\{ \frac{\tilde{r}_{\rm eq}}{\tilde{r}_{\rm T}} \right\}.
\label{eq:K_BH_small_mass_a_rS_on_WC}
\eeq
}
\item{$M_{1} \leq M_{\rm BH} \leq M_{\rm T}$
\beq
{\cal K}_{\rm BH} = \frac{7}{2} - \frac{3}{2 \tilde{r}_{\rm T}^{2}} +
3 \ln \! \left\{ \frac{\tilde{r}_{\rm J}}{\tilde{r}_{\rm T}} \right\} -
2 \left\{ \frac{\tilde{r}_{\rm J}}{\tilde{r}_{\rm eq}} \right\}^{\! 3/2}.
\label{eq:K_BH_medium_mass_rS_on_WC}
\eeq
}
\item{$M_{\rm T} \leq M_{\rm BH}$\\
As for the saturation plateau, three cases can be identified. We start with $1 \leq \tilde{r}_{\rm H}$ for which we get
\beq
{\cal K}_{\rm BH} = \left\{ \frac{M_{\rm T}}{M_{\rm BH}} \right\}^{\! 2}
\left\{ \frac{7}{2} - \frac{3}{2 \tilde{r}_{\rm H}^{2}} - 2 \left( {\tilde{r}_{\rm H}}/{\tilde{r}_{\rm E}} \right)^{3/2} \right\}.
\label{eq:K_BH_large_mass_a_rS_on_WC}
\eeq
For heavier black holes, $\tilde{r}_{\rm H}$ is smaller than $1$ while the minispike radius $\tilde{r}_{\rm E} \equiv \tilde{r}_{\rm eq}$ exceeds the Schwarzschild radius of the central object. This yields
\beq
{\cal K}_{\rm BH} = 2 \left\{ \frac{M_{\rm T}}{M_{\rm BH}} \right\}^{\! 2}
\tilde{r}_{\rm H}^{3/2}
\left\{ 1 - \frac{1}{\tilde{r}_{\rm eq}^{3/2}} \right\}.
\label{eq:K_BH_very_large_mass_a_rS_on_WC}
\eeq
In the last case where $\tilde{r}_{\rm eq} \leq 1$, the integral ${\cal K}_{\rm BH}$ vanishes.
}
\end{itemize}

\section{Qualitative comparisons with other results}
\label{append:comparison}
%
\begin{figure}[h!]
\centering
\includegraphics[width=0.495\textwidth]{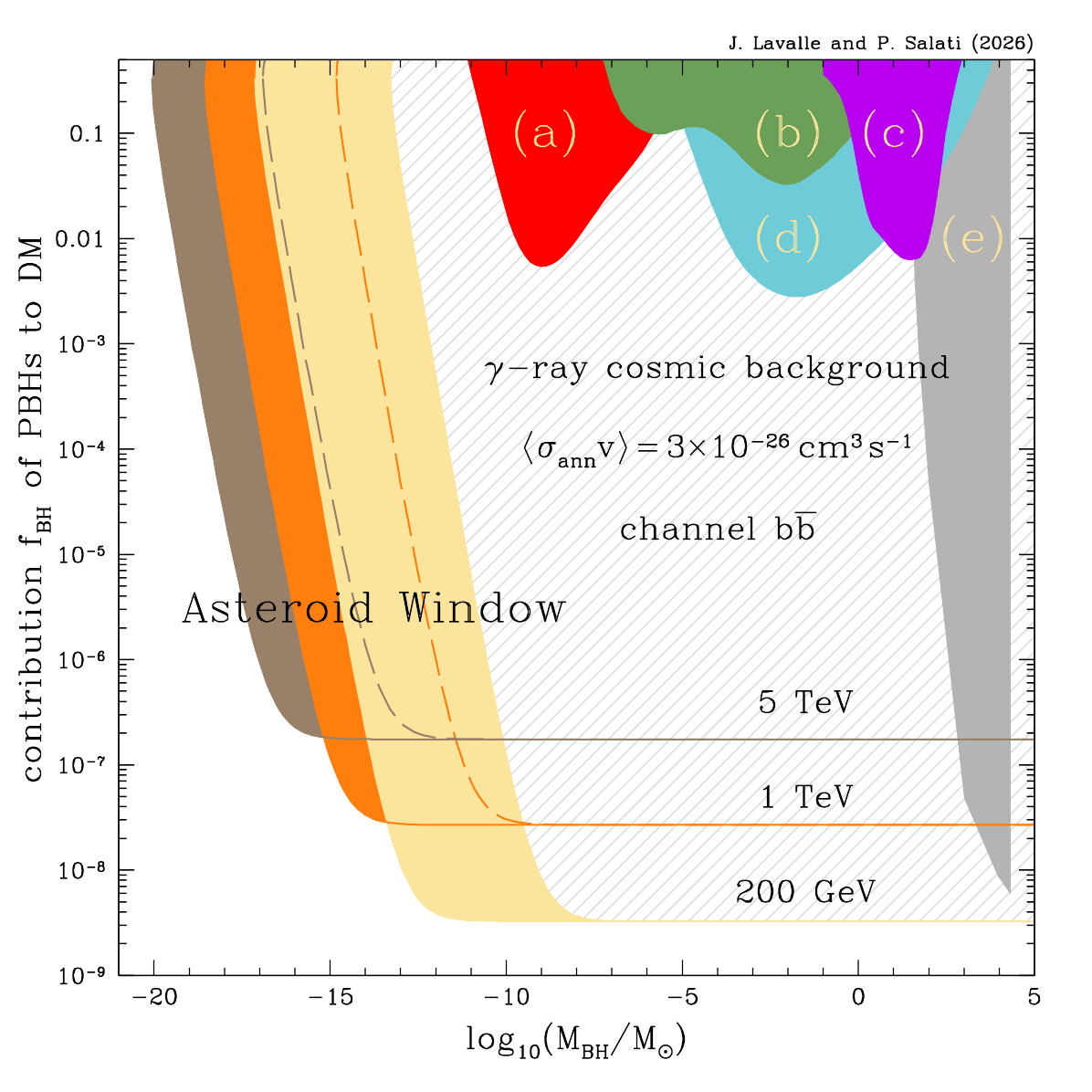}
\includegraphics[width=0.495\textwidth]{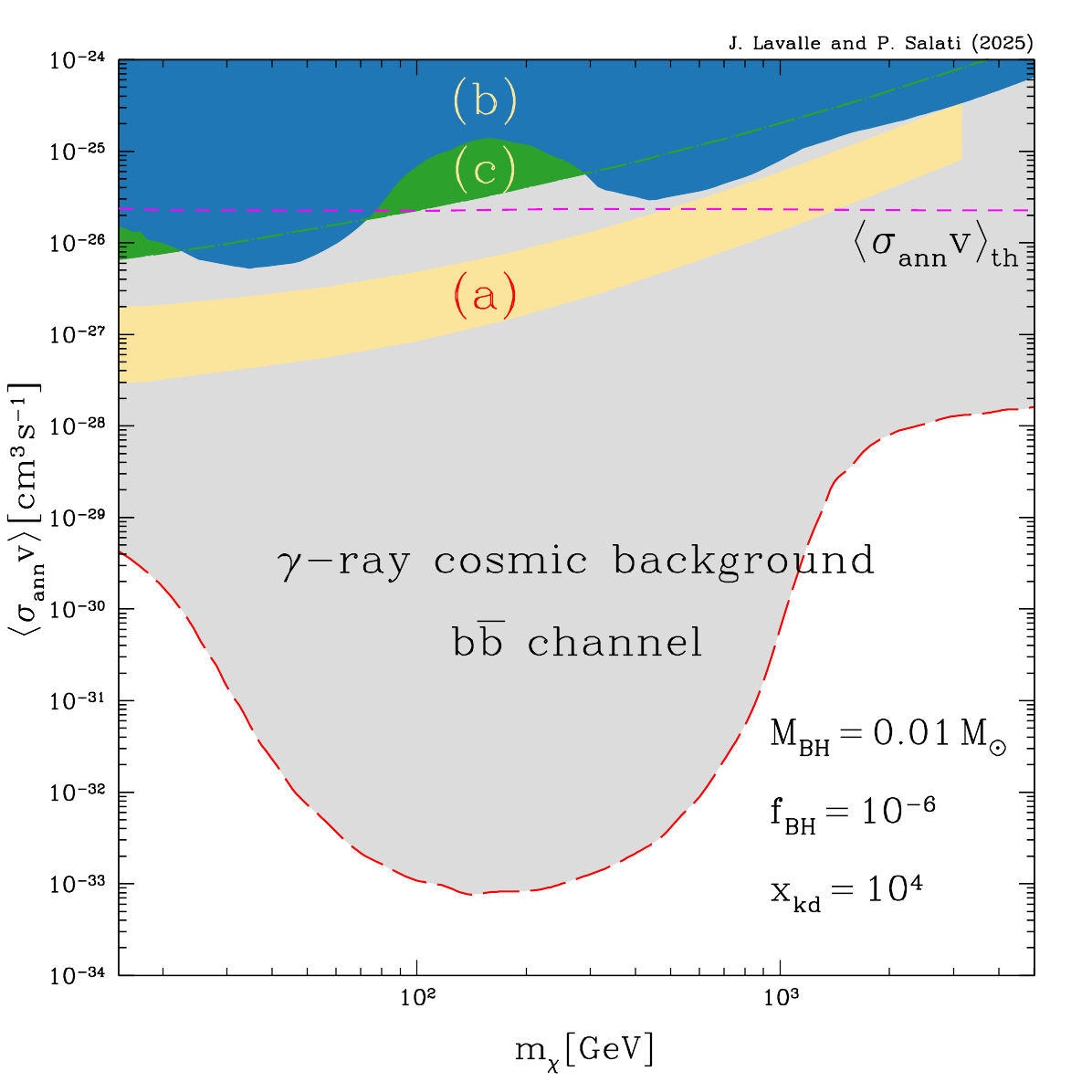}
\caption{
In the {\bf left panel}, our gamma-ray bounds on the PBH fraction are compared with other probes. The excluded regions (a), (b), (c), (d) and (e) correspond respectively to HSC~\cite{CroonEtAl2020}, OGLE~\cite{MrozEtAl2024}, EROS2~\cite{TisserandEtAl2007}, gravitational wave experiments~\cite{BoybeyiEtAl2024} and CMB distortion constraints based on accretion \cite{SerpicoEtAl2020}.
PBHs set upper limits on $f_{\rm BH}$ orders of magnitude more stringent than these experiments. For heavy black holes, the CMB distortion constraints are comparable.
In the {\bf right panel}, our limits on the annihilation cross-section $\asv$ are compared with other results. 
The yellow band (a) are radio constraints derived from the EMU survey~\cite{Regis:2021glv}. The blue region (b) is excluded by measurements of the cosmic-ray antiproton flux~\cite{Calore:2022stf}. We also display in domain (c) bounds from dwarf spheroidal galaxies (dSph) derived with Fermi-LAT~\cite{Fermi-LAT:2016uux}. The limits set on WIMP properties by a very modest population of sub-solar PBHs are here again much more constraining than the conventional bounds.
\label{fig:comparison_of_bounds}}
\end{figure}
%
This appendix section is dedicated to a qualitative comparison of our approximate bounds with other results in the literature --- this comparison is essentially condensed in Fig.~\ref{fig:comparison_of_bounds}. To this purpose, we recycle the left panel of Fig.~\ref{fig:gamma_CMB_bounds_on_f_BH} showing the extragalactic gamma-ray limit on the fraction of PBHs assuming fixed properties for WIMPs, and the right panel of Fig.~\ref{fig:gamma_sigav_vs_m_chi_tau_bb}, showing bounds on the annihilation cross section of WIMPs assuming fixed properties for PBHs.
In the left panel of Fig.~\ref{fig:comparison_of_bounds}, we compare our gamma-ray bounds on the PBH fraction with other probes from microlensing surveys (EROS2 \cite{TisserandEtAl2007}, OGLE \cite{MrozEtAl2024}, HSC \cite{CroonEtAl2020}, and combined EROS2-MACHO \cite{BlaineauEtAl2022}), gravitational wave experiments \cite{BoybeyiEtAl2024}, and CMB distortion constraints based on accretion \cite{SerpicoEtAl2020}\footnote{We used the data provided by the PBHbounds Github page \cite{PBHbounds}.} 
In the right panel of Fig.~\ref{fig:comparison_of_bounds}, we compare our gamma-ray bounds on the DM annihilation cros-section with the constraints from a radio survey of the LMC~\cite{Regis:2021glv}, a recent analysis on cosmic-ray antiprotons~\cite{Calore:2022stf} and observations of dwarf spheroidal galaxies (dSph) by Fermi-LAT~\cite{Fermi-LAT:2016uux}.

\vskip 0.1cm
The bounds set by PBHs on thermal DM are orders of magnitude more stringent than those derived in the literature.

\bibliographystyle{JHEP.bst}
\bibliography{PBH_DM_II.bib}
\end{document}